\documentclass[reprint,aip]{revtex4-2}
\usepackage{graphicx}% Include figure files
\usepackage{xcolor}
\usepackage{dcolumn}% Align table columns on decimal point
\usepackage{rotating}
\usepackage{makecell}
\usepackage{enumerate} 
\usepackage{bm}% bold math
\usepackage{bbm}
\usepackage{graphicx,amsmath}
\usepackage{tikz}
\usepackage[version=4]{mhchem}
\usetikzlibrary{quantikz}
\usepackage{subfigure}
\usepackage{booktabs}
\usepackage{siunitx}
\usepackage{multirow}
\usepackage{tabularx}
\usepackage{tikz}
\usepackage{circuitikz}
\usetikzlibrary{patterns}
\usetikzlibrary{ decorations.markings,  decorations.pathmorphing, decorations.pathreplacing, decorations.shapes, arrows.meta}
\newcommand{\Vdot}[1]{\draw[fill] #1 circle[radius=4pt];}

\newcommand{\Vline}[3]{\draw[shift={#1},rotate={#2},decorate,decoration={coil,aspect=0}] (0,0) -- (#3,0) ;}
\newcommand{\cmdot}[2][4]{\draw[shift={#2}, fill={rgb, 255:red, 208; green, 2; blue, 27}, fill opacity=1] (-#1pt,-#1pt) rectangle (4pt,4pt);}

\newcommand{\gfarrow}{ \arrow[thick,scale=-1.2]{Stealth[bend]} }
\newcommand{\gfarrowpos}{.43}

\newcommand{\gfline}[4][50]{\draw[shift={#2},rotate={#3},thick,postaction={decorate},decoration={markings,mark=at position \gfarrowpos with { \gfarrow }}] (0,0) to[out=-#1,in=#1-180] (#4,0);}
\newcommand{\phbbl}[4][50]{\gfline[#1]{#2}{#3}{#4};\draw[shift={#2},rotate={#3},thick,postaction={decorate},decoration={markings, mark=at position \gfarrowpos with { \gfarrow }}] (#4,0) to[out=180-#1,in=#1] (0,0);}
\newcommand{\ppbbl}[4][50]{\gfline[#1]{#2}{#3}{#4};\draw[shift={#2},rotate={#3},thick,postaction={decorate},decoration={markings,mark=at position \gfarrowpos with { \gfarrow }}] (0,0) to[out=#1,in=180-#1] (#4,0);}
\newcommand{\phbblc}[4][50]{\phbbl[#1]{#2}{#3}{#4} 
\pattern [shift={#2},rotate={#3},pattern=north east lines] (0,0) to[out=-#1,in=#1-180] (#4,0) to[in=#1,out=180-#1] cycle; }
\newcommand{\ppbblc}[4][50]{\ppbbl[#1]{#2}{#3}{#4} 
\pattern [shift={#2},rotate={#3},pattern=north east lines] (0,0) to[out=-#1,in=#1-180] (#4,0) to[in=#1,out=180-#1] cycle;}
\newcommand{\gfloop}[3]{\draw[shift={#1},rotate={#2},thick,postaction={decorate},decoration={markings,mark=at position 0.05 with { \gfarrow }}] (#3,0) circle[radius=#3];}

\newcommand{\ii}{\mathbbm{i}}
\newcommand{\hH}{\hat{H}}
\newcommand{\hU}{\hat{U}}
\newcommand{\hV}{\hat{V}}
\newcommand{\hp}{\hat{p}}
\newcommand{\hq}{\hat{q}}
\newcommand{\hr}{\hat{r}}
\newcommand{\hs}{\hat{s}}
\newcommand{\hx}{\hat{x}}
\newcommand{\hy}{\hat{y}}

\newcommand{\tr}{\mathrm{tr}}

\newcommand{\bC}{\mathbf{C}}

\newcommand{\bX}{\mathbf{X}}
\newcommand{\bY}{\mathbf{Y}}

\newcommand{\erias}[2]{\langle#1||#2\rangle}

\newcommand{\tord}{\mathcal{T}}
\newcommand{\Tau}[1]{\tord[#1]}

\newcommand{\thalf}{\frac{T}{2}}

\begin{document}

%\preprint{APS/123-QED}

\title{
A unified diagrammatic formulation of single-reference and multi-reference 
random phase approximations: the particle-hole and particle-particle channels
}
\author{Yuqi Wang}
\author{Wei-Hai Fang}
\author{Zhendong Li}
\email{zhendongli@bnu.edu.cn}
\address{Key Laboratory of Theoretical and Computational Photochemistry, Ministry of Education, College of Chemistry, Beijing Normal University, Beijing 100875, China}

\begin{abstract}
A diagrammatic multi-reference generalization of many-body perturbation theory was recently introduced 
[J. Phys. Chem. Lett., 2025, 16, 3047]. 
   This framework allows us to extend single-reference (SR) Green's function methods defined at the diagrammatic level 
naturally into multi-reference case, 
as previously exemplified by the formulation of multi-reference direct random phase approximation (MR-dRPA) 
and the multi-reference second-order screened exchange approximation (MR-SOSEX). 
In this work, we further elaborate this framework and
use it to develop MR generalizations of two other RPA variants, namely, 
particle-hole (ph) RPA with exchange (MR-RPAx) and particle-particle RPA (MR-ppRPA). 
We define these two MR generalizations by infinite order resummations of the generalized `ring' and `ladder' diagrams 
with antisymmetrized interaction vertices, respectively,
which incorporate the contributions from the active-space connected two-body Green's functions.
As for MR-dRPA, we derive unified sets of equations that hold at both SR and MR levels for RPAx and ppRPA, 
respectively.
We perform numerical studies of prototypical systems using the three MR-RPA methods and 
carry out a perturbative analysis to gain a deeper understanding of their behaviors.
We find that error cancellation between the second and third orders is a key factor for both SR-RPA and MR-RPA. 
In addition, we observe that MR-phRPA (MR-dRPA and MR-RPAx) and MR-ppRPA tend to overestimate and underestimate correlation energies, 
respectively, suggesting that a better accuracy can be achieved by further combining these two channels in the future.
\end{abstract}

\maketitle

\section{Introduction}
Accurate prediction of ground-state energies of strongly correlated electronic systems 
remains a significant challenge in quantum chemistry and many-body physics. 
Traditional single-reference (SR) perturbation theory, 
which employs a quadratic zeroth-order Hamiltonian ($\hH_0$) and a single-determinant reference, 
fails in the presence of strong correlation. 
Such failure motivates the development of multi-reference perturbation theories (MRPT) 
with a multi-determinant reference\cite{park_multireference_2020},
including the second-order complete active space perturbation theory (CASPT2)\cite{andersson_secondorder_1992} 
and second-order $N$-electron valence state perturbation theory (NEVPT2)\cite{angeli_introduction_2001} 
as two of the most popular methods. 
Developing nonperturbative methods that include \emph{infinite order} contributions is highly nontrivial. 
Traditionally,
such methods are mainly developed from a multi-reference coupled cluster (MRCC) perspective 
using time-independent wavefunction formulation\cite{lyakh_multireference_2012,evangelista_perspective_2018,adam2025multireference}.
Recently, we tackled this problem from a different perspective 
by developing a diagrammatic generalization of the traditional many-body perturbation theory (MBPT) 
based on time-dependent Green's functions\cite{fetter_quantum_1971,negele_quantum_1998,martin_interacting_2016} 
for interacting $\hH_0$\cite{wang_generalized_2025}. 
The key idea is to introduce generalized Feynman diagrams, which can involve cumulant (or connected) Green's functions, 
derived from the cumulant expansion of time-ordered many-body Green's functions\cite{negele_quantum_1998}.
This development opens up the possibility of developing multi-reference methods beyond the second order 
by partially resumming certain types of diagrams to infinite order, 
analogous to the single-reference case\cite{fetter_quantum_1971,martin_interacting_2016}.

As a concrete application of this theoretical framework, 
we formulated a multi-reference generalization of the random phase approximation (RPA) 
in terms of diagrammatic resummation. 
In the language of Feynman diagrams, 
the standard single-reference RPA, 
which has been successfully applied in both molecular systems\cite{furche_molecular_2001,zhu_range-separated_2010,eshuis_parameter-free_2011,paier_assessment_2012} 
and condensed-phase systems\cite{harl_cohesive_2008,harl_accurate_2009,lu_ab_2009,ren_exploring_2009,schimka_accurate_2010,harl_assessing_2010,lebegue_cohesive_2010,mittendorfer_graphene_2011,olsen_dispersive_2011,casadei_density-functional_2012,neuhauser_expeditious_2013,casadei_density_2016,schafer_ab_2018}, 
is formulated by resumming `ring' diagrams to infinite order\cite{pines_collective_1952,bohm_collective_1953,gell-mann_correlation_1957,goldstone_derivation_1957,mclachlan_time-dependent_1964,furche_developing_2008,hesselmann_random-phase_2011,ren_random-phase_2012,chen_random-phase_2017}. 
Following the same spirit, 
our multi-reference RPA (MR-RPA) is naturally defined by 
replacing the standard ring diagrams with the generalized ones\cite{wang_generalized_2025}. 
This distinguishes it from other MR generalizations of RPA from different perspectives,
such as the equation of motion (EOM) of excitation operators\cite{chatterjee_excitation_2012,pernal_intergeminal_2014, pastorczak_correlation_2018, pernal_electron_2018,guo_spinless_2024,tucholska_duality_2024} 
and the ring coupled cluster theory\cite{szabados_ring_2017,margocsy_ring_2020}.
This MR-RPA method delivers promising results in the description 
of the bond-breaking processes of prototypical molecules, where single-reference RPA fails miserably. 
The use of diagrams offers the possibility of systematic improvement by adding more diagrams, 
which is demonstrated by the multi-reference generalization of 
the second-order screened exchange (MR-SOSEX) developed in the same work\cite{wang_generalized_2025}.

The MR-RPA previously developed targets the particle-hole (ph) channel and neglects all the exchange terms, 
and thus will be referred to more precisely as multi-reference direct RPA (MR-dRPA) in the present work. 
While it improves single-reference dRPA (SR-dRPA) significantly in the presence of strong correlation, 
it also inherit drawbacks from SR-dRPA, such as the self-interaction error (SIE) due to the lack of correct fermionic antisymmetry, 
and too negative correlation energies\cite{mori-sanchez_failure_2012}.
In the single-reference case, efforts have been made to go beyond dRPA by adding corrections containing exchange terms\cite{gruneis_making_2009, heselmann_third-order_2011,bates_communication_2013, chen_performance_2018, hummel_screened_2019}, 
introducing single excitations\cite{ren_beyond_2011}, 
combining with density functional theories\cite{fuchs_accurate_2002,jiang_random-phase-approximation-based_2007,toulouse_adiabatic-connection_2009,hesselmann_random_2010,toulouse_range-separated_2010,paier_hybrid_2010,heselmann_correct_2011,trushin_toward_2021,riemelmoser_machine_2023}, 
or exploring other RPA variants\cite{klopper_spin_2011,angyan_correlation_2011,heselmann_random-phase-approximation_2012,eshuis_electron_2012,scuseria_particle-particle_2013,van_aggelen_exchange-correlation_2013,van_aggelen_exchange-correlation_2014, mussard_spin-unrestricted_2015,tahir_comparing_2019}. 
Specifically, the RPAx method directly includes exchange terms by using antisymmetrizing Coulomb interactions\cite{mclachlan_time-dependent_1964,angyan_correlation_2011,heselmann_third-order_2011,eshuis_electron_2012,heselmann_random-phase-approximation_2012}, 
which resolves the SIE in one-electron systems. 
Unfortunately, RPAx often either suffers from the triplet instability or gives correlation energies even more negative than dRPA, 
limiting its range of applicability. 
On the other hand, the particle-particle RPA (ppRPA), 
which instead targets the particle-particle and hole-hole channels, 
has been established as a promising alternative to the particle-hole RPA (phRPA)\cite{van_aggelen_exchange-correlation_2013,van_aggelen_exchange-correlation_2014}. 
In contrast to the phRPA variants, ppRPA preserves the correct antisymmetry, 
and does not suffer from instability. 
A series of successful applications of ppRPA has been reported\cite{yang_double_2013,yang_singlettriplet_2015,yang_charge_2017,li_accurate_2024,li_particleparticle_2024,yu_accurate_2025}.
In this work, we develop multi-reference generalizations of RPAx and ppRPA, 
termed as MR-RPAx and MR-ppRPA, respectively, 
via resummations of generalized Feynman diagrams following our previous work\cite{wang_generalized_2025},
and compare their performances against MR-dRPA.

The remaining part of this article is organized as follows. 
We will first derive the expressions for the MR-RPAx and MR-ppRPA correlation energies %using a complete active space (CAS) reference
in Sec. \ref{sec:theory}. 
In addition, to gain a deeper understanding of the performances of different RPA variants, 
we also develop a perturbative analysis of the RPA correlation energies.
In Sec. \ref{sec:results}, 
we apply the MR-RPA methods to prototypical systems to investigate their performances.
Conclusions are drawn in Sec. \ref{sec:conclusion} 
and future prospects on further improving the accuracy are highlighted.

% However, there is still large room for better accuracy, due to the lack of higher order components. One major reason for the difficulty of advancing MRPT is that $\hH_0$ in this case is not quadratic, but interacting (e.g. the Dyall Hamiltonian\cite{dyall_choice_1995}), which makes its inverse not directly available,\cite{sokolov_time-dependent_2016} and causes the failure of Wick's theorem.\cite{wick_evaluation_1950}

\section{Theory}\label{sec:theory}
\subsection{Recapitulation of the
generalized MBPT and MR-dRPA
}

We briefly recapitulate the generalized MBPT and the MR-dRPA formulation introduced in our previous work\cite{wang_generalized_2025}.
We assume the total second-quantized electronic Hamiltonian is partitioned in a general way as
\begin{align}
\hH &= \hH_0 + \hV , \label{eq:partition}\\
\hH_0 &= h_{pq}\hat{p}^\dagger \hat{q} + \frac{1}{2}h_{pr,qs}\hat{p}^\dagger\hat{q}^\dagger \hat{s}\hat{r},
\label{eq:PT-H0}\\
\hV %&= \hV_1 + \hV_2\nonumber\\
&= v_{pq}\hat{p}^\dagger\hat{q} + \frac{1}{2}v_{pr,qs}\hat{p}^\dagger\hat{q}^\dagger\hat{s}\hat{r}\nonumber\\
&= v_{pq}\hat{p}^\dagger\hat{q} + \frac{1}{4}\bar{v}_{pr,qs}\hat{p}^\dagger\hat{q}^\dagger\hat{s}\hat{r},\label{eq:PT-V}
\end{align}
where the Einstein summation convention has been used for repeated indices. 
Here, $h_{pq}$ ($v_{pq}$) is the zeroth- (first-) order one-electron interaction,
$h_{pr,qs}$ ($v_{pr,qs}$) the zeroth- (first-) order two-electron interaction,
and $\bar{v}_{pr,qs}=v_{pr,qs}-v_{ps,qr}$ is the antisymmetrized first-order two-electron interaction.
$\hat{p}^{(\dagger)}$ is the fermionic annihilation (creation) operator for the $p$-th spin-orbital.
The standard MBPT using a quadratic $\hH_0$ corresponds to setting $h_{pr,qs}$ to zero in the above equations.
The energy shift for a non-degenerate ground state,
viz., the difference between the lowest eigenstate energies of $\hH$ and $\hH_0$,
whose ground states are $|\Psi_0\rangle$ and $|\Phi_0\rangle$, respectively, can be written as\cite{negele_quantum_1998}
\begin{align}
\Delta E = \lim_{T\rightarrow\infty}\frac{\ii}{T}\ln \langle\hat{U}(\frac{T}{2},-\frac{T}{2})\rangle_0,
\end{align}
where $\langle \hU(\frac{T}{2},-\frac{T}{2})\rangle_0$ is a shorthand notation for
$\langle\Phi_0|\hU(\frac{T}{2},-\frac{T}{2})|\Phi_0\rangle$,
and $\hat{U}$ is the time-evolution operator in the interaction picture
\begin{align}
    \hat{U}(\frac{T}{2},-\frac{T}{2})&=\mathcal{T}\exp\left(-\ii \int_{-\thalf}^{\thalf} \hat{V}(t)dt\right), \label{eq:evolution}\\
    \hat{V}(t)&=e^{\ii\hat{H}_0 t}\hat{V} e^{-\ii\hat{H}_0t}. \label{eq:V-evolve}
\end{align}
Here, $\mathcal{T}$ is the time-ordering operator 
and the time variable $t$ is understood to be on a contour $\{t\equiv(1-\ii0^+)\tilde{t}:\tilde{t}\in\mathbbm{R}\}$. 
Expanding Eq. \eqref{eq:evolution} in $\hat{V}$, the $n$-th order energy is found as
\begin{align}
\Delta E^{(n)} =& \lim_{T\rightarrow\infty}\frac{\ii}{T}\frac{(-\ii)^n}{n!}\int_{-T/2}^{T/2}dt_1\int_{-T/2}^{T/2}dt_2\dots\int_{-T/2}^{T/2}dt_n\nonumber\\
&\langle \hV(t_1)\hV(t_2)\cdots \hV(t_n) \rangle_{0,\text{linked}} \label{eq:En}.
\end{align}

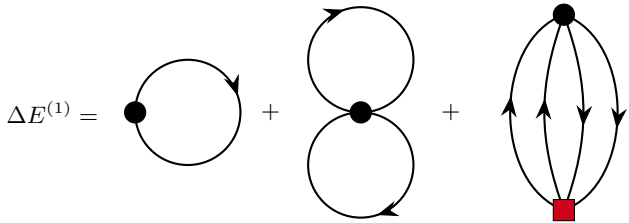
\begin{figure}[!htb]
    \centering
    \begin{tikzpicture}%[scale=.9]
    \node at (-1.1,.5) {$\Delta E^{(1)} = $};
    
    \gfloop{(0,.5)}{0}{.7}
    \Vdot{(0,.5)}

    \node at (1.8,.5) {$+$};
    
    \gfloop{(3,.5)}{90}{.7}
    \gfloop{(3,.5)}{-90}{.7}
    \Vdot{(3,.5)}

    \node at (4.2,.5) {$+$};
    
    \phbbl[70]{(5.7,-.8)}{90}{2.6}
    \phbbl[20]{(5.7,-.8)}{90}{2.6}
    \cmdot{(5.7,-.8)}\Vdot{(5.7,1.8)}
    \end{tikzpicture}
    \caption{Generalized Feynman diagrams for the first order energy $\Delta E^{(1)}$ in Eq. \eqref{eq:e1}. 
    The vertices depicted as black dots with two and four legs 
    represent the first-order one-electron and antisymmetrized two-electron interactions,\cite{hugenholtz_perturbation_1957} viz., the first and second terms of Eq. \eqref{eq:PT-V}, respectively. 
    The arrowed lines connecting such vertices represent zeroth-order Green's functions $G^0_{pq}$. 
    The red squares with four legs represent a two-body cumulant, e.g., $G^{0,c}_{pq,rs}$.}
    \label{fig:1st-energy}
\end{figure}

In standard MBPT, Wick's theorem\cite{wick_evaluation_1950} is 
employed to further expand Eq. \eqref{eq:En} into products of time-ordered one-body Green's functions,
which can then be represented compactly by introducing Feynman or Goldstone diagrams\cite{negele_quantum_1998}. The subscript
`linked' in Eq. \eqref{eq:En} means that only the linked diagrams are retained\cite{goldstone_derivation_1957}. 
However, for an interacting $\hH_0$, the standard Wick's theorem does not hold.
In this case, we can use the cumulant expansion of time-ordered Green's functions in place of Wick's theorem\cite{negele_quantum_1998,metzner_linked-cluster_1991,wang_generalized_2025}. 
The mathematical details of the cumulant expansion are given in the Supplemental Material.
Here, we only illustrate it for the two-body Green's function,
% \begin{align}
%     \tordexpt{\hp(t_1)\hq^\dagger(t_2)}_0 = \tordexpt{\hp(t_1)\hq^\dagger(t_2)}_{0,c},
% \end{align}
% \begin{align}
%     &\quad \tordexpt{\hr(t_1)\hs(t_2)\hq^\dagger(t_4)\hp^\dagger(t_3)}_0 \nonumber\\
%     &= \tordexpt{\hr(t_1)\hs(t_2)\hq^\dagger(t_4)\hp^\dagger(t_3)}_{0,c} \nonumber\\
%     &\quad - \tordexpt{\hr(t_1)\hq^\dagger(t_4)}_0\tordexpt{\hs(t_2)\hp^\dagger(t_3)}_0 \nonumber\\
%     &\quad + \tordexpt{\hr(t_1)\hp^\dagger(t_3)}_0\tordexpt{\hs(t_2)\hq^\dagger(t_4)}_0,
%     \label{eq:cmlt-v2}
% \end{align}
% where $\tord$ is the time-ordering operator putting larger time to the left. Eq. \eqref{eq:cmlt-v2} can be expressed more compactly with the notation $\ii G^0_{pq}(t_1,t_2)\equiv\tordexpt{\hp(t_1)\hq^\dagger(t_2)}_0$ and $(-1)G^{0,(c)}_{rs,pq}(t_1,t_2,t_3,t_4) \equiv \tordexpt{\hr(t_1)\hs(t_2)\hq^\dagger(t_4)\hp^\dagger(t_3)}_{0,(c)}$,
\begin{align}
    &\quad G^0_{rs,pq}(t_1,t_2,t_3,t_4)\nonumber\\
    &= G^{0,c}_{rs,pq}(t_1,t_2,t_3,t_4) \nonumber\\
    &\quad- G^0_{rq}(t_1,t_4)G^0_{sp}(t_2,t_3) + G^0_{rp}(t_1,t_3)G^0_{sq}(t_2,t_4),\label{eq:g2decomp}
\end{align}
where $G^{0,c}_{rs,pq}(t_1,t_2,t_3,t_4)$ represents the connected
two-body Green's function\cite{negele_quantum_1998}. 
Using Eq. \eqref{eq:g2decomp}, 
the first-order energy $\Delta E^{(1)}$ in Eq. \eqref{eq:En} 
can be written as
\begin{align}
    \Delta E^{(1)} &= (-\ii)v_{pq}G^0_{qp}(t,t^+)\nonumber\\
    &\qquad + \frac{1}{2}\bar{v}_{pr,qs}G^0_{rq}(t,t^+)G^0_{sp}(t,t^+) \nonumber\\
    &\qquad - \frac{1}{4}\bar{v}_{pr,qs}G^{0,c}_{rs,pq}(t,t^+,t^{+++},t^{++}),\label{eq:e1}
\end{align}
where $t^+$ is a shorthand notation for $t + 0^+$.
These three terms can be represented diagrammatically as Fig. \ref{fig:1st-energy}.
The red square in the last diagram represents the cumulant, $G^{0,c}_{rs,pq}$, 
which does not appear in standard MBPT with a quadratic $\hH_0$ and plays a similar role 
as the density cumulant in the extended Wick's theorem by
Kutzelnigg and Mukherjee\cite{kutzelnigg1997normal}.
The second-order energy diagrams can be enumerated in a similar way, where the summation of linked diagrams 
reproduces the second-order energy derived from the standard Rayleigh-Schr\"{o}dinger perturbation theory, 
which will be elaborated in details elsewhere.

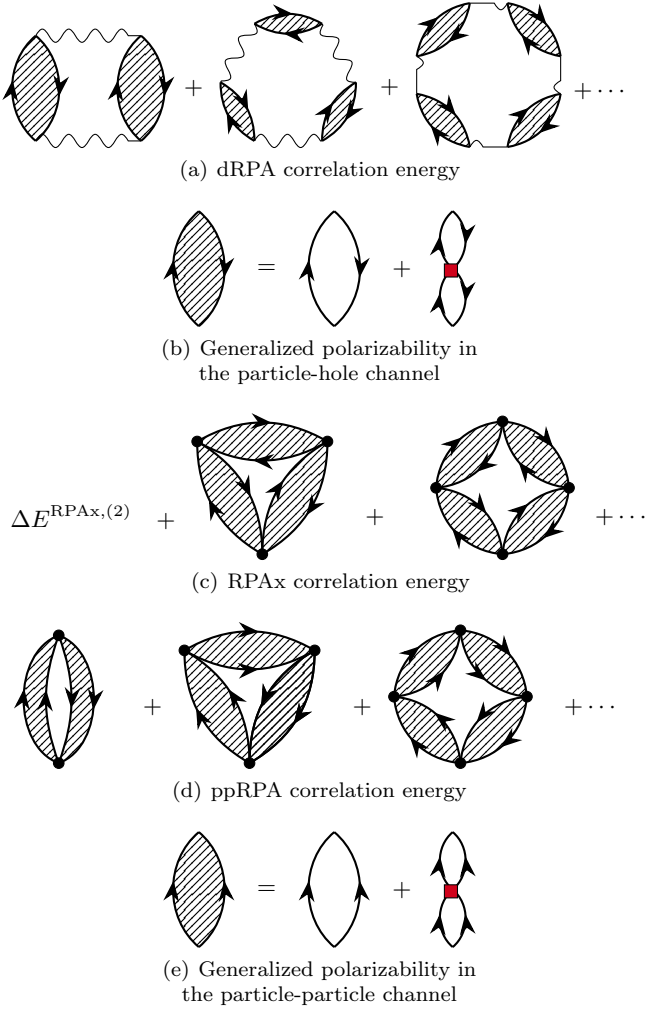
\begin{figure}[htb]
\subfigure[dRPA correlation energy]{
\begin{tikzpicture}[scale=.7]
    %\node at (-2,1) {$\Delta E^{\text{MR-RPA}} =$};
    \phbblc{(0,0)}{90}{2}
    \phbblc{(2,0)}{90}{2}
    \Vline{(0,0)}{0}{2}
    \Vline{(0,2)}{0}{2}
    \node at (3,1) {$+$};
\end{tikzpicture}
\hfil
\begin{tikzpicture}[scale=.6]
    \Vline{(0,0)}{0}{1.5}
    \phbblc[30]{(1.5,0)}{60}{1.5}
    \Vline{(1.5+.75,.75*1.732)}{120}{1.5}
    \phbblc[30]{(1.5,1.5*1.732)}{180}{1.5}
    \Vline{(0,1.5*1.732)}{-120}{1.5}
    \phbblc[30]{(-.75,.75*1.732)}{-60}{1.5}
    \node at (3,1.2) {$+$};
\end{tikzpicture}
\hfil
\begin{tikzpicture}[scale=.5]
    \Vline{(0,0)}{0}{1}
    \phbblc[35]{(1,0)}{45}{2}
    \Vline{(1+1.414,1.414)}{90}{1}
    \phbblc[35]{(1+1.414,2.414)}{135}{2}
    \Vline{(1,2.414+1.414)}{180}{1}
    \phbblc[35]{(0,2.414+1.414)}{-135}{2}
    \Vline{(-1.414,2.414)}{-90}{1}
    \phbblc[35]{(-1.414,1.414)}{-45}{2}
    \node at (3.5,1.5) {$+\cdots$};
\end{tikzpicture}
\label{fig:resum-drpa}
}

    \subfigure[Generalized polarizability in the particle-hole channel]{
    \begin{tikzpicture}[scale=.45]
        \phbblc{(0.,0)}{-90}{3.4}
        \node at (2,-1.7) {$=$};
        \phbbl{(4.,0)}{-90}{3.4}
        \node at (6.,-1.7) {$+$};
        \phbbl{(7.5,0)}{-90}{1.7}
        \phbbl{(7.5,-1.7)}{-90}{1.7}
        \cmdot[7]{(7.5,-1.7)}
    \end{tikzpicture}\label{fig:MR-ring}
    }
    
\subfigure[RPAx correlation energy]{
\begin{tikzpicture}[scale=.5]
    \node at (-4.5,1) {$\Delta E^{\text{RPAx,(2)}}\quad+$};
    \phbblc[30]{(0,0)}{60}{3.464}
    \phbblc[30]{(1.732,3)}{180}{3.464}
    \phbblc[30]{(-1.732,3)}{-60}{3.464}
    \Vdot{(0,0)}\Vdot{(1.732,3)}\Vdot{(-1.732,3)}
    \node at (3.,1) {$+$};
\end{tikzpicture}
\hspace{8pt}
\begin{tikzpicture}[scale=.5]
    \phbblc[40]{(0,0)}{45}{2.5}
    \phbblc[40]{(1.414*1.25,1.414*1.25)}{135}{2.5}
    \phbblc[40]{(0,2*1.414*1.25)}{-135}{2.5}
    \phbblc[40]{(-1.414*1.25,1.414*1.25)}{-45}{2.5}
    
    \Vdot{(0,0)}\Vdot{(1.414*1.25,1.414*1.25)}\Vdot{(0,2*1.414*1.25)}\Vdot{(-1.414*1.25,1.414*1.25)}
    \node at (3.2,1) {$+\cdots$};
\end{tikzpicture}\label{fig:MR-RPAx-3}
}

\subfigure[ppRPA correlation energy]{
\begin{tikzpicture}[scale=.5]
    \phbbl[70]{(-1.5,0)}{90}{3.4}
    \phbbl[20]{(-1.5,0)}{90}{3.4}
    \pattern [shift={(-1.5,0)},rotate=90,pattern=north east lines] (0,0) to[out=-70,in=-110] (3.4,0) to[out=-160,in=-20] cycle;
    \pattern [shift={(-1.5,0)},rotate=90,pattern=north east lines] (0,0) to[out=70,in=110] (3.4,0) to[out=160,in=20] cycle;
    \Vdot{(-1.5,0)}\Vdot{(-1.5,3.4)}
    \node at (1,1.5) {$+$};
\end{tikzpicture}
\hfill
\begin{tikzpicture}[scale=.5]
    \ppbblc[30]{(0,0)}{60}{3.464}
    \ppbblc[30]{(0,0)}{60}{3.464}
    \ppbblc[30]{(1.732,3)}{180}{3.464}
    \ppbblc[30]{(-1.732,3)}{-60}{3.464}
    \Vdot{(0,0)}\Vdot{(1.732,3)}\Vdot{(-1.732,3)}
    \node at (3,1.5) {$+$};
\end{tikzpicture}
\hfill
\begin{tikzpicture}[scale=.5]
    \ppbblc[40]{(0,0)}{45}{2.5}
    \ppbblc[40]{(1.414*1.25,1.414*1.25)}{135}{2.5}
    \ppbblc[40]{(0,2*1.414*1.25)}{-135}{2.5}
    \ppbblc[40]{(-1.414*1.25,1.414*1.25)}{-45}{2.5}
    
    \Vdot{(0,0)}\Vdot{(1.414*1.25,1.414*1.25)}\Vdot{(0,2*1.414*1.25)}\Vdot{(-1.414*1.25,1.414*1.25)}
    \node at (3.5,1.5) {$+\cdots$};
\end{tikzpicture}\label{fig:ppRPA-resum}
}

    \subfigure[Generalized polarizability in the particle-particle channel ]{
    \begin{tikzpicture}[scale=.45]
        \ppbblc{(0.,0)}{-90}{3.4}
        \node at (2,-1.7) {$=$};
        \ppbbl{(4.,0)}{-90}{3.4}
        \node at (6.,-1.7) {$+$};
        \ppbbl{(7.5,0)}{-90}{1.7}
        \ppbbl{(7.5,-1.7)}{-90}{1.7}
        \cmdot[7]{(7.5,-1.7)}
    \end{tikzpicture}\label{fig:MR-ladder}
    }

\caption{Resummation of Feynman diagrams for RPA correlation energies.
The wiggly lines in (a) represent the first-order two-electron interactions $v_{pr,qs}$. The black dots, arrowed lines, and red squares have the same meanings as in Fig. \ref{fig:1st-energy}.
}
\end{figure}

One of the advantages of this diagrammatic formulation is that it allows us
to include high-order perturbative contributions via diagrammatic resummation as in the single-reference case.
For instance, the MR-dRPA correlation energy can be defined by a resummation of generalized `ring' diagrams to infinite order,
as shown in Fig. \ref{fig:resum-drpa}. 
In such diagrams, the interaction lines are connected by a generalized `bubble' diagram, 
illustrated by Fig. \ref{fig:MR-ring}, 
which corresponds to the first two terms of Eq. \eqref{eq:g2decomp} in the limit $t_3 = t_1^+$ and $t_4 = t_2^+$, viz.,
\begin{align}
\ii\Pi^0_{pr,qs}(t_1,t_2) \equiv G_{rq}^0(t_1,t_2^+)G_{sp}^0(t_2,t_1^+)-G_{rs,pq}^{0,c}(t_1,t_2,t_1^+,t_2^+).
\end{align}
This quantity can be interpreted as a generalized polarizability, 
whose contributions to the $n$-th order energy in Eq. \eqref{eq:En} can be expressed as\cite{wang_generalized_2025}
\begin{align}
    \Delta E^{(n),\text{ring}} = -\frac{1}{2\pi}\frac{1}{2n}\int_{-\infty}^{\infty}d\omega \,\mathrm{tr}\left(\left[\mathbf{v}\mathbf{\Pi}^0(\ii\omega)\right]^n\right),
\end{align}
where $\frac{1}{2n}$ is the symmetry factor of the $n$-th order `ring' diagram in Fig. \ref{fig:resum-drpa}. 
Then, MR-dRPA correlation energy defined by Fig. \ref{fig:resum-drpa} is simply
\begin{align}
    \Delta E^{\text{dRPA}} \equiv \sum_{n\ge 2}\Delta E^{(n),\text{ring}}.
\end{align}
In Ref. \cite{wang_generalized_2025}, 
we derived three mathematically equivalent expressions for $\Delta E^{\text{dRPA}}$.
The first one involves an imaginary-frequency integration
\begin{align}
\Delta E^{\mathrm{dRPA}}
=\int_{-\infty}^{\infty}\,
\frac{d\omega}{2\pi}
\frac{1}{2}\tr\left[\ln\left(\mathbf{I}-\mathbf{v}\mathbf{\Pi}^0(\ii\omega)\right)+\mathbf{v}\mathbf{\Pi}^0(\ii\omega)\right],\label{eq:ecor-drpa-log}
\end{align}
which is the starting point for low-scaling formulation.
The second one is the `plasmon formula'
\begin{align}
    \Delta E^{\mathrm{dRPA}} = \frac{1}{2} \left(\tr(\boldsymbol{\Omega}) - \tr (\mathbf{A})\right), \label{eq:ecor-plas-drpa}
\end{align}
where $\boldsymbol{\Omega}$ needs to be solved from a non-Hermitian generalized eigenvalue problem,
\begin{align}
    \begin{bmatrix}
\mathbf{A} & \mathbf{B}\\
\mathbf{B}^* & \mathbf{A}^*
\end{bmatrix}
\begin{bmatrix}
\mathbf{X} & \mathbf{Y}^*\\
\mathbf{Y} & \mathbf{X}^*
\end{bmatrix}
=\begin{bmatrix}
    \mathbf{I} & \mathbf{0}\\
    \mathbf{0} & -\mathbf{I}
\end{bmatrix}
\begin{bmatrix}
\mathbf{X} & \mathbf{Y}^*\\
\mathbf{Y} & \mathbf{X}^*
\end{bmatrix}\begin{bmatrix}
        \boldsymbol{\Omega} & \mathbf{0}\\
        \mathbf{0} & -\boldsymbol{\Omega}
    \end{bmatrix},\label{eq:drpa-diag}
\end{align}
whose building blocks, $\mathbf{A}$ and $\mathbf{B}$, are defined by\cite{wang_generalized_2025}
\begin{align}
A_{LR}&=\omega_L\delta_{LR} + \langle \Phi_L|\hat{p}^\dagger \hat{r}|\Phi_0\rangle v_{pr,qs} \langle \Phi_0|\hat{q}^\dagger \hat{s}|\Phi_R\rangle,\label{eq:mat-ph-A-d}\\
B_{LR}&=\langle \Phi_L|\hat{p}^\dagger \hat{r}|\Phi_0\rangle v_{pr,qs} \langle \Phi_R|\hat{q}^\dagger \hat{s}|\Phi_0\rangle,\label{eq:mat-ph-B-d}
\end{align}
where $|\Phi_L\rangle$ represents a zeroth-order excited state (with the same number of electrons with the ground state), 
and $\omega_L = E_L^{(0)}-E_0^{(0)}$ is the corresponding zeroth-order excitation energy. 
Eq. \eqref{eq:drpa-diag} exhibits a paired structure in the eigenvalues, i.e., $\boldsymbol{\Omega}$ and $-\boldsymbol{\Omega}$.
To the best of our understandings, 
these expressions cannot be derived from the EOM approach\cite{rowe_equations--motion_1968} unless $|\Phi_0\rangle$ is
a single Slater determinant. The third one is the ring coupled cluster like formula
\begin{align}
    \Delta E^{\text{dRPA}} &= \frac{1}{2}\tr(\mathbf{BT}),\label{eq:ecor-cc-drpa}
\end{align}
where the amplitude $\mathbf{T}$ needs to be solved from
a Riccati equation
\begin{align}
\mathbf{B}^* + \mathbf{A}^*\mathbf{T} &+ \mathbf{T} \mathbf{A}
+\mathbf{TBT} = \mathbf{0}.\label{eq:drpa-cc}
\end{align}
A distinct feature of our MR-dRPA formulation\cite{wang_generalized_2025} 
is the seamless connection to the standard single reference theory, as it is derived following the same diagrammatic resummation as SR-dRPA, 
with only the definition of diagrams being generalized. 
As a result, 
Eqs. \eqref{eq:ecor-drpa-log}-\eqref{eq:drpa-cc} all share the same mathematical structure 
with their SR counterparts\cite{ren_random-phase_2012,furche_developing_2008,scuseria_ground_2008}, 
making the standard SR-dRPA a special case of our generalized theory.

\subsection{Multi-reference particle-hole random phase approximation with exchange}
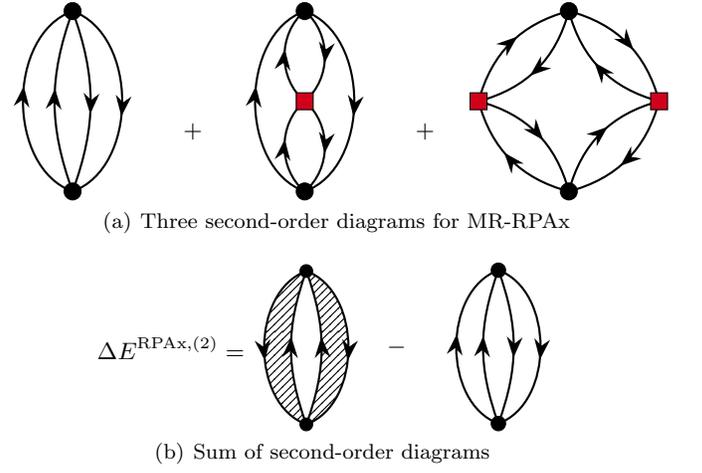
\begin{figure}[hbt]
\subfigure[Three second-order diagrams for MR-RPAx]{
\begin{tikzpicture}[scale=.8]
    \phbbl[70]{(0,0)}{90}{3}
    \phbbl[20]{(0,0)}{90}{3}
    
    \Vdot{(0,0)}\Vdot{(0,3)}
    \node at (2.,1) {$+$};
\end{tikzpicture}
\hspace{4pt}
\begin{tikzpicture}[scale=.8]
    \phbbl[70]{(0,0)}{90}{3}
    \phbbl{(0,0)}{90}{1.5}
    \phbbl{(0,1.5)}{90}{1.5}
    
    \Vdot{(0,0)}\Vdot{(0,3)}\cmdot{(0,1.5)}
    \node at (2.,1) {$+$};
\end{tikzpicture}
\hspace{4pt}
\begin{tikzpicture}[scale=.8]
    \phbbl[30]{(0,0)}{45}{1.5*1.414}
    \phbbl[30]{(0,0)}{135}{1.5*1.414}
    \phbbl[30]{(1.5,1.5)}{135}{1.5*1.414}
    \phbbl[30]{(-1.5,1.5)}{45}{1.5*1.414}
    
    \Vdot{(0,0)}\Vdot{(0,3)}\cmdot{(1.5,1.5)}\cmdot{(-1.5,1.5)}
\end{tikzpicture}\label{fig:MR-RPAx-o2}
}

\subfigure[Sum of second-order diagrams]{
\begin{tikzpicture}[scale=.6]
    \node at (-4,1.7) {$\Delta E^{\text{RPAx,(2)}}= $};
    %\phbblc[70]{(-1.5,0)}{90}{3.4}
    %\phbblc[20]{(-1.5,0)}{90}{3.4}
    \gfline[70]{(-1.,0)}{90}{3.4}
    \gfline[-70]{(-1.,0)}{90}{3.4}
    \gfline[-20]{(-1.,3.4)}{-90}{3.4}
    \gfline[20]{(-1.,3.4)}{-90}{3.4}

    \pattern [shift={(-1.,0)},rotate=90,pattern=north east lines] (0,0) to[out=-70,in=-110] (3.4,0) to[out=-160,in=-20] cycle;

    \pattern [shift={(-1.,0)},rotate=90,pattern=north east lines] (0,0) to[out=70,in=110] (3.4,0) to[out=160,in=20] cycle;
    
    \Vdot{(-1.,0)}\Vdot{(-1.,3.4)}
    \node at (1,1.7) {$-$};
\end{tikzpicture}
\hspace{4pt}
\begin{tikzpicture}[scale=.68]
    \phbbl[70]{(0,0)}{90}{3}
    \phbbl[20]{(0,0)}{90}{3}
    
    \Vdot{(0,0)}\Vdot{(0,3)}
\end{tikzpicture}\label{fig:MR-RPAx-o2-comb}
}

%\subfigure[Resummation from the third order to infinite]{
%\begin{tikzpicture}[scale=.6]
%    \phbblc[30]{(0,0)}{60}{3.464}
%    \phbblc[30]{(1.732,3)}{180}{3.464}
%    \phbblc[30]{(-1.732,3)}{-60}{3.464}
%    \Vdot{(0,0)}\Vdot{(1.732,3)}\Vdot{(-1.732,3)}
%    \node at (3.,1) {$+$};
%\end{tikzpicture}
%\hspace{8pt}
%\begin{tikzpicture}[scale=.6]
%    \phbblc[40]{(0,0)}{45}{2.5}
%    \phbblc[40]{(1.414*1.25,1.414*1.25)}{135}{2.5}
%    \phbblc[40]{(0,2*1.414*1.25)}{-135}{2.5}
%%    \phbblc[40]{(-1.414*1.25,1.414*1.25)}{-45}{2.5}
    
%    \Vdot{(0,0)}\Vdot{(1.414*1.25,1.414*1.25)}\Vdot{(0,2*1.414*1.25)}\Vdot{(-1.414*1.25,1.414*1.25)}
%    \node at (3.2,1) {$+\cdots$};
%\end{tikzpicture}
%}
\caption{Special treatment at the second order for MR-RPAx. 
The black dots, arrowed lines, and red squares have the same meanings as in Fig. \ref{fig:1st-energy}.}
\label{fig:RPAx}
\end{figure}

%\subsubsection{General theory}
In this section, we extend the above derivation of MR-dRPA to MR-RPAx. 
The MR-RPAx correlation energy is defined as a resummation of the generalized `ring'
diagrams with antisymmetrized Coulomb interactions, see Fig. \ref{fig:MR-RPAx-3}.
The use of antisymmetrized vertices removes the SIE in one-electron systems, e.g., \ce{H} atom or \ce{H2+}. 
However, the second-order term requires a special attention.
The three second-order diagrams for MR-RPAx, shown in Fig. \ref{fig:MR-RPAx-o2}, 
cannot be combined together in terms of the generalized polarizability shown in  Fig. \ref{fig:MR-ring}, 
as their symmetry factors are $\frac{1}{8}$ (due to the presence of two equivalent pairs of lines), 
$\frac{1}{2}$ and $\frac{1}{4}$, respectively, 
while a combination using Fig. \ref{fig:MR-ring} would expect them to be $\frac{1}{4}, \frac{1}{2}$ and $\frac{1}{4}$, respectively.
Therefore, a subtraction of the first diagram is needed, as shown in Fig. \ref{fig:MR-RPAx-o2-comb}. 
A similar subtraction also appears in SR-RPAx\cite{fukuda_linearized_1964}.
The third- and higher-order diagrams are free of such issue.

Following the procedure for deriving MR-dRPA\cite{wang_generalized_2025}, 
we can find the algebraic expression for the MR-RPAx correlation energy as
\begin{align}
\Delta E^{\mathrm{RPAx}} =\int_{-\infty}^{\infty}&
\frac{d\omega}{2\pi}\,
\frac{1}{2}\tr\left[\ln\left(\mathbf{I}-\mathbf{\bar{v}}\mathbf{\Pi}^0(\ii\omega)\right)+\mathbf{\bar{v}}\mathbf{\Pi}^0(\ii\omega)\right]\nonumber\\
&- \Delta E^{(2),a},\label{eq:ecorr-rpax}
\end{align}
where the last term $\Delta E^{(2),a}$ corresponds to the second term in Fig. \ref{fig:MR-RPAx-o2-comb}.
To derive its explicit expression, we introduce
\begin{align}
    \ii\Pi^{0,a}_{pr,qs}(t_1,t_2) \equiv G_{rq}^0(t_1,t_2)G_{sp}^0(t_2,t_1), \label{eq:pi0-a}
\end{align}
and its Fourier transform
\begin{align}
    &\Pi^{0,a}_{pr,qs}(\omega) = \int_{-\infty}^{\infty} dt \, e^{\ii \omega t}\Pi^{0,a}_{pr,qs}(t,0)\nonumber\\
    &= \sum_{PH}\frac{\langle \Phi_0 |\hp^\dagger|\Phi_H^{N-1}\rangle\langle \Phi_0|\hr|\Phi^{N+1}_P\rangle \langle \Phi^{N+1}_P|\hq^\dagger|\Phi_0\rangle\langle \Phi^{N-1}_H |\hs|\Phi_0\rangle}{\omega - \omega^{N+1}_{P}-\omega^{N-1}_{H} + \ii 0^+} \nonumber\\
    &\quad- \sum_{PH}\frac{\langle \Phi^{N+1}_P | \hp^\dagger|\Phi_0\rangle\langle \Phi^{N-1}_I|\hr|\Phi_0\rangle\langle \Phi_0|\hq^\dagger|\Phi^{N-1}_H\rangle\langle \Phi_0 |\hs|\Phi^{N+1}_P\rangle }{\omega + \omega^{N+1}_{P}+\omega^{N-1}_{H} - \ii 0^+},
\end{align}
where $|\Phi_P^{N+1}\rangle$ and $|\Phi_H^{N-1}\rangle$ are the zeroth-order eigenstates 
with $N+1$ and $N-1$ electrons (with $N$ being the number of electrons of the ground state), respectively, 
viz., $\hH_0 |\Phi_P^{N+1}\rangle = E_P^{N+1,(0)} |\Phi_P^{+1}\rangle$, 
$\hH_0 |\Phi_H^{N-1}\rangle = E_H^{N-1,(0)} |\Phi_H^{N-1}\rangle$, 
$\omega^{N+1}_P = E_P^{N+1,(0)} - E_0^{(0)}$, 
and $\omega^{N-1}_H = E_H^{N-1,(0)} - E_0^{(0)}$. 
Then, $\Delta E^{(2),a}$ can be expressed in a form similar to the second-order M\o ller-Plesset (MP2) correlation energy,
\begin{align}
\Delta E^{(2),a}=&-\frac{1}{8}
\int_{-\infty}^{\infty}
\frac{d\omega}{2\pi}\,
\tr[\bar{\mathbf{v}}\mathbf{\Pi}^{0,a}(\ii\omega)\bar{\mathbf{v}}\mathbf{\Pi}^{0,a}(\ii\omega)]\nonumber\\
=&
-\frac{1}{4}\sum_{PQHI}\frac{|V_{PH,QI}|^2}{\omega^{N+1}_{P}+\omega^{N-1}_{H}+\omega^{N+1}_{Q}+\omega^{N-1}_{I}},\label{eq:e2-pi0-a}
\end{align}
with $V_{PH,QI}$ defined as
\begin{align}
    V_{PH,QI} \equiv &\langle \Phi^{N+1}_{P}| \hp^\dagger|\Phi_0\rangle
   \langle \Phi^{N-1}_{H}|\hr|\Phi_0\rangle \bar{v}_{pr,qs}\nonumber\\
   &\quad \langle \Phi^{N+1}_{Q}|\hq^\dagger|\Phi_0\rangle\langle \Phi^{N-1}_{I}|\hs|\Phi_0\rangle.
\end{align}
Eq. \eqref{eq:e2-pi0-a} reduces to the MP2 correlation energy in the single-reference limit.

Except for the presence of $\Delta E^{(2),a}$, the mathematical form of the MR-RPAx correlation energy 
\eqref{eq:ecorr-rpax} is similar to the MR-dRPA correlation energy \eqref{eq:ecor-drpa-log}, 
but with the antisymmtrizing interactions. 
Consequently, we can introduce a non-Hermitian generalized eigenvalue problem
\begin{align}
    \begin{bmatrix}
\mathbf{\bar{A}} & \mathbf{\bar{B}}\\
\mathbf{\bar{B}}^* & \mathbf{\bar{A}}^*
\end{bmatrix}
\begin{bmatrix}
\mathbf{X} & \mathbf{Y}^*\\
\mathbf{Y} & \mathbf{X}^*
\end{bmatrix}
=\begin{bmatrix}
    \mathbf{I} & \mathbf{0}\\
    \mathbf{0} & -\mathbf{I}
\end{bmatrix}
\begin{bmatrix}
\mathbf{X} & \mathbf{Y}^*\\
\mathbf{Y} & \mathbf{X}^*
\end{bmatrix}\begin{bmatrix}
        \boldsymbol{\bar{\Omega}} & \mathbf{0}\\
        \mathbf{0} & -\boldsymbol{\bar{\Omega}}
    \end{bmatrix},\label{eq:ph-diag}
\end{align}
whose building blocks, $\mathbf{\bar{A}}$ and $\mathbf{\bar{B}}$, are defined by
\begin{align}
\bar{A}_{LR}&=\omega_L\delta_{LR} + \langle \Phi_L|\hat{p}^\dagger \hat{r}|\Phi_0\rangle \bar{v}_{pr,qs} \langle \Phi_0|\hat{q}^\dagger \hat{s}|\Phi_R\rangle,\label{eq:mat-ph-A}\\
\bar{B}_{LR}&=\langle \Phi_L|\hat{p}^\dagger \hat{r}|\Phi_0\rangle \bar{v}_{pr,qs} \langle \Phi_R|\hat{q}^\dagger \hat{s}|\Phi_0\rangle,\label{eq:mat-ph-B}
\end{align}
which are the counterparts of $\mathbf{A}$ \eqref{eq:mat-ph-A-d}
and $\mathbf{B}$ \eqref{eq:mat-ph-B-d} in MR-dRPA 
with antisymmetrized interactions.
Eq. \eqref{eq:ph-diag} has the same mathematical structure as Eq. \eqref{eq:drpa-diag}, 
and thus also has paired eigenvalues, i.e., $\boldsymbol{\bar{\Omega}}$ and $-\boldsymbol{\bar{\Omega}}$. 
The MR-RPAx correlation energy can then be expressed in a `plasmon' formula,
\begin{align}
    \Delta E^{\mathrm{RPAx}} = \frac{1}{2} \left(\tr(\boldsymbol{\bar{\Omega}}) - \tr (\mathbf{\bar{A}})\right) - \Delta E^{(2),a}.\label{eq:ecor-rpax-plasmon}
\end{align}
This equation holds for both SR- and MR-RPAx. 
%Note that this expression is not consistent with those in literature, which only contain the first term and undergo a long debate on whether prefactor should be $\frac{1}{2}$ or $\frac{1}{4}$.\cite{mclachlan_time-dependent_1964,scuseria_ground_2008,angyan_correlation_2011,eshuis_electron_2012,tahir_comparing_2019} The reason is that traditionally RPAx equations are derived from the equation of motion (EOM) of excitation operators, which doesn't give a correlation itself, and the correlation energy has to be defined externally using the adiabatic connection fluctuation-dissipation theorem (ACFDT) or coupled cluster theory. In contrast, diagrammatic resummation gives an unique definition for the correlation energy and removes such ambiguity.
Following the same derivation in MR-dRPA, a coupled cluster like form equivalent to Eqs. \eqref{eq:ecorr-rpax} and \eqref{eq:ecor-rpax-plasmon} can also be derived as
\begin{align}
    \Delta E^{\text{RPAx}} &= \frac{1}{2}\tr(\mathbf{\bar{B}T}) - \Delta E^{(2),a},\label{eq:ecor-cc-rpax}
\end{align}  
where $\mathbf{T} \equiv \mathbf{Y}\mathbf{X}^{-1}$ is to be solved from
\begin{align}
    \mathbf{\bar{B}}^* + \mathbf{\bar{A}}^*\mathbf{T} &+ \mathbf{T} \mathbf{\bar{A}}
+\mathbf{T\bar{B}T} = \mathbf{0}.\label{eq:rpax-cc}
\end{align}
These equations differ from those in MR-dRPA (Eqs. \eqref{eq:ecor-cc-drpa} and \eqref{eq:drpa-cc}) only by the replacement of $\mathbf{A}$ and $\mathbf{B}$ with $\mathbf{\bar{A}}$ and $\mathbf{\bar{B}}$, along with the correction term for the second term shown in Fig. \ref{fig:MR-RPAx-o2-comb}.

The above MR-RPAx formulation is valid for any partition of the Hamiltonian as Eqs. \eqref{eq:partition}-\eqref{eq:PT-V}. 
However, to develop an accurate and efficient method, 
an appropriate choice of $\hH_0$ and the reference state (viz. $|\Phi_0\rangle$) is essential. 
Our previous choice for MR-dRPA, that is, the Dyall Hamiltonian as $\hH_0$ and the complete active space self-consistent field (CASSCF) wavefunction\cite{roos_complete_1980} as $|\Phi_0\rangle$, 
is also employed in this work for MR-RPAx.
Specifically, the spin-orbitals are classified into three categories: 
(i) core (closed shell) orbitals, labeled by $\{i,j,k,\dots\}$; 
(ii) active orbitals, labeled by $\{w, x, y,\dots\}$; and 
(iii) virtual (unoccupied) orbitals, labeled by $\{a, b, c,\dots\}$. 
The Dyall Hamiltonian\cite{dyall_choice_1995}
is defined as a sum of the inactive and active parts,
\begin{align}
    \hH_{\text{Dyall}} =& \hH_{\text{inact}}+\hH_{\text{act}} ,\label{eq:Hdyall-1main}\\
    \hH_{\text{inact}} =& \epsilon_i \hat{i}^{\dagger}\hat{i} + \epsilon_a \hat{a}^{\dagger}\hat{a} ,\label{eq:Hdyall-2main}\\
    \hH_{\text{act}} =& h^{\text{eff}}_{xy}\hat{x}^{\dagger}\hat{y} + \frac{1}{4}
    \langle xy||zw\rangle \hat{x}^{\dagger}\hat{y}^{\dagger}\hat{w}\hat{z}.\label{eq:Hdyall-3main}
\end{align}
Here, $\epsilon_i$ and $\epsilon_a$ are 
the canonical orbital energies generated by the mean-field of core and active orbitals, viz.,
\begin{align}
     \epsilon_i \delta_{ij} =& h_{ij} + \langle ik||jk \rangle + \langle ix||jy\rangle\gamma_{xy},\\
     \epsilon_a \delta_{ab} =& h_{ab} + \langle ak||bk \rangle + \langle ax||by\rangle\gamma_{xy}, 
\end{align}
where $\langle pq||rs \rangle$ denotes an antisymmetrized two-electron Coulomb integral, 
and $\gamma_{xy}$ is the one-particle density matrix in the active space. 
In Eq. \eqref{eq:Hdyall-3main},
$h^{\text{eff}}_{xy}$ is a mean-field generated by the core orbitals only,
\begin{align}
    h^{\text{eff}}_{xy} = h_{xy} + \langle xk||yk \rangle.
\end{align}
The CASSCF wavefunction can be written as a product, 
$|\Phi_0\rangle=|\Theta_0\rangle|\Xi_0^{N_{\text{act}}}\rangle$, 
where $|\Theta_0\rangle$ is the inactive part, which is simply a single Slater determinant, 
and $|\Xi_0^{N_{\text{act}}}\rangle$ is the active part, 
which is a multi-determinant wavefunction that describes the strong correlation in an active space 
with $N_{\text{act}}$ active electrons distributed in $M_{\text{act}}$ active spatial orbitals, 
denoted as CAS($N_{\text{act}},M_{\text{act}}$). 
Using $\hH_{\text{Dyall}}$ as $\hH_0$ and CASSCF as $|\Phi_0\rangle$, 
MR-RPAx involves
the same types of excitations as in MR-dRPA
that can couple with $|\Phi_0\rangle$
through Eqs. \eqref{eq:mat-ph-A} and \eqref{eq:mat-ph-B},
\begin{equation}\begin{aligned}
|\Phi_L\rangle \in \begin{cases}
    |\Theta_i^a\rangle|\Xi_0^{N_{\text{act}}}\rangle \quad (|\Theta_i^a\rangle = \hat{a}^\dagger\hat{i}|\Theta_0\rangle)\,, \\
    |\Theta_i\rangle|\Xi_\lambda^{N_{\text{act}}+1}\rangle \quad (|\Theta_i\rangle = \hat{i}|\Theta_0\rangle)\,, \\
    |\Theta^a\rangle|\Xi_\lambda^{N_{\text{act}}-1}\rangle \quad (|\Theta^a\rangle = \hat{a}^\dagger|\Theta_0\rangle) \,, \\\
    |\Theta_0\rangle|\Xi_{\lambda>0}^{N_{\text{act}}}\rangle,
\end{cases}
\end{aligned}\label{eq:CAS-ex-ph}\end{equation}
leading to a $4\times4$ block structure of $\mathbf{\bar{A}}$ and $\mathbf{\bar{B}}$.
Detailed expressions of their matrix elements are given in the Supplemental Material.

\subsection{Multi-reference particle-particle random phase approximation}
%\subsubsection{General theory}
In this section, we derive MR-ppRPA by generalizing the above derivations to the particle-particle channel. 
To begin with, the Hamiltonian is partitioned as
\begin{align}
    \hH &= (\hH_0- \mu \hat{N}) + \hV ,\\
    \hV &= v_{pq}\hat{p}^\dagger\hat{q} + \frac{1}{2}g_{pq,rs}\hat{p}^\dagger\hat{q}^\dagger\hat{s}\hat{r}\nonumber\\
    &= v_{pq}\hat{p}^\dagger\hat{q} + \frac{1}{4}\bar{g}_{pq,rs}\hat{p}^\dagger\hat{q}^\dagger\hat{s}\hat{r},
\end{align}
where $\hat{N}\equiv\sum_p \hat{p}^\dagger\hat{p}$ is the number operator, 
and a chemical potential $\mu$ is introduced to adjust the number of electrons in the ground state, 
as in the single reference ppRPA theory\cite{van_aggelen_exchange-correlation_2013,van_aggelen_exchange-correlation_2014}. 
Since $[\hH_0,\hat{N}]=0$ and $[\hV,\hat{N}]=0$, % in quantum chemistry, 
$\hH_0$ and $\hH_0-\mu\hat{N}$ generates the same $\hV(t)$. 
Thus, the correlation energy is not affected by the chemical potential, 
as long as the zeroth-order ground state stays the same.
Note that although $\hat{V}$ is the same operator as that in Eq. \eqref{eq:PT-V}, 
the indices in $g_{pq,rs}$ and $\bar{g}_{pq,rs} \equiv g_{pq,rs}-g_{pq,sr}$ are arranged in a different order 
from that in $v_{pr,qs}$ and $\bar{v}_{pr,qs}$, 
for the sake of writing the succeeding formulae in a matrix product form.

To define the MR-ppRPA correlation energy, 
we introduce the generalized `ladder' diagram defined by Figs. \ref{fig:ppRPA-resum} and \ref{fig:MR-ladder}. 
The generalized polarizability in the particle-particle channel, as shown in Fig. \ref{fig:MR-ladder}, corresponds
to setting $t_2=t_1^+$ and $t_3=t_4^+$ in Eq. \eqref{eq:g2decomp}, viz.,
\begin{align}
     \ii K^0_{rs,pq}(t_1,t_2) &\equiv G^0_{rq}(t_1,t_2)G^0_{sp}(t_1,t_2) \nonumber\\
     &\quad - G^0_{rp}(t_1,t_2)G_{sq}^0(t_1,t_2) \nonumber\\
       &\quad - G^{0,c}_{rs,pq}(t_1,t_1^+,t_2^+,t_2)\nonumber\\
       &=\langle \Tau{\hr(t_1)\hs(t_1^+)\hq^\dagger(t_2)\hp^\dagger(t_2^+)} \rangle_0,\label{eq:F0}
\end{align}
where $\hp^{(\dagger)}(t) \equiv e^{\ii(\hH_0-\mu\hat{N})t}\hp^{(\dagger)}e^{-\ii(\hH_0-\mu\hat{N})t}$. 
The $n$-th order ppRPA energy can then be written as
\begin{align}
\Delta &E^{(n),\text{ladder}}= \lim_{T\rightarrow\infty}\frac{\ii}{T}\frac{1}{n}\int_{-T/2}^{T/2}dt_1\int_{-T/2}^{T/2}dt_2\dots\int_{-T/2}^{T/2}dt_n\nonumber\\
&\tr\left(\left[\frac{1}{4}\mathbf{\bar{g}}\mathbf{K}^0(t_1,t_2)\frac{1}{4}\mathbf{\bar{g}}\mathbf{K}^0(t_2,t_3)\cdots\frac{1}{4}\mathbf{\bar{g}}\mathbf{K}^0(t_n,t_1)\right]^n\right),\label{eq:Epp-t}
\end{align}
where $\frac{1}{n}$ results from the symmetry factor of the $n$-th order ppRPA diagram in Fig. \ref{fig:ppRPA-resum}. 
As Eq. \eqref{eq:F0} obeys $K^0_{rs,pq}(t_1,t_2)=K^0_{rs,pq}(t_1-t_2,0)$, 
we introduce its Fourier transform as
\begin{align}
    K_{rs,pq}^0(\omega) &\equiv \int_{-\infty}^{\infty} dt \, e^{\ii\omega t}K_{rs,pq}(t,0)\nonumber\\
    &=\sum_P\frac{\langle \Phi_0|\hs\hr|\Phi^{N+2}_P\rangle\langle \Phi^{N+2}_P| \hp^\dagger\hq^\dagger |\Phi_0\rangle}{\omega-(\omega^{N+2}_P-2\mu)+\ii 0^+}\nonumber\\
    &\quad -\sum_H \frac{\langle \Phi^{N-2}_H|\hs\hr|\Phi_0\rangle\langle \Phi_0|\hp^\dagger \hq^\dagger |\Phi^{N-2}_H\rangle}{\omega+(\omega^{N-2}_H+2\mu)-\ii 0^+},\label{eq:pi-0}
\end{align}
where $|\Phi^{N+2}_P\rangle$ ($|\Phi^{N-2}_H\rangle$) is a zeroth-order eigenstate with $N+2$ ($N-2$) electrons 
(with $N$ being the number of electrons in $|\Phi_0\rangle$), 
viz., $\hH_0|\Phi^{N+2}_P\rangle = E^{N+2,(0)}_{P}|\Phi^{N+2}_P\rangle$, 
$\hH_0|\Phi^{N-2}_H\rangle = E^{N-2,(0)}_H|\Phi^{N-2}_H\rangle$, 
with the corresponding zeroth-order excitation energies for adding and removing two electrons denoted by 
$\omega^{N+2}_{P} = E^{N+2,(0)}_{P}-E^{(0)}_0$ and $\omega^{N-2}_{H} = E^{N-2,(0)}_{H}-E^{(0)}_0$, respectively. 
Using Eq. \eqref{eq:pi-0}, Eq. \eqref{eq:Epp-t} can be converted to an equivalent real-frequency formula ,
\begin{align} 
\Delta E^{(n),\text{ladder}} =\frac{\ii}{2\pi}\frac{1}{n}\int_{-\infty}^{\infty}d\omega \, \mathrm{tr}\left(\left[\frac{1}{4}\mathbf{\bar{g}}\mathbf{K}^0(\omega)\right]^n\right),\label{eq:Epp-w-real}
\end{align}
or an imaginary-frequency formula using the analytical structure of $\mathbf{K}(\omega)$, 
given that $\omega^{N+2}_P - 2\mu$ and $\omega^{N-2}_H + 2\mu$ are positive,
\begin{align}
    \Delta E^{(n),\text{ladder}}=-\frac{1}{2\pi}\frac{1}{n}\int_{-\infty}^{\infty}d\omega \, \mathrm{tr}\left(\left[\frac{1}{4}\mathbf{\bar{g}}\mathbf{K}^0(\ii\omega)\right]^n\right),\label{eq:Epp-w-imag}
\end{align}
which is more amenable for numerical integration.
The MR-ppRPA correlation energy is then found by summing it from the second order
to infinite order
\begin{align}
    \Delta E^{\text{ppRPA}} &\equiv \sum_{n\ge 2}-\frac{1}{2\pi}\frac{1}{n}\int_{-\infty}^{\infty}d\omega \, \mathrm{tr}\left(\left[\frac{1}{4}\mathbf{\bar{g}}\mathbf{K}^0(\ii\omega)\right]^n\right)\nonumber\\
    &= \int_{-\infty}^{\infty}\frac{d\omega}{2\pi}\,\mathrm{tr}\left[\ln\left(\mathbf{I}-\frac{1}{4}\mathbf{\bar{g}K}^0(\ii\omega)\right) + \frac{1}{4}\mathbf{\bar{g}K}^0(\ii\omega)\right]. \label{eq:ecor-pprpa-sum}
\end{align}

As shown in the Supplemental Material, the integration in Eq. \eqref{eq:ecor-pprpa-sum} can be 
carried out analytically using techniques similar to those developed in the single reference case\cite{peng_equivalence_2013}, 
which requires solving the following non-Hermitian generalized eigenvalue problem
\begin{align}
    \begin{bmatrix}
        \mathbf{A}^+ & \mathbf{C}\\
        \mathbf{C}^\dagger & \mathbf{A}^-
    \end{bmatrix}\begin{bmatrix}
        \mathbf{X}^+ & \mathbf{Y}^-  \\ 
        \mathbf{Y}^+ & \mathbf{X}^-
    \end{bmatrix} = \begin{bmatrix}
        \mathbf{I}^+ & \mathbf{0}\\
        \mathbf{0} & -\mathbf{I}^{-}
    \end{bmatrix}\begin{bmatrix}
        \mathbf{X}^+ & \mathbf{Y}^-  \\ 
        \mathbf{Y}^+ & \mathbf{X}^-
    \end{bmatrix}\begin{bmatrix}
        \boldsymbol{\Omega}^+ & \mathbf{0}\\
        \mathbf{0} & \boldsymbol{\Omega}^-
    \end{bmatrix}, \label{eq:pp-diag}
\end{align}
with the building blocks defined as
\begin{align}
    A^+_{PQ} &= (\omega^{N+2}_{P}-2\mu)\delta_{PQ}\nonumber\\
    &\quad +\frac{1}{4}\langle \Phi^{N+2}_P| \hp^\dagger \hq^\dagger |\Phi_0\rangle \bar{g}_{pq,rs}\langle \Phi_0|\hs\hr|\Phi^{N+2}_Q\rangle,\label{eq:mat-pp-1}\\
    A^-_{HI} &= (\omega^{N-2}_{H}+2\mu)\delta_{HI}\nonumber\\
    &\quad +\frac{1}{4}\langle \Phi_0|\hp^\dagger \hq^\dagger |\Phi^{N-2}_H\rangle \bar{g}_{pq,rs} \langle \Phi^{N-2}_I|\hs\hr|\Phi_0\rangle,\label{eq:mat-pp-3}\\
    C_{PH} &= \frac{1}{4}\langle \Phi^{N+2}_P|\hp^\dagger \hq^\dagger |\Phi_0\rangle \bar{g}_{pq,rs}\langle \Phi^{N-2}_H|\hs\hr|\Phi_0\rangle,\label{eq:mat-pp-2}
\end{align}
where $\boldsymbol{\Omega}^{+(-)}$ is a diagonal matrix containing positive (negative) eigenvalues.
The matrices $\mathbf{A}^+$ and $\mathbf{A}^-$ have dimensions as $N_{pp}\times N_{pp}$ and $N_{hh}\times N_{hh}$, respectively, 
where $N_{pp}$ and $N_{hh}$ denote the number of $(N+2)$- and $(N-2)$-electron states, respectively. 
The identity matrices
$\mathbf{I}^+$ and $\mathbf{I}^-$ have the same dimensions as $\mathbf{A}^+$ and $\mathbf{A}^-$, respectively. 
The derivation for Eq. \eqref{eq:pp-diag} (see Supplemental Material for details) puts two restrictions on the chemical potential, $\mu$. First, it should make the matrix $\begin{bmatrix}
        \mathbf{A}^+ & \mathbf{C}\\
        \mathbf{C}^\dagger & \mathbf{A}^-
\end{bmatrix}$ positive-definite,
so that $\boldsymbol{\Omega}^{+}$ and $\boldsymbol{\Omega}^{-}$ have the same dimensions as $\mathbf{A}^+$ and $\mathbf{A}^-$, respectively, 
as has been proved in the single reference case\cite{peng_equivalence_2013}.
Second, $\omega^{N+2}_A - 2\mu$ and $\omega^{N-2}_I + 2\mu$ should be positive, 
to enable the usage of contour techniques to integrate the frequency.
Under these conditions,
MR-ppRPA correlation energy can be written in two equivalent forms
\begin{align}
    \Delta E^{\text{ppRPA}} = \tr(\boldsymbol{\Omega}^+) - \tr(\mathbf{A}^+) = -\tr(\boldsymbol{\Omega}^-) - \tr(\mathbf{A}^-). \label{eq:ecor-plas-pprpa}
\end{align}
Eq. \eqref{eq:ecor-plas-pprpa} shows that the chemical potential does not affect the final correlation energy. 
A reasonable choice for $\mu$ adopted in this work is
\begin{align}
    \mu = \frac{1}{2}\left(\min \{\omega^{N+1}_A\}-\min \{\omega^{N-1}_I\}\right).
\end{align}
It is generalized from the single reference counterpart 
$\frac{1}{2}(\varepsilon_{\mathrm{HOMO}}+
\varepsilon_{\mathrm{LUMO}})$\cite{van_aggelen_exchange-correlation_2013,van_aggelen_exchange-correlation_2014,tahir_comparing_2019}, 
which puts the chemical potential in the middle of the highest-occupied and lowest-unoccupied molecular orbital energies.

As for MR-phRPA, MR-ppRPA can also be converted to a coupled cluster form. 
By introducing two matrices, 
$\mathbf{U} \equiv \bY^+(\bX^+)^{-1}$ and $\mathbf{R} \equiv \mathbf{X^+}\boldsymbol{\Omega}^+ (\mathbf{X}^+)^{-1}$, 
the positive branch of Eq. \eqref{eq:pp-diag} can be rewritten into two equations,
\begin{align}
    \mathbf{A}^+ + \mathbf{CU} &= \mathbf{R},\\
    -\mathbf{C} - \mathbf{A^-U} &= \mathbf{U}\mathbf{R}.
\end{align}
Substituting the first equation into the second to eliminate $\mathbf{R}$, 
we reach a coupled cluster like equation for $\mathbf{U}$
\begin{align}
    \bC^\dagger+\mathbf{A^-\mathbf{U}}+\mathbf{U}\mathbf{A}^++\mathbf{U}\mathbf{C}\mathbf{U} =\mathbf{0}.\label{eq:cc-pp}
\end{align}
Meanwhile, using the invariant property of trace, 
$\sum_{\mu} \Omega_{\mu}^+ = \tr (\boldsymbol{\Omega^+}) = \tr(\mathbf{R})$,
we find the MR-ppRPA correlation energy can be expressed as
\begin{align}
    \Delta E^{\text{ppRPA}} = \tr(\mathbf{R}) - \tr(\mathbf{A}^+) = \tr(\mathbf{CU}).\label{eq:ecor-cc-pp}
\end{align}
Again, our diagrammatic formulation distinguishes it from previous EOM-based multireference generalization of ppRPA\cite{tucholska_duality_2024}.

\begin{figure*}[t]
    \centering
    \subfigure[$(N+2)$-electron states.]{
    \includegraphics[width=0.4\textwidth]{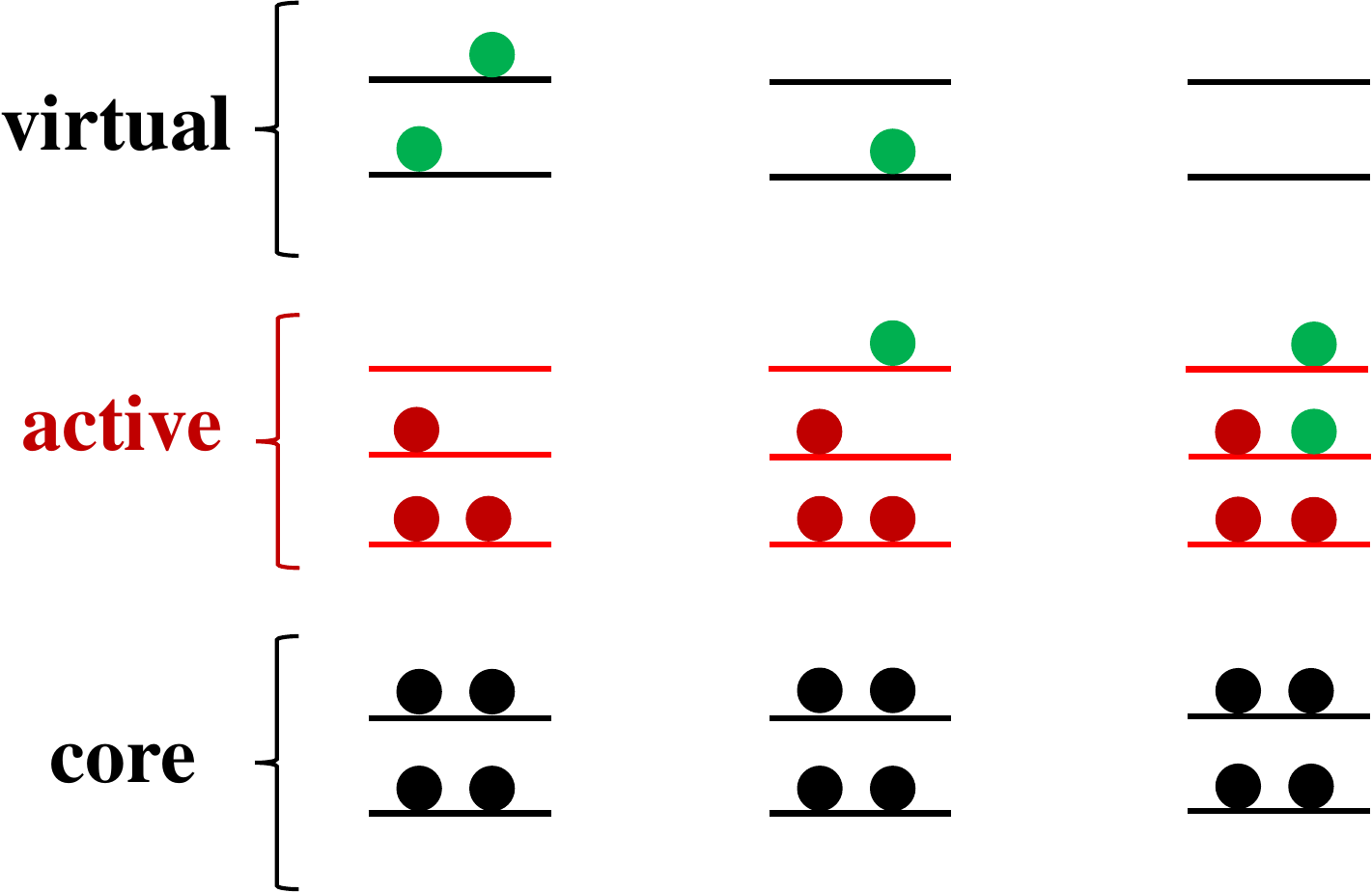}
    }
    \hfil
    \subfigure[$(N-2)$-electron states.]{
    \includegraphics[width=0.4\textwidth]{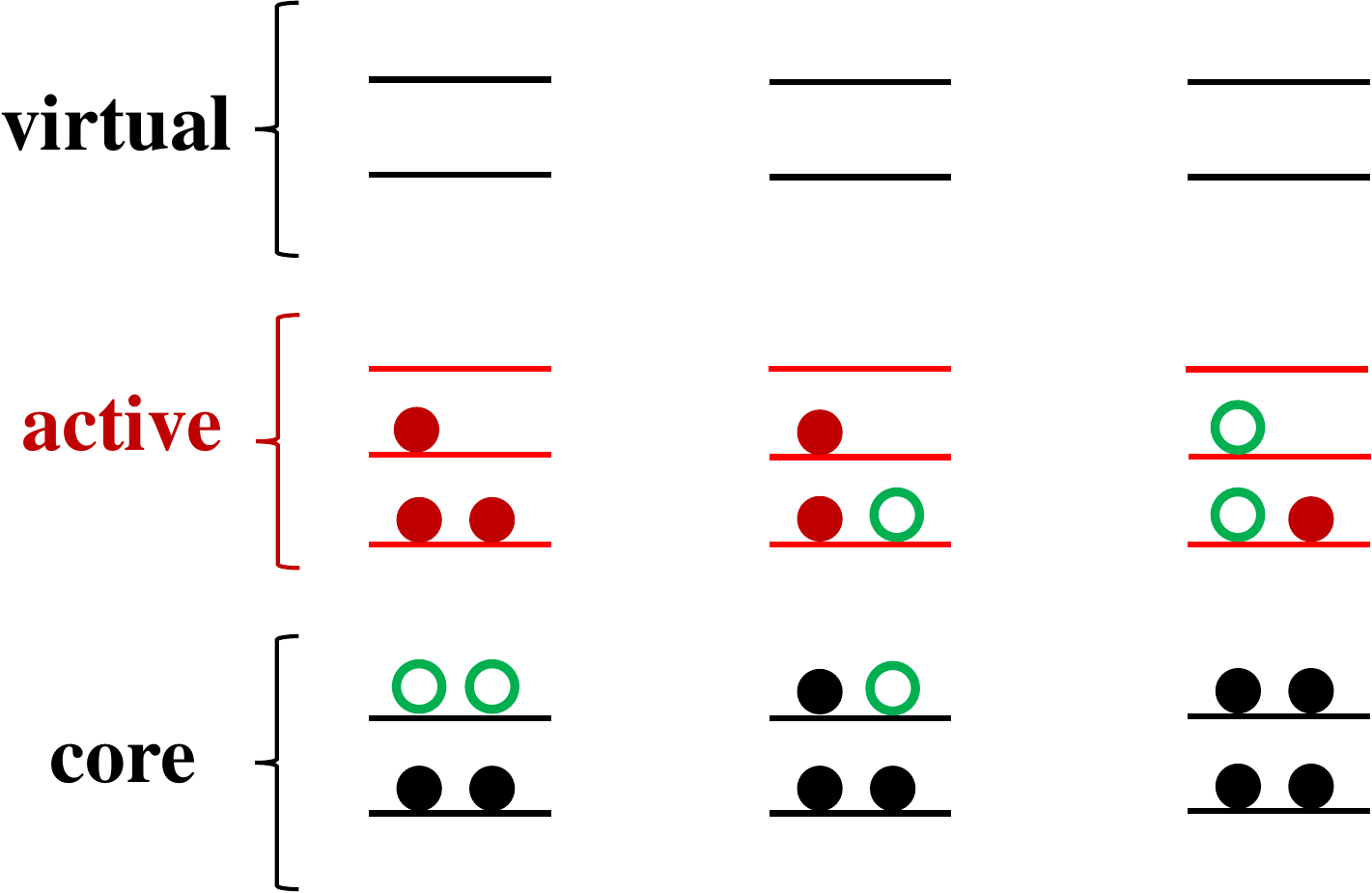}
    }
    \caption{
    The zeroth-order $(N+2)$- and $(N-2)$-electron states in MR-ppRPA that can couple with the CASSCF reference.
    The green filled circle means an electron is created, 
    while the green empty circle means an electron is annihilated.}
    \label{fig:pp-and-hh-CAS}
\end{figure*}

With the CASSCF reference, 
the zeroth-order $(N+2)$-electron states $|\Phi^{+2}_A\rangle$ 
that can couple with the reference state $|\Phi_0\rangle$ through Eqs. \eqref{eq:mat-pp-1}-\eqref{eq:mat-pp-2}
can be categorized into three classes, viz.,
\begin{equation} \begin{aligned}
|\Phi_A^{+2}\rangle \in \begin{cases}
    |\Theta^{ab}\rangle|\Xi^{N_{\text{act}}}_0\rangle,\, (|\Theta^{ab}\rangle = \hat{a}^\dagger \hat{b}^\dagger |\Theta_0\rangle, a>b), \\
    |\Theta^{a}\rangle|\Xi_\lambda^{N_{\text{act}}+1}\rangle,\\
    |\Theta_0\rangle|\Xi_\lambda^{N_{\text{act}}+2}\rangle.
\end{cases}
\end{aligned}\label{eq:CAS-ex-pp}\end{equation}
Likewise, the zeroth-order $(N-2)$-electron states $|\Phi^{-2}_I\rangle$ coupled with $|\Phi_0\rangle$ can be classified into the following three classes, viz.,
\begin{equation}\begin{aligned}
|\Phi^{-2}_I\rangle \in \begin{cases}
    |\Theta_{ij}\rangle|\Xi^{N_{\text{act}}}_0\rangle,\, (|\Theta_{ij}\rangle = \hat{j}\hat{i}|\Theta_0\rangle, i>j), \\
    |\Theta_{i}\rangle|\Xi_\lambda^{N_{\text{act}}-1}\rangle, \\
    |\Theta_0\rangle|\Xi_\lambda^{N_{\text{act}}-2}\rangle.
\end{cases}
\end{aligned}\label{eq:CAS-ex-hh}\end{equation}
Fig. \ref{fig:pp-and-hh-CAS} illustrates schematically the relevant excitations in Eqs. \eqref{eq:CAS-ex-pp} and \eqref{eq:CAS-ex-hh}. 
The matrices $\mathbf{A}^+, \mathbf{A}^-$ and $\mathbf{C}$ with CASSCF reference can then be written as $3\times 3$ block matrices.
Detailed expressions for their elements are presented in the Supplemental Material.

\subsection{Perturbative analysis}\label{sec:theory-pt}
To gain a deeper understanding of the behavior of the diagrammatic resummation, 
we perform a perturbative analysis of the MR-RPA correlation energies starting from the coupled cluster like equations 
(Eqs. \eqref{eq:drpa-cc}, \eqref{eq:rpax-cc}, and \eqref{eq:cc-pp}). 
For MR-ppRPA, we make a perturbative expansion for $\mathbf{U}$ in the orders of the perturbation $\hat{V}$,
\begin{align}
    \mathbf{U} &= \mathbf{U}^{(1)} + \mathbf{U}^{(2)} + \cdots. \label{eq:ampl-pp-pt}
\end{align}
Recognizing that $\mathbf{C}$ in Eq. \eqref{eq:mat-pp-2} is a first-order quantity, 
and the first and second terms of $\mathbf{A}^+$ \eqref{eq:mat-pp-1} and $\mathbf{A}^-$ \eqref{eq:mat-pp-3} 
are of zeroth and first orders, respectively, 
we can establish a recursive equation for $\mathbf{U}^{(n)}$
using Eq. \eqref{eq:cc-pp},
\begin{align}
    U_{HP}^{(n)} = -\frac{1}{\omega^{N-2}_H+\omega^{N+2}_P}&\left[
    \bC^\dagger\delta_{n1}+\mathbf{V}^-\mathbf{U}^{(n-1)}+\mathbf{U}^{(n-1)}\mathbf{V}^+\right.\nonumber\\
    &\left. +\sum_{i=1}^{n-2}\mathbf{U}^{(i)}\mathbf{C}\mathbf{U}^{(n-1-i)}\right]_{HP},\label{eq:ccpt-pp}
\end{align}
where the elements of $\mathbf{V}^+$ and $\mathbf{V}^-$ 
are the second terms in Eqs. \eqref{eq:mat-pp-1} and \eqref{eq:mat-pp-3}, respectively. 
Eq. \eqref{eq:ccpt-pp} allows $\mathbf{U}^{(n)}$ to be solved recursively starting from the first order,
\begin{align}
    U_{HP}^{(1)} = -\frac{C_{PH}^*}{\omega^{N-2}_H + \omega^{N+2}_P}.
\end{align}
Using Eq. \eqref{eq:ecor-cc-pp}, we find the $(n+1)$-th order ppRPA correlation energy as
\begin{align}
    \Delta E^{\text{ppRPA},(n+1)} = \tr\left(\mathbf{C}\mathbf{U}^{(n)} \right).
\end{align}

For MR-dRPA and MR-RPAx, perturbative analysis can be conducted in the same way. 
With $\mathbf{T} = \sum_{n=1}^{\infty}\mathbf{T}^{(n)}$ for MR-RPAx,
the resulting equations are summarized as follows:
\begin{align}
    \Delta E^{\text{RPAx},(n+1)} = \frac{1}{2}\tr\left(\mathbf{\bar{B}T}^{(n)}\right)- \Delta E^{(2),a}\delta_{n1},
\end{align}
\begin{align}
    T^{(n)}_{LR} = -\frac{1}{\omega_L+\omega_R}&\left[\mathbf{\bar{B}}^*\delta_{n1} + \mathbf{\bar{V}}^*\mathbf{T}^{(n-1)} + \mathbf{T}^{(n-1)}\mathbf{\bar{V}}\right.\nonumber\\
    &\left. +\sum_{i=1}^{n-2}\mathbf{T}^{(i)}\mathbf{\bar{B}}\mathbf{T}^{(n-1-i)}\right]_{LR},
\end{align}
\begin{align}
    T^{(1)}_{LR} = -\frac{\bar{B}^*_{LR}}{\omega_L+\omega_R},
\end{align}
with $\bar{V}_{LR}$ being the second term of Eq. \eqref{eq:mat-ph-A}. 
Likewise, the resulting equations for MR-dRPA are
\begin{align}
    \Delta E^{\text{dRPA},(n+1)} = \frac{1}{2}\tr\left(\mathbf{BT}^{(n)}\right),
\end{align}
\begin{align}
    T^{(n)}_{LR} = -\frac{1}{\omega_L+\omega_R}&\left[\mathbf{B}^*\delta_{n1} + \mathbf{V}^*\mathbf{T}^{(n-1)} + \mathbf{T}^{(n-1)}\mathbf{V}\right.\nonumber\\
    &\left. +\sum_{i=1}^{n-2}\mathbf{T}^{(i)}\mathbf{B}\mathbf{T}^{(n-1-i)}\right]_{LR},
\end{align}
\begin{align}
    T^{(1)}_{LR} = -\frac{B^*_{LR}}{\omega_L+\omega_R},
\end{align}
with $V_{LR}$ being the second term of Eq. \eqref{eq:mat-ph-A-d}.

\section{Results and discussion}\label{sec:results}
We have implemented MR-RPAx and MR-ppRPA based on the \textsc{PySCF} package\cite{sun_recent_2020}. 
For comparison, we use the previously obtained MR-dRPA\cite{wang_generalized_2025}, spin-adapted \textit{ab initio} density matrix renormalization group (DMRG)\cite{white_density_1992,chan_density_2011,xiang_distributed_2024} 
and strongly-contracted NEVPT2\cite{angeli_introduction_2001} (SC-NEVPT2) results.
The \textsc{MOKIT}\cite{zou_molecular_2024} program was used to prepare the initial active orbitals for \ce{H2O}.
Since we mainly focus on the performances of the proposed MR-RPAx and MR-ppRPA in this work, 
we use the plasmon formulae for computing the correlation energies, 
while low-scaling formulations for large molecules will be described elsewhere.

\subsection{Size-extensivity}
MR-dRPA, MR-RPAx and MR-ppRPA only resum linked diagrams, and hence the correlation energies should naturally be size-extensive (i.e., scale linearly with the system size),
due to the property of linked diagrams.
We illustrate this feature with a \ce{Li4} model, 
where two \ce{Li2} are separated by 1000\,\AA. 
The \ce{Li-Li} bond in \ce{Li2} is set as 3\,\AA.
Results calculated using the cc-pVDZ basis set are shown in Table \ref{tab:Li4}, which demonstrates the size-extensivity
of all MR-RPA variants.

\begin{table}[hbt]
    \renewcommand{\arraystretch}{1.2}
    \setlength{\tabcolsep}{2mm}
    \caption{
Size-extensivity test for multi-reference methods using the \ce{Li4} model. 
Energies (in Hartree) are calculated with the cc-pVDZ basis set. 
The active spaces (i.e., CAS(2,2) for \ce{Li2} and CAS(4,4) for \ce{Li4}) contain $\sigma$ bonding orbitals and their corresponding anti-bonding
orbitals.
    }
    \centering

    \scalebox{0.95}{
    
    \begin{tabular}{cccc}
    \hline\hline
      Method       &  2$\times E$(\ce{Li2}) & $E$(\ce{Li2}$\cdots$\ce{Li2}) & Difference \\
    \hline
      CASSCF       &  \num{-29.76044742}     & \num{-29.76044742 }   &  \num{1e-10}   \\
      SC-NEVPT2    &  \num{-0.01551936 }     & \num{-0.01551937  }   &  \num{8e-9}  \\
      MR-dRPA      &  \num{-0.04806161 }     & \num{-0.04806160  }   &  \num{5e-9}   \\
      MR-RPAx      &  \num{-0.15236894}      & \num{-0.15236893}     &  \num{9e-9}   \\
      MR-ppRPA     &  \num{-0.01209069}      & \num{-0.01209069}     &  \num{1e-10}   \\
    \hline\hline
    \end{tabular}
    }
    \label{tab:Li4}
\end{table}

%For size-consistency, defined as that the energy of a separated dimer is equal to the sum of two isolated monomers, the problem is divided in two cases. In the cases where the separated dimer is in a product state (e.g. the \ce{Li4} model), size-consistency is also fulfilled as a property of linked diagrams.\cite{hanrath_concepts_2009} In the cases where the separated dimer is in an entangled state (e.g. \ce{N2} dissociated into two \ce{N} atoms), it raises a question on how to calculate the correlation energies of open-shell systems with non-singlet ground states. The simplest scheme is to directly apply the plasmon (Eqs. \eqref{eq:ecor-plas-drpa}, \eqref{eq:ecor-rpax-plasmon}, and \eqref{eq:ecor-plas-pprpa}) or CC formulae (Eqs. \eqref{eq:ecor-cc-drpa}, \eqref{eq:ecor-cc-rpax}, and \eqref{eq:ecor-cc-pp}). However, for MR-RPAx and MR-ppRPA, this scheme breaks the spin-multiplet, viz. the three components of a triplet state differ in energies. As a result, size-consistency cannot be well defined for MR-RPAx and MR-ppRPA. This problem may be solved by using the singlet-embedding technique for the non-singlet correlation energies.\cite{tatsuaki_interaction-round--face_2000} MR-dRPA is free of this issue, and size-consistency is observed numerically. A more detailed discussion will be presented elsewhere.

\subsection{Potential energy curves}

\begin{figure*}[htb]
\includegraphics[width=0.98\textwidth]{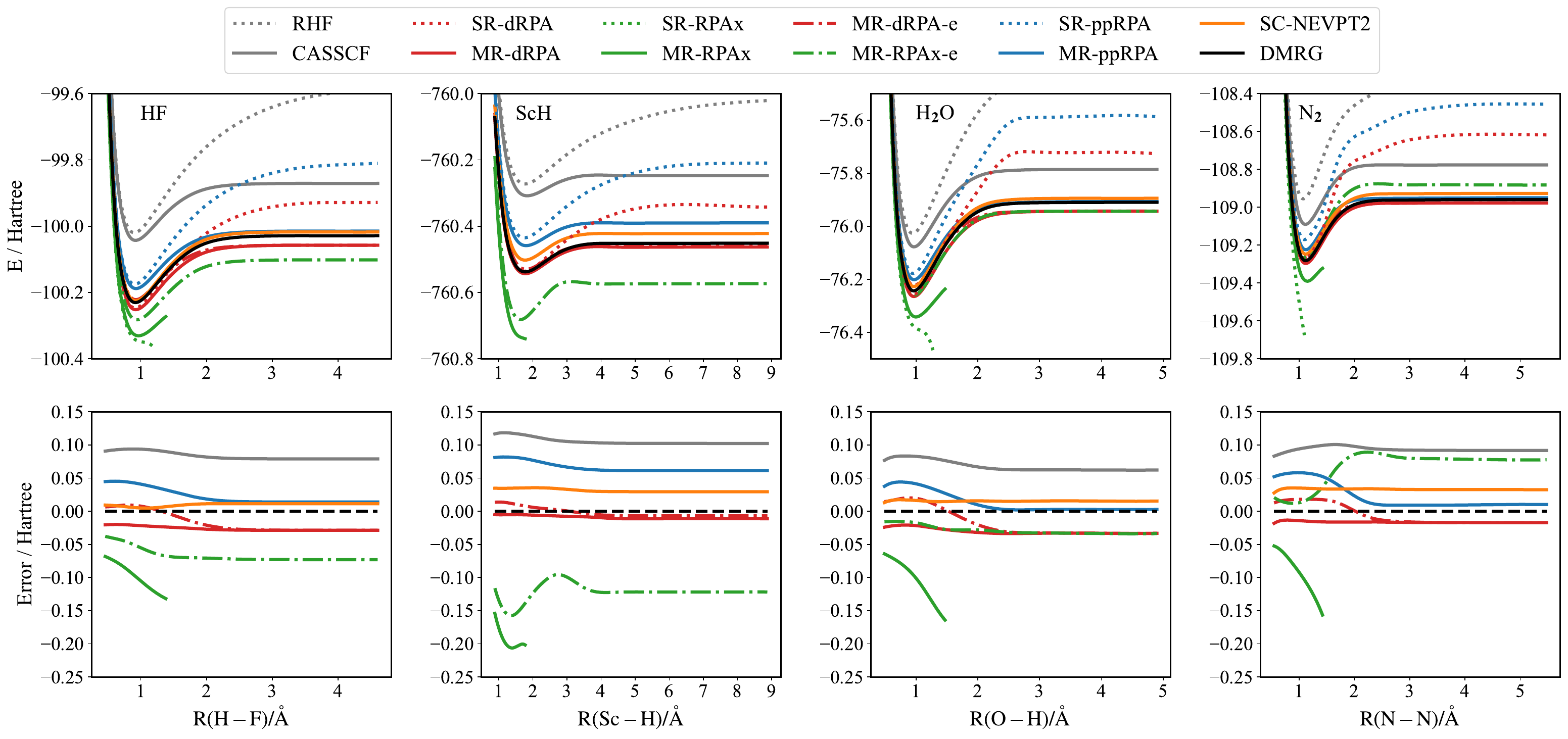}
\caption{PECs of \ce{HF}, \ce{ScH}, \ce{H2O} and \ce{N2} calculated using different methods with the cc-pVDZ basis set. Errors of multi-reference results with respect to the nearly exact DMRG results\cite{wang_generalized_2025} are shown below. Note that the CASSCF errors are scaled by \num{0.4} to fit into the same figure for comparison with other methods.}\label{fig:PECs}
\end{figure*}

%\begin{figure}[htb]
%    \centering
%
%    \subfigure[Absolute error in milli-Hartree (mH) of the dissociation energies calculated by multi-reference methods]{
%    \includegraphics[width=.98\linewidth]{figures/Edis-4-mol.pdf}
%    }
%
%    \subfigure[Non-parallel errors (NPEs) in milli-Hartree (mH) of the dissociation PECs calculated by multi-reference methods]{
%    \includegraphics[width=.98\linewidth]{figures/NPE-4-mol.pdf}
%    }
%
%    \caption{Comparison of different multi-reference methods with respect to the nearly exact DMRG results using the cc-pVDZ basis set}
%    \label{fig:HF-ScH-H2O-N2}
%\end{figure}

We apply MR-RPAx and MR-ppRPA to compute potential energy curves (PECs) 
of the previously investigated molecules\cite{wang_generalized_2025} (\ce{HF}, \ce{ScH}, \ce{H2O}, and \ce{N2}) 
using the cc-pVDZ\cite{dunning_gaussian_1989} basis set. 
The nearly exact spin-adapted DMRG results are employed as reference, and SC-NEVPT2 results are also presented for comparison in Fig. \ref{fig:PECs}.
Detailed numerical results are shown in the Supplemental Material.
Again, we find that all SR-RPA methods, including SR-dRPA, 
SR-RPAx, and SR-ppRPA, fail at stretched geometries, 
indicating the breakdown of standard single reference perturbation theory based on the single determinant reference. 
Both MR-dRPA and MR-ppRPA resolve this issue 
by including the strong correlation in the active space at the zeroth order,
and also show significant improvements over the CASSCF reference 
by adding the missing dynamic correlation.
For three (\ce{HF}, \ce{H2O} and \ce{N2}) of the four investigated molecules,
MR-ppRPA displays the best accuracy at the dissociated limit among the MR-RPA methods 
and surpasses SC-NEVPT2 for \ce{H2O} and \ce{N2} at large bond distances.
However, around the equilibrium geometries, MR-ppRPA tends to underestimate the correlation energy.

In contrast, MR-RPAx fails at stretched geometries due to the instability,
where Eq. \eqref{eq:ph-diag} gives imaginary roots, like its single reference counterpart (SR-RPAx).
However, we find that the instability problem of MR-RPAx can be largely rescued 
by neglecting the screening from the excitation within the active space,
that is, only retaining the first three types of excitations in Eq. \eqref{eq:CAS-ex-ph}.
%The key argument behind is that with a properly selected active space, the instability should be an active-space behavior.
We denote the MR-dRPA and MR-RPAx methods after such treatment as MR-dRPA-e and MR-RPAx-e, respectively.
Figure \ref{fig:PECs} shows that MR-RPAx-e avoids the instability problem across all the investigated bond lengths, 
and gives a qualitatively correct description of the molecular dissociation.
For MR-dRPA-e, we find that it agrees well with MR-dRPA at longer bond distances,
but gives higher energies than MR-dRPA at the equilibrium geometries due to the removal of the active space screening.
%As a result, the PES of MR-dRPA-e is nearly parallel to that of MR-ppRPA.
Compared to the nearly exact DMRG results, 
we find that both MR-dRPA and MR-RPAx overestimate the correlation energies, 
while MR-ppRPA underestimates the correlation energies.

\subsection{Perturbative analysis}
\begin{table*}[hbt]
    \renewcommand{\arraystretch}{1.5}
    \setlength{\tabcolsep}{2.5mm}
    \centering
    \caption{Perturbative analysis of various SR- and MR-RPA methods for the correlation energies (in Hartree) 
    up to the fifth order for the \ce{HF} molecule with the cc-pVDZ basis set. 
    Correlation energies are defined with the RHF and CASSCF references for SR and MR methods, respectively.}
    
\begin{tabular}{c|cc|ccccc}
    \hline\hline
    \makecell{$R(\text{H-F})/R_0$\\$(R_0=0.92$\,\AA)} & \multicolumn{2}{c|}{Method} & $\Delta E$ & $\Delta E^{(2)}$ & $\Delta E^{(3)}$ & $\Delta E^{(4)}$ & $\Delta E^{(5)}$ \\
    \hline \multirow{6}{*}{1.0}
   & \multirow{3}{*}{SR-} & dRPA & \num{-0.227763} &\num{-0.299221} & \num{0.111245}   & \num{-0.068298} & \num{0.051618} \\
   & &RPAx & \num{-0.323674} &\num{-0.203910} & \num{-0.068568}  & \num{-0.033057} & \num{-0.008748} \\
   & &ppRPA & \num{-0.155538} &\num{-0.203910} & \num{0.065891}   &  \num{-0.024878} & \num{0.010938} \\
   \cline{2-8}
   & \multirow{3}{*}{MR-} &dRPA & \num{-0.208960} &\num{-0.264927} & \num{0.084039}   & \num{-0.045821} & \num{0.030583} \\
   & &RPAx  & \num{-0.286035} &\num{-0.190258} & \num{-0.058419}  & \num{-0.026825} & \num{-0.005491} \\
   & &ppRPA & \num{-0.145296} &\num{-0.186275} & \num{0.054066}  & \num{-0.017656} & \num{0.006276} \\
    \hline \multirow{6}{*}{3.0}
   & \multirow{3}{*}{SR-}  &dRPA & \num{-0.291682} &\num{-0.519458} &  \num{0.658132}  &  \num{-1.534139}   &   \num{4.322695}  \\
   & &RPAx  & / &\num{-0.319826} &  \num{-0.320402} &  \num{-0.699236}   &   \num{-1.162561}  \\
   & &ppRPA & \num{-0.193691} &\num{-0.319826} &  \num{0.312202}  &  \num{-0.578482}   &   \num{1.312409}  \\\cline{2-8}
   & \multirow{3}{*}{MR-}  & dRPA  & \num{-0.186816} &\num{-0.232932} &  \num{0.065333}  &  \num{-0.029478}   &   \num{0.016512}  \\ 
   & &RPAx  & / &\num{-0.190790} &  \num{-0.072174} &  \num{-0.382439}   &   \num{-0.185910} \\
   & &ppRPA & \num{-0.144077} &\num{-0.182104} &  \num{0.049215}  &  \num{-0.014997}   &   \num{0.004873}  \\
    \hline\hline
\end{tabular}
    \label{tab:HF-RPA-low-orders}
\end{table*}

We analyze single- and multi-reference RPA correlation energies using the perturbative analysis developed in Sec. \ref{sec:theory-pt}, 
in order to check the convergence behavior of the diagrammatic resummation and gain a deeper understanding of the performances of the three MR-RPA variants.
The perturbation expansion of the RPA correlation energies up to
the fifth order for the \ce{HF} molecule, 
calculated using the cc-pVDZ basis set at both the equilibrium and stretched geometries, 
are summarized in Table \ref{tab:HF-RPA-low-orders}.
For each RPA variant, 
we find that the corresponding multi-reference theory exhibits faster convergence with respect to 
the perturbation order than its single reference counterpart, 
especially at the stretched geometry, 
indicating the importance of using an interacting $\hH_0$ in this case. At the stretched geometry,
%, MR correlation energies at the third and higher orders can be smaller by magnitudes. 
the correlation energies are found to be divergent in the perturbation series for all the three SR-RPA variants, 
as the magnitude of $\Delta E^{(n)}$ is increasingly large,
indicating the breakdown of standard MBPT in such case. 
MR-dRPA and MR-ppRPA resolves the divergence, 
while MR-RPAx still suffers from it. 
Among the three MR-RPA variants, 
MR-RPAx shows poorest behavior at the stretched geometry where the instability happens, 
leading to imaginary roots in Eq. \eqref{eq:ph-diag}. 
This may also be connected to the observation that for both SR-RPAx and MR-RPAx, 
$\Delta E^{(n)}$ is negative at each order and becomes diverging at large $n$.
In contrast, $\Delta E^{(n)}$ for MR-dRPA and MR-ppRPA series are of alternating signs, 
and the magnitude is decreasing as $n$ increases.

As shown in Table \ref{tab:HF-RPA-low-orders}, the SR- and MR-dRPA correlation energies beyond the second order 
are all positive, viz., $\Delta E - \Delta E^{(2)}>0$.
Therefore, the overestimation of correlation energies in dRPA is mainly due to the lack of
exchange at the second order. However, the reason for the overestimation of correlation energies in RPAx
is different, which is mainly due to the negative contributions at each order.
Therefore, we can attribute the better accuracy of dRPA to the error cancellation between the second and higher orders.
For SR-ppRPA, the second order energy is exactly the MP2 correlation energy, 
which is also the second-order SR-RPAx energy. 
However, the second-order MR-ppRPA energy is not identical to 
the second-order perturbation energy based on $\hH_{\text{Dyall}}$. 
For comparison, 
the SC-NEVPT2 correlation energies at $R/R_0=1.0$ and $3.0$ are \num{-0.182330} and \num{-0.146828} Hartrees, 
respectively.
The latter is much higher than the second-order MR-ppRPA energy. 
Therefore,
the good accuracy of MR-ppRPA at stretched geometries can be attributed to 
the error cancellations between the second and higher orders.

\section{Conclusion}\label{sec:conclusion}
In this work, we introduce two new multi-reference methods, namely, MR-RPAx and MR-ppRPA, 
for the electron correlation energies, 
generalizing our previously developed diagrammatic approach for MR-dRPA\cite{wang_generalized_2025}.
Three equivalent mathematical expressions for the correlation energy, 
i.e., the imaginary-frequency formula, plasmon formula, and coupled cluster like formula, 
are derived for all the three RPA variants. 
We numerically compare the three MR-RPA methods and their single-reference counterparts for prototypical molecules. 
We find that MR-dRPA offers the most balanced treatment for the PECs among all the RPA methods, 
although MR-ppRPA tends to perform better at the dissociated limit.
A perturbative analysis reveals that a major reason for such numerical behaviors of
MR-dRPA and MR-ppRPA are the error cancellations between the second and higher orders.
We observe that MR-phRPA (MR-dRPA or MR-RPAx) and MR-ppRPA overestimate and underestimate the correlation energies, 
respectively. 
This suggests that combining these two channels\cite{tahir_comparing_2019,scuseria_particle-particle_2013}
can potentially deliver more accurate energies. 
While the present study only focuses on systems with singlet ground states,
extension to open-shell systems with nonsinglet ground states is another interesting
direction. Moreover, to treat systems with large active spaces,
contraction approximations can be adopted in the RPA equation.
Work along these lines is being undertaken in our laboratory.

%However, this is a non-trivial task, as straightforward combination of the ph and pp channels failed in the single reference theory.\cite{tahir_comparing_2019} Coupling the two channels in a coupled cluster manner may result in a more balanced description of the correlation energy.\cite{scuseria_particle-particle_2013}

\section*{Acknowledgment}
This work was supported by the Innovation Program for Quantum Science and Technology (Grant No. 2023ZD0300200) and the Fundamental Research Funds for the Central Universities.

\let\oldaddcontentsline\addcontentsline% Store \addcontentsline
\renewcommand{\addcontentsline}[3]{}% Make \addcontentsline a no-op

\bibliographystyle{apsrev4-2}
\bibliography{mrrpa}

\begin{thebibliography}{96}%
\makeatletter
\providecommand \@ifxundefined [1]{%
 \@ifx{#1\undefined}
}%
\providecommand \@ifnum [1]{%
 \ifnum #1\expandafter \@firstoftwo
 \else \expandafter \@secondoftwo
 \fi
}%
\providecommand \@ifx [1]{%
 \ifx #1\expandafter \@firstoftwo
 \else \expandafter \@secondoftwo
 \fi
}%
\providecommand \natexlab [1]{#1}%
\providecommand \enquote  [1]{``#1''}%
\providecommand \bibnamefont  [1]{#1}%
\providecommand \bibfnamefont [1]{#1}%
\providecommand \citenamefont [1]{#1}%
\providecommand \href@noop [0]{\@secondoftwo}%
\providecommand \href [0]{\begingroup \@sanitize@url \@href}%
\providecommand \@href[1]{\@@startlink{#1}\@@href}%
\providecommand \@@href[1]{\endgroup#1\@@endlink}%
\providecommand \@sanitize@url [0]{\catcode `\\12\catcode `\$12\catcode
  `\&12\catcode `\#12\catcode `\^12\catcode `\_12\catcode `\%12\relax}%
\providecommand \@@startlink[1]{}%
\providecommand \@@endlink[0]{}%
\providecommand \url  [0]{\begingroup\@sanitize@url \@url }%
\providecommand \@url [1]{\endgroup\@href {#1}{\urlprefix }}%
\providecommand \urlprefix  [0]{URL }%
\providecommand \Eprint [0]{\href }%
\providecommand \doibase [0]{https://doi.org/}%
\providecommand \selectlanguage [0]{\@gobble}%
\providecommand \bibinfo  [0]{\@secondoftwo}%
\providecommand \bibfield  [0]{\@secondoftwo}%
\providecommand \translation [1]{[#1]}%
\providecommand \BibitemOpen [0]{}%
\providecommand \bibitemStop [0]{}%
\providecommand \bibitemNoStop [0]{.\EOS\space}%
\providecommand \EOS [0]{\spacefactor3000\relax}%
\providecommand \BibitemShut  [1]{\csname bibitem#1\endcsname}%
\let\auto@bib@innerbib\@empty
%</preamble>
\bibitem [{\citenamefont {Park}\ \emph {et~al.}(2020)\citenamefont {Park},
  \citenamefont {Al-Saadon}, \citenamefont {MacLeod}, \citenamefont
  {Shiozaki},\ and\ \citenamefont {Vlaisavljevich}}]{park_multireference_2020}%
  \BibitemOpen
  \bibfield  {author} {\bibinfo {author} {\bibfnamefont {J.~W.}\ \bibnamefont
  {Park}}, \bibinfo {author} {\bibfnamefont {R.}~\bibnamefont {Al-Saadon}},
  \bibinfo {author} {\bibfnamefont {M.~K.}\ \bibnamefont {MacLeod}}, \bibinfo
  {author} {\bibfnamefont {T.}~\bibnamefont {Shiozaki}},\ and\ \bibinfo
  {author} {\bibfnamefont {B.}~\bibnamefont {Vlaisavljevich}},\ }\href
  {https://doi.org/10.1021/acs.chemrev.9b00496} {\bibfield  {journal} {\bibinfo
   {journal} {Chem. Rev.}\ }\textbf {\bibinfo {volume} {120}},\ \bibinfo
  {pages} {5878} (\bibinfo {year} {2020})}\BibitemShut {NoStop}%
\bibitem [{\citenamefont {Andersson}\ \emph {et~al.}(1992)\citenamefont
  {Andersson}, \citenamefont {Malmqvist},\ and\ \citenamefont
  {Roos}}]{andersson_secondorder_1992}%
  \BibitemOpen
  \bibfield  {author} {\bibinfo {author} {\bibfnamefont {K.}~\bibnamefont
  {Andersson}}, \bibinfo {author} {\bibfnamefont {P.}~\bibnamefont
  {Malmqvist}},\ and\ \bibinfo {author} {\bibfnamefont {B.~O.}\ \bibnamefont
  {Roos}},\ }\href {https://doi.org/10.1063/1.462209} {\bibfield  {journal}
  {\bibinfo  {journal} {J. Chem. Phys.}\ }\textbf {\bibinfo {volume} {96}},\
  \bibinfo {pages} {1218} (\bibinfo {year} {1992})}\BibitemShut {NoStop}%
\bibitem [{\citenamefont {Angeli}\ \emph {et~al.}(2001)\citenamefont {Angeli},
  \citenamefont {Cimiraglia}, \citenamefont {Evangelisti}, \citenamefont
  {Leininger},\ and\ \citenamefont {Malrieu}}]{angeli_introduction_2001}%
  \BibitemOpen
  \bibfield  {author} {\bibinfo {author} {\bibfnamefont {C.}~\bibnamefont
  {Angeli}}, \bibinfo {author} {\bibfnamefont {R.}~\bibnamefont {Cimiraglia}},
  \bibinfo {author} {\bibfnamefont {S.}~\bibnamefont {Evangelisti}}, \bibinfo
  {author} {\bibfnamefont {T.}~\bibnamefont {Leininger}},\ and\ \bibinfo
  {author} {\bibfnamefont {J.-P.}\ \bibnamefont {Malrieu}},\ }\href
  {https://doi.org/10.1063/1.1361246} {\bibfield  {journal} {\bibinfo
  {journal} {J. Chem. Phys.}\ }\textbf {\bibinfo {volume} {114}},\ \bibinfo
  {pages} {10252} (\bibinfo {year} {2001})}\BibitemShut {NoStop}%
\bibitem [{\citenamefont {Lyakh}\ \emph {et~al.}(2012)\citenamefont {Lyakh},
  \citenamefont {Musiał}, \citenamefont {Lotrich},\ and\ \citenamefont
  {Bartlett}}]{lyakh_multireference_2012}%
  \BibitemOpen
  \bibfield  {author} {\bibinfo {author} {\bibfnamefont {D.~I.}\ \bibnamefont
  {Lyakh}}, \bibinfo {author} {\bibfnamefont {M.}~\bibnamefont {Musiał}},
  \bibinfo {author} {\bibfnamefont {V.~F.}\ \bibnamefont {Lotrich}},\ and\
  \bibinfo {author} {\bibfnamefont {R.~J.}\ \bibnamefont {Bartlett}},\ }\href
  {https://doi.org/10.1021/cr2001417} {\bibfield  {journal} {\bibinfo
  {journal} {Chem. Rev.}\ }\textbf {\bibinfo {volume} {112}},\ \bibinfo {pages}
  {182} (\bibinfo {year} {2012})}\BibitemShut {NoStop}%
\bibitem [{\citenamefont {Evangelista}(2018)}]{evangelista_perspective_2018}%
  \BibitemOpen
  \bibfield  {author} {\bibinfo {author} {\bibfnamefont {F.~A.}\ \bibnamefont
  {Evangelista}},\ }\href {https://doi.org/10.1063/1.5039496} {\bibfield
  {journal} {\bibinfo  {journal} {J. Chem. Phys.}\ }\textbf {\bibinfo {volume}
  {149}},\ \bibinfo {pages} {030901} (\bibinfo {year} {2018})}\BibitemShut
  {NoStop}%
\bibitem [{\citenamefont {Adam}\ \emph {et~al.}(2025)\citenamefont {Adam},
  \citenamefont {Waigum},\ and\ \citenamefont
  {K{\"o}hn}}]{adam2025multireference}%
  \BibitemOpen
  \bibfield  {author} {\bibinfo {author} {\bibfnamefont {R.~G.}\ \bibnamefont
  {Adam}}, \bibinfo {author} {\bibfnamefont {A.}~\bibnamefont {Waigum}},\ and\
  \bibinfo {author} {\bibfnamefont {A.}~\bibnamefont {K{\"o}hn}},\ }\href@noop
  {} {\bibfield  {journal} {\bibinfo  {journal} {WIREs Comput. Mol. Sci.}\
  }\textbf {\bibinfo {volume} {15}},\ \bibinfo {pages} {e70023} (\bibinfo
  {year} {2025})}\BibitemShut {NoStop}%
\bibitem [{\citenamefont {Fetter}\ and\ \citenamefont
  {Walecka}(1971)}]{fetter_quantum_1971}%
  \BibitemOpen
  \bibfield  {author} {\bibinfo {author} {\bibfnamefont {A.}~\bibnamefont
  {Fetter}}\ and\ \bibinfo {author} {\bibfnamefont {J.}~\bibnamefont
  {Walecka}},\ }\href@noop {} {\emph {\bibinfo {title} {Quantum {Theory} of
  {Many}-{Particle} {System}}}},\ International {Series} in {Pure} and
  {Applied} {Physics}\ (\bibinfo  {publisher} {MacGraw-Hill},\ \bibinfo
  {address} {New York},\ \bibinfo {year} {1971})\BibitemShut {NoStop}%
\bibitem [{\citenamefont {Negele}\ and\ \citenamefont
  {Orland}(1998)}]{negele_quantum_1998}%
  \BibitemOpen
  \bibfield  {author} {\bibinfo {author} {\bibfnamefont {J.~W.}\ \bibnamefont
  {Negele}}\ and\ \bibinfo {author} {\bibfnamefont {H.}~\bibnamefont
  {Orland}},\ }\href {https://doi.org/10.1201/9780429497926} {\emph {\bibinfo
  {title} {Quantum {Many}-particle {Systems}}}}\ (\bibinfo  {publisher} {CRC
  Press},\ \bibinfo {address} {Boca Raton},\ \bibinfo {year}
  {1998})\BibitemShut {NoStop}%
\bibitem [{\citenamefont {Martin}\ \emph {et~al.}(2016)\citenamefont {Martin},
  \citenamefont {Reining},\ and\ \citenamefont
  {Ceperley}}]{martin_interacting_2016}%
  \BibitemOpen
  \bibfield  {author} {\bibinfo {author} {\bibfnamefont {R.~M.}\ \bibnamefont
  {Martin}}, \bibinfo {author} {\bibfnamefont {L.}~\bibnamefont {Reining}},\
  and\ \bibinfo {author} {\bibfnamefont {D.~M.}\ \bibnamefont {Ceperley}},\
  }\href@noop {} {\emph {\bibinfo {title} {Interacting {Electrons}: {Theory}
  and {Computational} {Approaches}}}}\ (\bibinfo  {publisher} {Cambridge
  University Press},\ \bibinfo {year} {2016})\BibitemShut {NoStop}%
\bibitem [{\citenamefont {Wang}\ \emph {et~al.}(2025)\citenamefont {Wang},
  \citenamefont {Fang},\ and\ \citenamefont {Li}}]{wang_generalized_2025}%
  \BibitemOpen
  \bibfield  {author} {\bibinfo {author} {\bibfnamefont {Y.}~\bibnamefont
  {Wang}}, \bibinfo {author} {\bibfnamefont {W.-H.}\ \bibnamefont {Fang}},\
  and\ \bibinfo {author} {\bibfnamefont {Z.}~\bibnamefont {Li}},\ }\href
  {https://doi.org/10.1021/acs.jpclett.5c00258} {\bibfield  {journal} {\bibinfo
   {journal} {J. Phys. Chem. Lett.}\ }\textbf {\bibinfo {volume} {16}},\
  \bibinfo {pages} {3047} (\bibinfo {year} {2025})}\BibitemShut {NoStop}%
\bibitem [{\citenamefont {Furche}(2001)}]{furche_molecular_2001}%
  \BibitemOpen
  \bibfield  {author} {\bibinfo {author} {\bibfnamefont {F.}~\bibnamefont
  {Furche}},\ }\href {https://doi.org/10.1103/PhysRevB.64.195120} {\bibfield
  {journal} {\bibinfo  {journal} {Phys. Rev. B}\ }\textbf {\bibinfo {volume}
  {64}},\ \bibinfo {pages} {195120} (\bibinfo {year} {2001})}\BibitemShut
  {NoStop}%
\bibitem [{\citenamefont {Zhu}\ \emph {et~al.}(2010)\citenamefont {Zhu},
  \citenamefont {Toulouse}, \citenamefont {Savin},\ and\ \citenamefont
  {Ángyán}}]{zhu_range-separated_2010}%
  \BibitemOpen
  \bibfield  {author} {\bibinfo {author} {\bibfnamefont {W.}~\bibnamefont
  {Zhu}}, \bibinfo {author} {\bibfnamefont {J.}~\bibnamefont {Toulouse}},
  \bibinfo {author} {\bibfnamefont {A.}~\bibnamefont {Savin}},\ and\ \bibinfo
  {author} {\bibfnamefont {J.~G.}\ \bibnamefont {Ángyán}},\ }\href
  {https://doi.org/10.1063/1.3431616} {\bibfield  {journal} {\bibinfo
  {journal} {J. Chem. Phys.}\ }\textbf {\bibinfo {volume} {132}},\ \bibinfo
  {pages} {244108} (\bibinfo {year} {2010})}\BibitemShut {NoStop}%
\bibitem [{\citenamefont {Eshuis}\ and\ \citenamefont
  {Furche}(2011)}]{eshuis_parameter-free_2011}%
  \BibitemOpen
  \bibfield  {author} {\bibinfo {author} {\bibfnamefont {H.}~\bibnamefont
  {Eshuis}}\ and\ \bibinfo {author} {\bibfnamefont {F.}~\bibnamefont
  {Furche}},\ }\href {https://doi.org/10.1021/jz200238f} {\bibfield  {journal}
  {\bibinfo  {journal} {J. Phys. Chem. Lett.}\ }\textbf {\bibinfo {volume}
  {2}},\ \bibinfo {pages} {983} (\bibinfo {year} {2011})}\BibitemShut {NoStop}%
\bibitem [{\citenamefont {Paier}\ \emph {et~al.}(2012)\citenamefont {Paier},
  \citenamefont {Ren}, \citenamefont {Rinke}, \citenamefont {Scuseria},
  \citenamefont {Grüneis}, \citenamefont {Kresse},\ and\ \citenamefont
  {Scheffler}}]{paier_assessment_2012}%
  \BibitemOpen
  \bibfield  {author} {\bibinfo {author} {\bibfnamefont {J.}~\bibnamefont
  {Paier}}, \bibinfo {author} {\bibfnamefont {X.}~\bibnamefont {Ren}}, \bibinfo
  {author} {\bibfnamefont {P.}~\bibnamefont {Rinke}}, \bibinfo {author}
  {\bibfnamefont {G.~E.}\ \bibnamefont {Scuseria}}, \bibinfo {author}
  {\bibfnamefont {A.}~\bibnamefont {Grüneis}}, \bibinfo {author}
  {\bibfnamefont {G.}~\bibnamefont {Kresse}},\ and\ \bibinfo {author}
  {\bibfnamefont {M.}~\bibnamefont {Scheffler}},\ }\href
  {https://doi.org/10.1088/1367-2630/14/4/043002} {\bibfield  {journal}
  {\bibinfo  {journal} {New J. Phys.}\ }\textbf {\bibinfo {volume} {14}},\
  \bibinfo {pages} {043002} (\bibinfo {year} {2012})}\BibitemShut {NoStop}%
\bibitem [{\citenamefont {Harl}\ and\ \citenamefont
  {Kresse}(2008)}]{harl_cohesive_2008}%
  \BibitemOpen
  \bibfield  {author} {\bibinfo {author} {\bibfnamefont {J.}~\bibnamefont
  {Harl}}\ and\ \bibinfo {author} {\bibfnamefont {G.}~\bibnamefont {Kresse}},\
  }\href {https://doi.org/10.1103/PhysRevB.77.045136} {\bibfield  {journal}
  {\bibinfo  {journal} {Phys. Rev. B}\ }\textbf {\bibinfo {volume} {77}},\
  \bibinfo {pages} {045136} (\bibinfo {year} {2008})}\BibitemShut {NoStop}%
\bibitem [{\citenamefont {Harl}\ and\ \citenamefont
  {Kresse}(2009)}]{harl_accurate_2009}%
  \BibitemOpen
  \bibfield  {author} {\bibinfo {author} {\bibfnamefont {J.}~\bibnamefont
  {Harl}}\ and\ \bibinfo {author} {\bibfnamefont {G.}~\bibnamefont {Kresse}},\
  }\href {https://doi.org/10.1103/PhysRevLett.103.056401} {\bibfield  {journal}
  {\bibinfo  {journal} {Phys. Rev. Lett.}\ }\textbf {\bibinfo {volume} {103}},\
  \bibinfo {pages} {056401} (\bibinfo {year} {2009})}\BibitemShut {NoStop}%
\bibitem [{\citenamefont {Lu}\ \emph {et~al.}(2009)\citenamefont {Lu},
  \citenamefont {Li}, \citenamefont {Rocca},\ and\ \citenamefont
  {Galli}}]{lu_ab_2009}%
  \BibitemOpen
  \bibfield  {author} {\bibinfo {author} {\bibfnamefont {D.}~\bibnamefont
  {Lu}}, \bibinfo {author} {\bibfnamefont {Y.}~\bibnamefont {Li}}, \bibinfo
  {author} {\bibfnamefont {D.}~\bibnamefont {Rocca}},\ and\ \bibinfo {author}
  {\bibfnamefont {G.}~\bibnamefont {Galli}},\ }\href
  {https://doi.org/10.1103/PhysRevLett.102.206411} {\bibfield  {journal}
  {\bibinfo  {journal} {Phys. Rev. Lett.}\ }\textbf {\bibinfo {volume} {102}},\
  \bibinfo {pages} {206411} (\bibinfo {year} {2009})}\BibitemShut {NoStop}%
\bibitem [{\citenamefont {Ren}\ \emph {et~al.}(2009)\citenamefont {Ren},
  \citenamefont {Rinke},\ and\ \citenamefont {Scheffler}}]{ren_exploring_2009}%
  \BibitemOpen
  \bibfield  {author} {\bibinfo {author} {\bibfnamefont {X.}~\bibnamefont
  {Ren}}, \bibinfo {author} {\bibfnamefont {P.}~\bibnamefont {Rinke}},\ and\
  \bibinfo {author} {\bibfnamefont {M.}~\bibnamefont {Scheffler}},\ }\href
  {https://doi.org/10.1103/PhysRevB.80.045402} {\bibfield  {journal} {\bibinfo
  {journal} {Phys. Rev. B}\ }\textbf {\bibinfo {volume} {80}},\ \bibinfo
  {pages} {045402} (\bibinfo {year} {2009})}\BibitemShut {NoStop}%
\bibitem [{\citenamefont {Schimka}\ \emph {et~al.}(2010)\citenamefont
  {Schimka}, \citenamefont {Harl}, \citenamefont {Stroppa}, \citenamefont
  {Grüneis}, \citenamefont {Marsman}, \citenamefont {Mittendorfer},\ and\
  \citenamefont {Kresse}}]{schimka_accurate_2010}%
  \BibitemOpen
  \bibfield  {author} {\bibinfo {author} {\bibfnamefont {L.}~\bibnamefont
  {Schimka}}, \bibinfo {author} {\bibfnamefont {J.}~\bibnamefont {Harl}},
  \bibinfo {author} {\bibfnamefont {A.}~\bibnamefont {Stroppa}}, \bibinfo
  {author} {\bibfnamefont {A.}~\bibnamefont {Grüneis}}, \bibinfo {author}
  {\bibfnamefont {M.}~\bibnamefont {Marsman}}, \bibinfo {author} {\bibfnamefont
  {F.}~\bibnamefont {Mittendorfer}},\ and\ \bibinfo {author} {\bibfnamefont
  {G.}~\bibnamefont {Kresse}},\ }\href {https://doi.org/10.1038/nmat2806}
  {\bibfield  {journal} {\bibinfo  {journal} {Nat. Mater.}\ }\textbf {\bibinfo
  {volume} {9}},\ \bibinfo {pages} {741} (\bibinfo {year} {2010})}\BibitemShut
  {NoStop}%
\bibitem [{\citenamefont {Harl}\ \emph {et~al.}(2010)\citenamefont {Harl},
  \citenamefont {Schimka},\ and\ \citenamefont {Kresse}}]{harl_assessing_2010}%
  \BibitemOpen
  \bibfield  {author} {\bibinfo {author} {\bibfnamefont {J.}~\bibnamefont
  {Harl}}, \bibinfo {author} {\bibfnamefont {L.}~\bibnamefont {Schimka}},\ and\
  \bibinfo {author} {\bibfnamefont {G.}~\bibnamefont {Kresse}},\ }\href
  {https://doi.org/10.1103/PhysRevB.81.115126} {\bibfield  {journal} {\bibinfo
  {journal} {Phys. Rev. B}\ }\textbf {\bibinfo {volume} {81}},\ \bibinfo
  {pages} {115126} (\bibinfo {year} {2010})}\BibitemShut {NoStop}%
\bibitem [{\citenamefont {Lebègue}\ \emph {et~al.}(2010)\citenamefont
  {Lebègue}, \citenamefont {Harl}, \citenamefont {Gould}, \citenamefont
  {Ángyán}, \citenamefont {Kresse},\ and\ \citenamefont
  {Dobson}}]{lebegue_cohesive_2010}%
  \BibitemOpen
  \bibfield  {author} {\bibinfo {author} {\bibfnamefont {S.}~\bibnamefont
  {Lebègue}}, \bibinfo {author} {\bibfnamefont {J.}~\bibnamefont {Harl}},
  \bibinfo {author} {\bibfnamefont {T.}~\bibnamefont {Gould}}, \bibinfo
  {author} {\bibfnamefont {J.~G.}\ \bibnamefont {Ángyán}}, \bibinfo {author}
  {\bibfnamefont {G.}~\bibnamefont {Kresse}},\ and\ \bibinfo {author}
  {\bibfnamefont {J.~F.}\ \bibnamefont {Dobson}},\ }\href
  {https://doi.org/10.1103/PhysRevLett.105.196401} {\bibfield  {journal}
  {\bibinfo  {journal} {Phys. Rev. Lett.}\ }\textbf {\bibinfo {volume} {105}},\
  \bibinfo {pages} {196401} (\bibinfo {year} {2010})}\BibitemShut {NoStop}%
\bibitem [{\citenamefont {Mittendorfer}\ \emph {et~al.}(2011)\citenamefont
  {Mittendorfer}, \citenamefont {Garhofer}, \citenamefont {Redinger},
  \citenamefont {Klimeš}, \citenamefont {Harl},\ and\ \citenamefont
  {Kresse}}]{mittendorfer_graphene_2011}%
  \BibitemOpen
  \bibfield  {author} {\bibinfo {author} {\bibfnamefont {F.}~\bibnamefont
  {Mittendorfer}}, \bibinfo {author} {\bibfnamefont {A.}~\bibnamefont
  {Garhofer}}, \bibinfo {author} {\bibfnamefont {J.}~\bibnamefont {Redinger}},
  \bibinfo {author} {\bibfnamefont {J.}~\bibnamefont {Klimeš}}, \bibinfo
  {author} {\bibfnamefont {J.}~\bibnamefont {Harl}},\ and\ \bibinfo {author}
  {\bibfnamefont {G.}~\bibnamefont {Kresse}},\ }\href
  {https://doi.org/10.1103/PhysRevB.84.201401} {\bibfield  {journal} {\bibinfo
  {journal} {Phys. Rev. B}\ }\textbf {\bibinfo {volume} {84}},\ \bibinfo
  {pages} {201401} (\bibinfo {year} {2011})}\BibitemShut {NoStop}%
\bibitem [{\citenamefont {Olsen}\ \emph {et~al.}(2011)\citenamefont {Olsen},
  \citenamefont {Yan}, \citenamefont {Mortensen},\ and\ \citenamefont
  {Thygesen}}]{olsen_dispersive_2011}%
  \BibitemOpen
  \bibfield  {author} {\bibinfo {author} {\bibfnamefont {T.}~\bibnamefont
  {Olsen}}, \bibinfo {author} {\bibfnamefont {J.}~\bibnamefont {Yan}}, \bibinfo
  {author} {\bibfnamefont {J.~J.}\ \bibnamefont {Mortensen}},\ and\ \bibinfo
  {author} {\bibfnamefont {K.~S.}\ \bibnamefont {Thygesen}},\ }\href
  {https://doi.org/10.1103/PhysRevLett.107.156401} {\bibfield  {journal}
  {\bibinfo  {journal} {Phys. Rev. Lett.}\ }\textbf {\bibinfo {volume} {107}},\
  \bibinfo {pages} {156401} (\bibinfo {year} {2011})}\BibitemShut {NoStop}%
\bibitem [{\citenamefont {Casadei}\ \emph {et~al.}(2012)\citenamefont
  {Casadei}, \citenamefont {Ren}, \citenamefont {Rinke}, \citenamefont
  {Rubio},\ and\ \citenamefont {Scheffler}}]{casadei_density-functional_2012}%
  \BibitemOpen
  \bibfield  {author} {\bibinfo {author} {\bibfnamefont {M.}~\bibnamefont
  {Casadei}}, \bibinfo {author} {\bibfnamefont {X.}~\bibnamefont {Ren}},
  \bibinfo {author} {\bibfnamefont {P.}~\bibnamefont {Rinke}}, \bibinfo
  {author} {\bibfnamefont {A.}~\bibnamefont {Rubio}},\ and\ \bibinfo {author}
  {\bibfnamefont {M.}~\bibnamefont {Scheffler}},\ }\href
  {https://doi.org/10.1103/PhysRevLett.109.146402} {\bibfield  {journal}
  {\bibinfo  {journal} {Phys. Rev. Lett.}\ }\textbf {\bibinfo {volume} {109}},\
  \bibinfo {pages} {146402} (\bibinfo {year} {2012})}\BibitemShut {NoStop}%
\bibitem [{\citenamefont {Neuhauser}\ \emph {et~al.}(2013)\citenamefont
  {Neuhauser}, \citenamefont {Rabani},\ and\ \citenamefont
  {Baer}}]{neuhauser_expeditious_2013}%
  \BibitemOpen
  \bibfield  {author} {\bibinfo {author} {\bibfnamefont {D.}~\bibnamefont
  {Neuhauser}}, \bibinfo {author} {\bibfnamefont {E.}~\bibnamefont {Rabani}},\
  and\ \bibinfo {author} {\bibfnamefont {R.}~\bibnamefont {Baer}},\ }\href
  {https://doi.org/10.1021/jz3021606} {\bibfield  {journal} {\bibinfo
  {journal} {J. Phys. Chem. Lett.}\ }\textbf {\bibinfo {volume} {4}},\ \bibinfo
  {pages} {1172} (\bibinfo {year} {2013})}\BibitemShut {NoStop}%
\bibitem [{\citenamefont {Casadei}\ \emph {et~al.}(2016)\citenamefont
  {Casadei}, \citenamefont {Ren}, \citenamefont {Rinke}, \citenamefont
  {Rubio},\ and\ \citenamefont {Scheffler}}]{casadei_density_2016}%
  \BibitemOpen
  \bibfield  {author} {\bibinfo {author} {\bibfnamefont {M.}~\bibnamefont
  {Casadei}}, \bibinfo {author} {\bibfnamefont {X.}~\bibnamefont {Ren}},
  \bibinfo {author} {\bibfnamefont {P.}~\bibnamefont {Rinke}}, \bibinfo
  {author} {\bibfnamefont {A.}~\bibnamefont {Rubio}},\ and\ \bibinfo {author}
  {\bibfnamefont {M.}~\bibnamefont {Scheffler}},\ }\href
  {https://doi.org/10.1103/PhysRevB.93.075153} {\bibfield  {journal} {\bibinfo
  {journal} {Phys. Rev. B}\ }\textbf {\bibinfo {volume} {93}},\ \bibinfo
  {pages} {075153} (\bibinfo {year} {2016})}\BibitemShut {NoStop}%
\bibitem [{\citenamefont {Schäfer}\ \emph {et~al.}(2018)\citenamefont
  {Schäfer}, \citenamefont {Fan}, \citenamefont {Grünwald},\ and\
  \citenamefont {Kresse}}]{schafer_ab_2018}%
  \BibitemOpen
  \bibfield  {author} {\bibinfo {author} {\bibfnamefont {T.}~\bibnamefont
  {Schäfer}}, \bibinfo {author} {\bibfnamefont {Z.}~\bibnamefont {Fan}},
  \bibinfo {author} {\bibfnamefont {M.}~\bibnamefont {Grünwald}},\ and\
  \bibinfo {author} {\bibfnamefont {G.}~\bibnamefont {Kresse}},\ }\href
  {https://doi.org/10.1103/PhysRevB.98.144103} {\bibfield  {journal} {\bibinfo
  {journal} {Phys. Rev. B}\ }\textbf {\bibinfo {volume} {98}},\ \bibinfo
  {pages} {144103} (\bibinfo {year} {2018})}\BibitemShut {NoStop}%
\bibitem [{\citenamefont {Pines}\ and\ \citenamefont
  {Bohm}(1952)}]{pines_collective_1952}%
  \BibitemOpen
  \bibfield  {author} {\bibinfo {author} {\bibfnamefont {D.}~\bibnamefont
  {Pines}}\ and\ \bibinfo {author} {\bibfnamefont {D.}~\bibnamefont {Bohm}},\
  }\href {https://doi.org/10.1103/PhysRev.85.338} {\bibfield  {journal}
  {\bibinfo  {journal} {Phys. Rev.}\ }\textbf {\bibinfo {volume} {85}},\
  \bibinfo {pages} {338} (\bibinfo {year} {1952})}\BibitemShut {NoStop}%
\bibitem [{\citenamefont {Bohm}\ and\ \citenamefont
  {Pines}(1953)}]{bohm_collective_1953}%
  \BibitemOpen
  \bibfield  {author} {\bibinfo {author} {\bibfnamefont {D.}~\bibnamefont
  {Bohm}}\ and\ \bibinfo {author} {\bibfnamefont {D.}~\bibnamefont {Pines}},\
  }\href {https://doi.org/10.1103/PhysRev.92.609} {\bibfield  {journal}
  {\bibinfo  {journal} {Phys. Rev.}\ }\textbf {\bibinfo {volume} {92}},\
  \bibinfo {pages} {609} (\bibinfo {year} {1953})}\BibitemShut {NoStop}%
\bibitem [{\citenamefont {Gell-Mann}\ and\ \citenamefont
  {Brueckner}(1957)}]{gell-mann_correlation_1957}%
  \BibitemOpen
  \bibfield  {author} {\bibinfo {author} {\bibfnamefont {M.}~\bibnamefont
  {Gell-Mann}}\ and\ \bibinfo {author} {\bibfnamefont {K.~A.}\ \bibnamefont
  {Brueckner}},\ }\href {https://doi.org/10.1103/PhysRev.106.364} {\bibfield
  {journal} {\bibinfo  {journal} {Phys. Rev.}\ }\textbf {\bibinfo {volume}
  {106}},\ \bibinfo {pages} {364} (\bibinfo {year} {1957})}\BibitemShut
  {NoStop}%
\bibitem [{\citenamefont {Goldstone}(1957)}]{goldstone_derivation_1957}%
  \BibitemOpen
  \bibfield  {author} {\bibinfo {author} {\bibfnamefont {J.}~\bibnamefont
  {Goldstone}},\ }\href {https://api.semanticscholar.org/CorpusID:123005551}
  {\bibfield  {journal} {\bibinfo  {journal} {Proc. R. Soc. A}\ }\textbf
  {\bibinfo {volume} {239}},\ \bibinfo {pages} {267 } (\bibinfo {year}
  {1957})}\BibitemShut {NoStop}%
\bibitem [{\citenamefont {McLachlan}\ and\ \citenamefont
  {Ball}(1964)}]{mclachlan_time-dependent_1964}%
  \BibitemOpen
  \bibfield  {author} {\bibinfo {author} {\bibfnamefont {A.~D.}\ \bibnamefont
  {McLachlan}}\ and\ \bibinfo {author} {\bibfnamefont {M.~A.}\ \bibnamefont
  {Ball}},\ }\href {https://doi.org/10.1103/RevModPhys.36.844} {\bibfield
  {journal} {\bibinfo  {journal} {Rev. Mod. Phys.}\ }\textbf {\bibinfo {volume}
  {36}},\ \bibinfo {pages} {844} (\bibinfo {year} {1964})}\BibitemShut
  {NoStop}%
\bibitem [{\citenamefont {Furche}(2008)}]{furche_developing_2008}%
  \BibitemOpen
  \bibfield  {author} {\bibinfo {author} {\bibfnamefont {F.}~\bibnamefont
  {Furche}},\ }\href {https://doi.org/10.1063/1.2977789} {\bibfield  {journal}
  {\bibinfo  {journal} {J. Chem. Phys.}\ }\textbf {\bibinfo {volume} {129}},\
  \bibinfo {pages} {114105} (\bibinfo {year} {2008})}\BibitemShut {NoStop}%
\bibitem [{\citenamefont {Hesselmann}\ and\ \citenamefont
  {Görling}(2011)}]{hesselmann_random-phase_2011}%
  \BibitemOpen
  \bibfield  {author} {\bibinfo {author} {\bibfnamefont {A.}~\bibnamefont
  {Hesselmann}}\ and\ \bibinfo {author} {\bibfnamefont {A.}~\bibnamefont
  {Görling}},\ }\href {https://doi.org/10.1080/00268976.2011.614282}
  {\bibfield  {journal} {\bibinfo  {journal} {Mol. Phys.}\ }\textbf {\bibinfo
  {volume} {109}},\ \bibinfo {pages} {2473} (\bibinfo {year}
  {2011})}\BibitemShut {NoStop}%
\bibitem [{\citenamefont {Ren}\ \emph {et~al.}(2012)\citenamefont {Ren},
  \citenamefont {Rinke}, \citenamefont {Joas},\ and\ \citenamefont
  {Scheffler}}]{ren_random-phase_2012}%
  \BibitemOpen
  \bibfield  {author} {\bibinfo {author} {\bibfnamefont {X.}~\bibnamefont
  {Ren}}, \bibinfo {author} {\bibfnamefont {P.}~\bibnamefont {Rinke}}, \bibinfo
  {author} {\bibfnamefont {C.}~\bibnamefont {Joas}},\ and\ \bibinfo {author}
  {\bibfnamefont {M.}~\bibnamefont {Scheffler}},\ }\href
  {https://doi.org/10.1007/s10853-012-6570-4} {\bibfield  {journal} {\bibinfo
  {journal} {J. Mater. Sci.}\ }\textbf {\bibinfo {volume} {47}},\ \bibinfo
  {pages} {7447} (\bibinfo {year} {2012})}\BibitemShut {NoStop}%
\bibitem [{\citenamefont {Chen}\ \emph {et~al.}(2017)\citenamefont {Chen},
  \citenamefont {Voora}, \citenamefont {Agee}, \citenamefont {Balasubramani},\
  and\ \citenamefont {Furche}}]{chen_random-phase_2017}%
  \BibitemOpen
  \bibfield  {author} {\bibinfo {author} {\bibfnamefont {G.~P.}\ \bibnamefont
  {Chen}}, \bibinfo {author} {\bibfnamefont {V.~K.}\ \bibnamefont {Voora}},
  \bibinfo {author} {\bibfnamefont {M.~M.}\ \bibnamefont {Agee}}, \bibinfo
  {author} {\bibfnamefont {S.~G.}\ \bibnamefont {Balasubramani}},\ and\
  \bibinfo {author} {\bibfnamefont {F.}~\bibnamefont {Furche}},\ }\href
  {https://doi.org/10.1146/annurev-physchem-040215-112308} {\bibfield
  {journal} {\bibinfo  {journal} {Annu. Rev. Phys. Chem.}\ }\textbf {\bibinfo
  {volume} {68}},\ \bibinfo {pages} {421} (\bibinfo {year} {2017})}\BibitemShut
  {NoStop}%
\bibitem [{\citenamefont {Chatterjee}\ and\ \citenamefont
  {Pernal}(2012)}]{chatterjee_excitation_2012}%
  \BibitemOpen
  \bibfield  {author} {\bibinfo {author} {\bibfnamefont {K.}~\bibnamefont
  {Chatterjee}}\ and\ \bibinfo {author} {\bibfnamefont {K.}~\bibnamefont
  {Pernal}},\ }\href {https://doi.org/10.1063/1.4766934} {\bibfield  {journal}
  {\bibinfo  {journal} {J. Chem. Phys.}\ }\textbf {\bibinfo {volume} {137}},\
  \bibinfo {pages} {204109} (\bibinfo {year} {2012})}\BibitemShut {NoStop}%
\bibitem [{\citenamefont {Pernal}(2014)}]{pernal_intergeminal_2014}%
  \BibitemOpen
  \bibfield  {author} {\bibinfo {author} {\bibfnamefont {K.}~\bibnamefont
  {Pernal}},\ }\href {https://doi.org/10.1021/ct500478t} {\bibfield  {journal}
  {\bibinfo  {journal} {J. Chem. Theory Comput.}\ }\textbf {\bibinfo {volume}
  {10}},\ \bibinfo {pages} {4332} (\bibinfo {year} {2014})}\BibitemShut
  {NoStop}%
\bibitem [{\citenamefont {Pastorczak}\ and\ \citenamefont
  {Pernal}(2018)}]{pastorczak_correlation_2018}%
  \BibitemOpen
  \bibfield  {author} {\bibinfo {author} {\bibfnamefont {E.}~\bibnamefont
  {Pastorczak}}\ and\ \bibinfo {author} {\bibfnamefont {K.}~\bibnamefont
  {Pernal}},\ }\href {https://doi.org/10.1021/acs.jctc.8b00213} {\bibfield
  {journal} {\bibinfo  {journal} {J. Chem. Theory Comput.}\ }\textbf {\bibinfo
  {volume} {14}},\ \bibinfo {pages} {3493} (\bibinfo {year}
  {2018})}\BibitemShut {NoStop}%
\bibitem [{\citenamefont {Pernal}(2018)}]{pernal_electron_2018}%
  \BibitemOpen
  \bibfield  {author} {\bibinfo {author} {\bibfnamefont {K.}~\bibnamefont
  {Pernal}},\ }\href {https://doi.org/10.1103/PhysRevLett.120.013001}
  {\bibfield  {journal} {\bibinfo  {journal} {Phys. Rev. Lett.}\ }\textbf
  {\bibinfo {volume} {120}},\ \bibinfo {pages} {013001} (\bibinfo {year}
  {2018})}\BibitemShut {NoStop}%
\bibitem [{\citenamefont {Guo}\ and\ \citenamefont
  {Pernal}(2024)}]{guo_spinless_2024}%
  \BibitemOpen
  \bibfield  {author} {\bibinfo {author} {\bibfnamefont {Y.}~\bibnamefont
  {Guo}}\ and\ \bibinfo {author} {\bibfnamefont {K.}~\bibnamefont {Pernal}},\
  }\href {https://doi.org/10.1039/D4FD00054D} {\bibfield  {journal} {\bibinfo
  {journal} {Faraday Discuss.}\ }\textbf {\bibinfo {volume} {254}},\ \bibinfo
  {pages} {332} (\bibinfo {year} {2024})}\BibitemShut {NoStop}%
\bibitem [{\citenamefont {Tucholska}\ \emph {et~al.}(2024)\citenamefont
  {Tucholska}, \citenamefont {Guo},\ and\ \citenamefont
  {Pernal}}]{tucholska_duality_2024}%
  \BibitemOpen
  \bibfield  {author} {\bibinfo {author} {\bibfnamefont {A.}~\bibnamefont
  {Tucholska}}, \bibinfo {author} {\bibfnamefont {Y.}~\bibnamefont {Guo}},\
  and\ \bibinfo {author} {\bibfnamefont {K.}~\bibnamefont {Pernal}},\ }\href
  {https://doi.org/10.1021/acs.jpclett.4c02788} {\bibfield  {journal} {\bibinfo
   {journal} {J. Phys. Chem. Lett.}\ }\textbf {\bibinfo {volume} {15}},\
  \bibinfo {pages} {12001} (\bibinfo {year} {2024})}\BibitemShut {NoStop}%
\bibitem [{\citenamefont {Szabados}\ and\ \citenamefont
  {Marg\'{o}csy}(2017)}]{szabados_ring_2017}%
  \BibitemOpen
  \bibfield  {author} {\bibinfo {author} {\bibfnamefont {{\'A}.}~\bibnamefont
  {Szabados}}\ and\ \bibinfo {author} {\bibfnamefont {{\'A}.}~\bibnamefont
  {Marg\'{o}csy}},\ }\href {https://doi.org/10.1080/00268976.2017.1317111}
  {\bibfield  {journal} {\bibinfo  {journal} {Mol. Phys.}\ }\textbf {\bibinfo
  {volume} {115}},\ \bibinfo {pages} {2731} (\bibinfo {year}
  {2017})}\BibitemShut {NoStop}%
\bibitem [{\citenamefont {Marg\'{o}csy}\ and\ \citenamefont
  {Szabados}(2020)}]{margocsy_ring_2020}%
  \BibitemOpen
  \bibfield  {author} {\bibinfo {author} {\bibfnamefont {{\'A}.}~\bibnamefont
  {Marg\'{o}csy}}\ and\ \bibinfo {author} {\bibfnamefont {{\'A}.}~\bibnamefont
  {Szabados}},\ }\href {https://doi.org/10.1063/5.0005075} {\bibfield
  {journal} {\bibinfo  {journal} {J. Chem. Phys.}\ }\textbf {\bibinfo {volume}
  {152}},\ \bibinfo {pages} {204114} (\bibinfo {year} {2020})}\BibitemShut
  {NoStop}%
\bibitem [{\citenamefont {Mori-Sánchez}\ \emph {et~al.}(2012)\citenamefont
  {Mori-Sánchez}, \citenamefont {Cohen},\ and\ \citenamefont
  {Yang}}]{mori-sanchez_failure_2012}%
  \BibitemOpen
  \bibfield  {author} {\bibinfo {author} {\bibfnamefont {P.}~\bibnamefont
  {Mori-Sánchez}}, \bibinfo {author} {\bibfnamefont {A.~J.}\ \bibnamefont
  {Cohen}},\ and\ \bibinfo {author} {\bibfnamefont {W.}~\bibnamefont {Yang}},\
  }\href {https://doi.org/10.1103/PhysRevA.85.042507} {\bibfield  {journal}
  {\bibinfo  {journal} {Phys. Rev. A}\ }\textbf {\bibinfo {volume} {85}},\
  \bibinfo {pages} {042507} (\bibinfo {year} {2012})}\BibitemShut {NoStop}%
\bibitem [{\citenamefont {Grüneis}\ \emph {et~al.}(2009)\citenamefont
  {Grüneis}, \citenamefont {Marsman}, \citenamefont {Harl}, \citenamefont
  {Schimka},\ and\ \citenamefont {Kresse}}]{gruneis_making_2009}%
  \BibitemOpen
  \bibfield  {author} {\bibinfo {author} {\bibfnamefont {A.}~\bibnamefont
  {Grüneis}}, \bibinfo {author} {\bibfnamefont {M.}~\bibnamefont {Marsman}},
  \bibinfo {author} {\bibfnamefont {J.}~\bibnamefont {Harl}}, \bibinfo {author}
  {\bibfnamefont {L.}~\bibnamefont {Schimka}},\ and\ \bibinfo {author}
  {\bibfnamefont {G.}~\bibnamefont {Kresse}},\ }\href
  {https://doi.org/10.1063/1.3250347} {\bibfield  {journal} {\bibinfo
  {journal} {J. Chem. Phys.}\ }\textbf {\bibinfo {volume} {131}},\ \bibinfo
  {pages} {154115} (\bibinfo {year} {2009})}\BibitemShut {NoStop}%
\bibitem [{\citenamefont {Heßelmann}(2011)}]{heselmann_third-order_2011}%
  \BibitemOpen
  \bibfield  {author} {\bibinfo {author} {\bibfnamefont {A.}~\bibnamefont
  {Heßelmann}},\ }\href {https://doi.org/10.1063/1.3590916} {\bibfield
  {journal} {\bibinfo  {journal} {J. Chem. Phys.}\ }\textbf {\bibinfo {volume}
  {134}},\ \bibinfo {pages} {204107} (\bibinfo {year} {2011})}\BibitemShut
  {NoStop}%
\bibitem [{\citenamefont {Bates}\ and\ \citenamefont
  {Furche}(2013)}]{bates_communication_2013}%
  \BibitemOpen
  \bibfield  {author} {\bibinfo {author} {\bibfnamefont {J.~E.}\ \bibnamefont
  {Bates}}\ and\ \bibinfo {author} {\bibfnamefont {F.}~\bibnamefont {Furche}},\
  }\href {https://doi.org/10.1063/1.4827254} {\bibfield  {journal} {\bibinfo
  {journal} {J. Chem. Phys.}\ }\textbf {\bibinfo {volume} {139}},\ \bibinfo
  {pages} {171103} (\bibinfo {year} {2013})}\BibitemShut {NoStop}%
\bibitem [{\citenamefont {Chen}\ \emph {et~al.}(2018)\citenamefont {Chen},
  \citenamefont {Agee},\ and\ \citenamefont {Furche}}]{chen_performance_2018}%
  \BibitemOpen
  \bibfield  {author} {\bibinfo {author} {\bibfnamefont {G.~P.}\ \bibnamefont
  {Chen}}, \bibinfo {author} {\bibfnamefont {M.~M.}\ \bibnamefont {Agee}},\
  and\ \bibinfo {author} {\bibfnamefont {F.}~\bibnamefont {Furche}},\ }\href
  {https://doi.org/10.1021/acs.jctc.8b00777} {\bibfield  {journal} {\bibinfo
  {journal} {J. Chem. Theory Comput.}\ }\textbf {\bibinfo {volume} {14}},\
  \bibinfo {pages} {5701} (\bibinfo {year} {2018})}\BibitemShut {NoStop}%
\bibitem [{\citenamefont {Hummel}\ \emph {et~al.}(2019)\citenamefont {Hummel},
  \citenamefont {Grüneis}, \citenamefont {Kresse},\ and\ \citenamefont
  {Ziesche}}]{hummel_screened_2019}%
  \BibitemOpen
  \bibfield  {author} {\bibinfo {author} {\bibfnamefont {F.}~\bibnamefont
  {Hummel}}, \bibinfo {author} {\bibfnamefont {A.}~\bibnamefont {Grüneis}},
  \bibinfo {author} {\bibfnamefont {G.}~\bibnamefont {Kresse}},\ and\ \bibinfo
  {author} {\bibfnamefont {P.}~\bibnamefont {Ziesche}},\ }\href
  {https://doi.org/10.1021/acs.jctc.8b01247} {\bibfield  {journal} {\bibinfo
  {journal} {J. Chem. Theory Comput.}\ }\textbf {\bibinfo {volume} {15}},\
  \bibinfo {pages} {3223} (\bibinfo {year} {2019})}\BibitemShut {NoStop}%
\bibitem [{\citenamefont {Ren}\ \emph {et~al.}(2011)\citenamefont {Ren},
  \citenamefont {Tkatchenko}, \citenamefont {Rinke},\ and\ \citenamefont
  {Scheffler}}]{ren_beyond_2011}%
  \BibitemOpen
  \bibfield  {author} {\bibinfo {author} {\bibfnamefont {X.}~\bibnamefont
  {Ren}}, \bibinfo {author} {\bibfnamefont {A.}~\bibnamefont {Tkatchenko}},
  \bibinfo {author} {\bibfnamefont {P.}~\bibnamefont {Rinke}},\ and\ \bibinfo
  {author} {\bibfnamefont {M.}~\bibnamefont {Scheffler}},\ }\href
  {https://doi.org/10.1103/PhysRevLett.106.153003} {\bibfield  {journal}
  {\bibinfo  {journal} {Phys. Rev. Lett.}\ }\textbf {\bibinfo {volume} {106}},\
  \bibinfo {pages} {153003} (\bibinfo {year} {2011})}\BibitemShut {NoStop}%
\bibitem [{\citenamefont {Fuchs}\ and\ \citenamefont
  {Gonze}(2002)}]{fuchs_accurate_2002}%
  \BibitemOpen
  \bibfield  {author} {\bibinfo {author} {\bibfnamefont {M.}~\bibnamefont
  {Fuchs}}\ and\ \bibinfo {author} {\bibfnamefont {X.}~\bibnamefont {Gonze}},\
  }\href {https://doi.org/10.1103/PhysRevB.65.235109} {\bibfield  {journal}
  {\bibinfo  {journal} {Phys. Rev. B}\ }\textbf {\bibinfo {volume} {65}},\
  \bibinfo {pages} {235109} (\bibinfo {year} {2002})}\BibitemShut {NoStop}%
\bibitem [{\citenamefont {Jiang}\ and\ \citenamefont
  {Engel}(2007)}]{jiang_random-phase-approximation-based_2007}%
  \BibitemOpen
  \bibfield  {author} {\bibinfo {author} {\bibfnamefont {H.}~\bibnamefont
  {Jiang}}\ and\ \bibinfo {author} {\bibfnamefont {E.}~\bibnamefont {Engel}},\
  }\href {https://doi.org/10.1063/1.2795707} {\bibfield  {journal} {\bibinfo
  {journal} {J. Chem. Phys.}\ }\textbf {\bibinfo {volume} {127}},\ \bibinfo
  {pages} {184108} (\bibinfo {year} {2007})}\BibitemShut {NoStop}%
\bibitem [{\citenamefont {Toulouse}\ \emph {et~al.}(2009)\citenamefont
  {Toulouse}, \citenamefont {Gerber}, \citenamefont {Jansen}, \citenamefont
  {Savin},\ and\ \citenamefont
  {Ángyán}}]{toulouse_adiabatic-connection_2009}%
  \BibitemOpen
  \bibfield  {author} {\bibinfo {author} {\bibfnamefont {J.}~\bibnamefont
  {Toulouse}}, \bibinfo {author} {\bibfnamefont {I.~C.}\ \bibnamefont
  {Gerber}}, \bibinfo {author} {\bibfnamefont {G.}~\bibnamefont {Jansen}},
  \bibinfo {author} {\bibfnamefont {A.}~\bibnamefont {Savin}},\ and\ \bibinfo
  {author} {\bibfnamefont {J.~G.}\ \bibnamefont {Ángyán}},\ }\href
  {https://doi.org/10.1103/PhysRevLett.102.096404} {\bibfield  {journal}
  {\bibinfo  {journal} {Phys. Rev. Lett.}\ }\textbf {\bibinfo {volume} {102}},\
  \bibinfo {pages} {096404} (\bibinfo {year} {2009})}\BibitemShut {NoStop}%
\bibitem [{\citenamefont {Hesselmann}\ and\ \citenamefont
  {Görling}(2010)}]{hesselmann_random_2010}%
  \BibitemOpen
  \bibfield  {author} {\bibinfo {author} {\bibfnamefont {A.}~\bibnamefont
  {Hesselmann}}\ and\ \bibinfo {author} {\bibfnamefont {A.}~\bibnamefont
  {Görling}},\ }\href {https://doi.org/10.1080/00268970903476662} {\bibfield
  {journal} {\bibinfo  {journal} {Mol. Phys.}\ }\textbf {\bibinfo {volume}
  {108}},\ \bibinfo {pages} {359} (\bibinfo {year} {2010})}\BibitemShut
  {NoStop}%
\bibitem [{\citenamefont {Toulouse}\ \emph {et~al.}(2010)\citenamefont
  {Toulouse}, \citenamefont {Zhu}, \citenamefont {Ángyán},\ and\
  \citenamefont {Savin}}]{toulouse_range-separated_2010}%
  \BibitemOpen
  \bibfield  {author} {\bibinfo {author} {\bibfnamefont {J.}~\bibnamefont
  {Toulouse}}, \bibinfo {author} {\bibfnamefont {W.}~\bibnamefont {Zhu}},
  \bibinfo {author} {\bibfnamefont {J.~G.}\ \bibnamefont {Ángyán}},\ and\
  \bibinfo {author} {\bibfnamefont {A.}~\bibnamefont {Savin}},\ }\href
  {https://doi.org/10.1103/PhysRevA.82.032502} {\bibfield  {journal} {\bibinfo
  {journal} {Phys. Rev. A}\ }\textbf {\bibinfo {volume} {82}},\ \bibinfo
  {pages} {032502} (\bibinfo {year} {2010})}\BibitemShut {NoStop}%
\bibitem [{\citenamefont {Paier}\ \emph {et~al.}(2010)\citenamefont {Paier},
  \citenamefont {Janesko}, \citenamefont {Henderson}, \citenamefont {Scuseria},
  \citenamefont {Grüneis},\ and\ \citenamefont {Kresse}}]{paier_hybrid_2010}%
  \BibitemOpen
  \bibfield  {author} {\bibinfo {author} {\bibfnamefont {J.}~\bibnamefont
  {Paier}}, \bibinfo {author} {\bibfnamefont {B.~G.}\ \bibnamefont {Janesko}},
  \bibinfo {author} {\bibfnamefont {T.~M.}\ \bibnamefont {Henderson}}, \bibinfo
  {author} {\bibfnamefont {G.~E.}\ \bibnamefont {Scuseria}}, \bibinfo {author}
  {\bibfnamefont {A.}~\bibnamefont {Grüneis}},\ and\ \bibinfo {author}
  {\bibfnamefont {G.}~\bibnamefont {Kresse}},\ }\href
  {https://doi.org/10.1063/1.3317437} {\bibfield  {journal} {\bibinfo
  {journal} {J. Chem. Phys.}\ }\textbf {\bibinfo {volume} {132}},\ \bibinfo
  {pages} {094103} (\bibinfo {year} {2010})}\BibitemShut {NoStop}%
\bibitem [{\citenamefont {Heßelmann}\ and\ \citenamefont
  {Görling}(2011)}]{heselmann_correct_2011}%
  \BibitemOpen
  \bibfield  {author} {\bibinfo {author} {\bibfnamefont {A.}~\bibnamefont
  {Heßelmann}}\ and\ \bibinfo {author} {\bibfnamefont {A.}~\bibnamefont
  {Görling}},\ }\href {https://doi.org/10.1103/PhysRevLett.106.093001}
  {\bibfield  {journal} {\bibinfo  {journal} {Phys. Rev. Lett.}\ }\textbf
  {\bibinfo {volume} {106}},\ \bibinfo {pages} {093001} (\bibinfo {year}
  {2011})}\BibitemShut {NoStop}%
\bibitem [{\citenamefont {Trushin}\ \emph {et~al.}(2021)\citenamefont
  {Trushin}, \citenamefont {Thierbach},\ and\ \citenamefont
  {Görling}}]{trushin_toward_2021}%
  \BibitemOpen
  \bibfield  {author} {\bibinfo {author} {\bibfnamefont {E.}~\bibnamefont
  {Trushin}}, \bibinfo {author} {\bibfnamefont {A.}~\bibnamefont {Thierbach}},\
  and\ \bibinfo {author} {\bibfnamefont {A.}~\bibnamefont {Görling}},\ }\href
  {https://doi.org/10.1063/5.0026849} {\bibfield  {journal} {\bibinfo
  {journal} {J. Chem. Phys.}\ }\textbf {\bibinfo {volume} {154}},\ \bibinfo
  {pages} {014104} (\bibinfo {year} {2021})}\BibitemShut {NoStop}%
\bibitem [{\citenamefont {Riemelmoser}\ \emph {et~al.}(2023)\citenamefont
  {Riemelmoser}, \citenamefont {Verdi}, \citenamefont {Kaltak},\ and\
  \citenamefont {Kresse}}]{riemelmoser_machine_2023}%
  \BibitemOpen
  \bibfield  {author} {\bibinfo {author} {\bibfnamefont {S.}~\bibnamefont
  {Riemelmoser}}, \bibinfo {author} {\bibfnamefont {C.}~\bibnamefont {Verdi}},
  \bibinfo {author} {\bibfnamefont {M.}~\bibnamefont {Kaltak}},\ and\ \bibinfo
  {author} {\bibfnamefont {G.}~\bibnamefont {Kresse}},\ }\href
  {https://doi.org/10.1021/acs.jctc.3c00848} {\bibfield  {journal} {\bibinfo
  {journal} {J. Chem. Theory Comput.}\ }\textbf {\bibinfo {volume} {19}},\
  \bibinfo {pages} {7287} (\bibinfo {year} {2023})}\BibitemShut {NoStop}%
\bibitem [{\citenamefont {Klopper}\ \emph {et~al.}(2011)\citenamefont
  {Klopper}, \citenamefont {Teale}, \citenamefont {Coriani}, \citenamefont
  {Pedersen},\ and\ \citenamefont {Helgaker}}]{klopper_spin_2011}%
  \BibitemOpen
  \bibfield  {author} {\bibinfo {author} {\bibfnamefont {W.}~\bibnamefont
  {Klopper}}, \bibinfo {author} {\bibfnamefont {A.~M.}\ \bibnamefont {Teale}},
  \bibinfo {author} {\bibfnamefont {S.}~\bibnamefont {Coriani}}, \bibinfo
  {author} {\bibfnamefont {T.~B.}\ \bibnamefont {Pedersen}},\ and\ \bibinfo
  {author} {\bibfnamefont {T.}~\bibnamefont {Helgaker}},\ }\href
  {https://doi.org/10.1016/j.cplett.2011.04.101} {\bibfield  {journal}
  {\bibinfo  {journal} {Chem. Phys. Lett.}\ }\textbf {\bibinfo {volume}
  {510}},\ \bibinfo {pages} {147} (\bibinfo {year} {2011})}\BibitemShut
  {NoStop}%
\bibitem [{\citenamefont {Ángyán}\ \emph {et~al.}(2011)\citenamefont
  {Ángyán}, \citenamefont {Liu}, \citenamefont {Toulouse},\ and\
  \citenamefont {Jansen}}]{angyan_correlation_2011}%
  \BibitemOpen
  \bibfield  {author} {\bibinfo {author} {\bibfnamefont {J.~G.}\ \bibnamefont
  {Ángyán}}, \bibinfo {author} {\bibfnamefont {R.-F.}\ \bibnamefont {Liu}},
  \bibinfo {author} {\bibfnamefont {J.}~\bibnamefont {Toulouse}},\ and\
  \bibinfo {author} {\bibfnamefont {G.}~\bibnamefont {Jansen}},\ }\href
  {https://doi.org/10.1021/ct200501r} {\bibfield  {journal} {\bibinfo
  {journal} {J. Chem. Theory Comput.}\ }\textbf {\bibinfo {volume} {7}},\
  \bibinfo {pages} {3116} (\bibinfo {year} {2011})}\BibitemShut {NoStop}%
\bibitem [{\citenamefont
  {Heßelmann}(2012)}]{heselmann_random-phase-approximation_2012}%
  \BibitemOpen
  \bibfield  {author} {\bibinfo {author} {\bibfnamefont {A.}~\bibnamefont
  {Heßelmann}},\ }\href {https://doi.org/10.1103/PhysRevA.85.012517}
  {\bibfield  {journal} {\bibinfo  {journal} {Phys. Rev. A}\ }\textbf {\bibinfo
  {volume} {85}},\ \bibinfo {pages} {012517} (\bibinfo {year}
  {2012})}\BibitemShut {NoStop}%
\bibitem [{\citenamefont {Eshuis}\ \emph {et~al.}(2012)\citenamefont {Eshuis},
  \citenamefont {Bates},\ and\ \citenamefont {Furche}}]{eshuis_electron_2012}%
  \BibitemOpen
  \bibfield  {author} {\bibinfo {author} {\bibfnamefont {H.}~\bibnamefont
  {Eshuis}}, \bibinfo {author} {\bibfnamefont {J.~E.}\ \bibnamefont {Bates}},\
  and\ \bibinfo {author} {\bibfnamefont {F.}~\bibnamefont {Furche}},\ }\href
  {https://doi.org/10.1007/s00214-011-1084-8} {\bibfield  {journal} {\bibinfo
  {journal} {Theor. Chem. Acc.}\ }\textbf {\bibinfo {volume} {131}},\ \bibinfo
  {pages} {1084} (\bibinfo {year} {2012})}\BibitemShut {NoStop}%
\bibitem [{\citenamefont {Scuseria}\ \emph {et~al.}(2013)\citenamefont
  {Scuseria}, \citenamefont {Henderson},\ and\ \citenamefont
  {Bulik}}]{scuseria_particle-particle_2013}%
  \BibitemOpen
  \bibfield  {author} {\bibinfo {author} {\bibfnamefont {G.~E.}\ \bibnamefont
  {Scuseria}}, \bibinfo {author} {\bibfnamefont {T.~M.}\ \bibnamefont
  {Henderson}},\ and\ \bibinfo {author} {\bibfnamefont {I.~W.}\ \bibnamefont
  {Bulik}},\ }\href {https://doi.org/10.1063/1.4820557} {\bibfield  {journal}
  {\bibinfo  {journal} {J. Chem. Phys.}\ }\textbf {\bibinfo {volume} {139}},\
  \bibinfo {pages} {104113} (\bibinfo {year} {2013})}\BibitemShut {NoStop}%
\bibitem [{\citenamefont {van Aggelen}\ \emph {et~al.}(2013)\citenamefont {van
  Aggelen}, \citenamefont {Yang},\ and\ \citenamefont
  {Yang}}]{van_aggelen_exchange-correlation_2013}%
  \BibitemOpen
  \bibfield  {author} {\bibinfo {author} {\bibfnamefont {H.}~\bibnamefont {van
  Aggelen}}, \bibinfo {author} {\bibfnamefont {Y.}~\bibnamefont {Yang}},\ and\
  \bibinfo {author} {\bibfnamefont {W.}~\bibnamefont {Yang}},\ }\href
  {https://doi.org/10.1103/PhysRevA.88.030501} {\bibfield  {journal} {\bibinfo
  {journal} {Phys. Rev. A}\ }\textbf {\bibinfo {volume} {88}},\ \bibinfo
  {pages} {030501} (\bibinfo {year} {2013})}\BibitemShut {NoStop}%
\bibitem [{\citenamefont {van Aggelen}\ \emph {et~al.}(2014)\citenamefont {van
  Aggelen}, \citenamefont {Yang},\ and\ \citenamefont
  {Yang}}]{van_aggelen_exchange-correlation_2014}%
  \BibitemOpen
  \bibfield  {author} {\bibinfo {author} {\bibfnamefont {H.}~\bibnamefont {van
  Aggelen}}, \bibinfo {author} {\bibfnamefont {Y.}~\bibnamefont {Yang}},\ and\
  \bibinfo {author} {\bibfnamefont {W.}~\bibnamefont {Yang}},\ }\href
  {https://doi.org/10.1063/1.4865816} {\bibfield  {journal} {\bibinfo
  {journal} {J. Chem. Phys.}\ }\textbf {\bibinfo {volume} {140}},\ \bibinfo
  {pages} {18A511} (\bibinfo {year} {2014})}\BibitemShut {NoStop}%
\bibitem [{\citenamefont {Mussard}\ \emph {et~al.}(2015)\citenamefont
  {Mussard}, \citenamefont {Reinhardt}, \citenamefont {Ángyán},\ and\
  \citenamefont {Toulouse}}]{mussard_spin-unrestricted_2015}%
  \BibitemOpen
  \bibfield  {author} {\bibinfo {author} {\bibfnamefont {B.}~\bibnamefont
  {Mussard}}, \bibinfo {author} {\bibfnamefont {P.}~\bibnamefont {Reinhardt}},
  \bibinfo {author} {\bibfnamefont {J.~G.}\ \bibnamefont {Ángyán}},\ and\
  \bibinfo {author} {\bibfnamefont {J.}~\bibnamefont {Toulouse}},\ }\href
  {https://doi.org/10.1063/1.4918710} {\bibfield  {journal} {\bibinfo
  {journal} {J. Chem. Phys.}\ }\textbf {\bibinfo {volume} {142}},\ \bibinfo
  {pages} {154123} (\bibinfo {year} {2015})}\BibitemShut {NoStop}%
\bibitem [{\citenamefont {Tahir}\ and\ \citenamefont
  {Ren}(2019)}]{tahir_comparing_2019}%
  \BibitemOpen
  \bibfield  {author} {\bibinfo {author} {\bibfnamefont {M.~N.}\ \bibnamefont
  {Tahir}}\ and\ \bibinfo {author} {\bibfnamefont {X.}~\bibnamefont {Ren}},\
  }\href {https://doi.org/10.1103/PhysRevB.99.195149} {\bibfield  {journal}
  {\bibinfo  {journal} {Phys. Rev. B}\ }\textbf {\bibinfo {volume} {99}},\
  \bibinfo {pages} {195149} (\bibinfo {year} {2019})}\BibitemShut {NoStop}%
\bibitem [{\citenamefont {Yang}\ \emph {et~al.}(2013)\citenamefont {Yang},
  \citenamefont {van Aggelen},\ and\ \citenamefont {Yang}}]{yang_double_2013}%
  \BibitemOpen
  \bibfield  {author} {\bibinfo {author} {\bibfnamefont {Y.}~\bibnamefont
  {Yang}}, \bibinfo {author} {\bibfnamefont {H.}~\bibnamefont {van Aggelen}},\
  and\ \bibinfo {author} {\bibfnamefont {W.}~\bibnamefont {Yang}},\ }\href
  {https://doi.org/10.1063/1.4834875} {\bibfield  {journal} {\bibinfo
  {journal} {J. Chem. Phys.}\ }\textbf {\bibinfo {volume} {139}},\ \bibinfo
  {pages} {224105} (\bibinfo {year} {2013})}\BibitemShut {NoStop}%
\bibitem [{\citenamefont {Yang}\ \emph {et~al.}(2015)\citenamefont {Yang},
  \citenamefont {Peng}, \citenamefont {Davidson},\ and\ \citenamefont
  {Yang}}]{yang_singlettriplet_2015}%
  \BibitemOpen
  \bibfield  {author} {\bibinfo {author} {\bibfnamefont {Y.}~\bibnamefont
  {Yang}}, \bibinfo {author} {\bibfnamefont {D.}~\bibnamefont {Peng}}, \bibinfo
  {author} {\bibfnamefont {E.~R.}\ \bibnamefont {Davidson}},\ and\ \bibinfo
  {author} {\bibfnamefont {W.}~\bibnamefont {Yang}},\ }\href
  {https://doi.org/10.1021/jp512727a} {\bibfield  {journal} {\bibinfo
  {journal} {J. Phys. Chem. A}\ }\textbf {\bibinfo {volume} {119}},\ \bibinfo
  {pages} {4923} (\bibinfo {year} {2015})}\BibitemShut {NoStop}%
\bibitem [{\citenamefont {Yang}\ \emph {et~al.}(2017)\citenamefont {Yang},
  \citenamefont {Dominguez}, \citenamefont {Zhang}, \citenamefont {Lutsker},
  \citenamefont {Niehaus}, \citenamefont {Frauenheim},\ and\ \citenamefont
  {Yang}}]{yang_charge_2017}%
  \BibitemOpen
  \bibfield  {author} {\bibinfo {author} {\bibfnamefont {Y.}~\bibnamefont
  {Yang}}, \bibinfo {author} {\bibfnamefont {A.}~\bibnamefont {Dominguez}},
  \bibinfo {author} {\bibfnamefont {D.}~\bibnamefont {Zhang}}, \bibinfo
  {author} {\bibfnamefont {V.}~\bibnamefont {Lutsker}}, \bibinfo {author}
  {\bibfnamefont {T.~A.}\ \bibnamefont {Niehaus}}, \bibinfo {author}
  {\bibfnamefont {T.}~\bibnamefont {Frauenheim}},\ and\ \bibinfo {author}
  {\bibfnamefont {W.}~\bibnamefont {Yang}},\ }\href
  {https://doi.org/10.1063/1.4977928} {\bibfield  {journal} {\bibinfo
  {journal} {J. Chem. Phys.}\ }\textbf {\bibinfo {volume} {146}},\ \bibinfo
  {pages} {124104} (\bibinfo {year} {2017})}\BibitemShut {NoStop}%
\bibitem [{\citenamefont {Li}\ \emph {et~al.}(2024{\natexlab{a}})\citenamefont
  {Li}, \citenamefont {Jin}, \citenamefont {Yu}, \citenamefont {Yang},\ and\
  \citenamefont {Zhu}}]{li_accurate_2024}%
  \BibitemOpen
  \bibfield  {author} {\bibinfo {author} {\bibfnamefont {J.}~\bibnamefont
  {Li}}, \bibinfo {author} {\bibfnamefont {Y.}~\bibnamefont {Jin}}, \bibinfo
  {author} {\bibfnamefont {J.}~\bibnamefont {Yu}}, \bibinfo {author}
  {\bibfnamefont {W.}~\bibnamefont {Yang}},\ and\ \bibinfo {author}
  {\bibfnamefont {T.}~\bibnamefont {Zhu}},\ }\href
  {https://doi.org/10.1021/acs.jpclett.4c00184} {\bibfield  {journal} {\bibinfo
   {journal} {J. Phys. Chem. Lett.}\ }\textbf {\bibinfo {volume} {15}},\
  \bibinfo {pages} {2757} (\bibinfo {year} {2024}{\natexlab{a}})}\BibitemShut
  {NoStop}%
\bibitem [{\citenamefont {Li}\ \emph {et~al.}(2024{\natexlab{b}})\citenamefont
  {Li}, \citenamefont {Jin}, \citenamefont {Yu}, \citenamefont {Yang},\ and\
  \citenamefont {Zhu}}]{li_particleparticle_2024}%
  \BibitemOpen
  \bibfield  {author} {\bibinfo {author} {\bibfnamefont {J.}~\bibnamefont
  {Li}}, \bibinfo {author} {\bibfnamefont {Y.}~\bibnamefont {Jin}}, \bibinfo
  {author} {\bibfnamefont {J.}~\bibnamefont {Yu}}, \bibinfo {author}
  {\bibfnamefont {W.}~\bibnamefont {Yang}},\ and\ \bibinfo {author}
  {\bibfnamefont {T.}~\bibnamefont {Zhu}},\ }\href
  {https://doi.org/10.1021/acs.jctc.4c00829} {\bibfield  {journal} {\bibinfo
  {journal} {J. Chem. Theory Comput.}\ }\textbf {\bibinfo {volume} {20}},\
  \bibinfo {pages} {7979} (\bibinfo {year} {2024}{\natexlab{b}})}\BibitemShut
  {NoStop}%
\bibitem [{\citenamefont {Yu}\ \emph {et~al.}(2025)\citenamefont {Yu},
  \citenamefont {Li}, \citenamefont {Zhu},\ and\ \citenamefont
  {Yang}}]{yu_accurate_2025}%
  \BibitemOpen
  \bibfield  {author} {\bibinfo {author} {\bibfnamefont {J.}~\bibnamefont
  {Yu}}, \bibinfo {author} {\bibfnamefont {J.}~\bibnamefont {Li}}, \bibinfo
  {author} {\bibfnamefont {T.}~\bibnamefont {Zhu}},\ and\ \bibinfo {author}
  {\bibfnamefont {W.}~\bibnamefont {Yang}},\ }\href
  {https://doi.org/10.1063/5.0251418} {\bibfield  {journal} {\bibinfo
  {journal} {J. Chem. Phys.}\ }\textbf {\bibinfo {volume} {162}},\ \bibinfo
  {pages} {094101} (\bibinfo {year} {2025})}\BibitemShut {NoStop}%
\bibitem [{\citenamefont {Wick}(1950)}]{wick_evaluation_1950}%
  \BibitemOpen
  \bibfield  {author} {\bibinfo {author} {\bibfnamefont {G.~C.}\ \bibnamefont
  {Wick}},\ }\href {https://doi.org/10.1103/PhysRev.80.268} {\bibfield
  {journal} {\bibinfo  {journal} {Phys. Rev.}\ }\textbf {\bibinfo {volume}
  {80}},\ \bibinfo {pages} {268} (\bibinfo {year} {1950})}\BibitemShut
  {NoStop}%
\bibitem [{\citenamefont {Metzner}(1991)}]{metzner_linked-cluster_1991}%
  \BibitemOpen
  \bibfield  {author} {\bibinfo {author} {\bibfnamefont {W.}~\bibnamefont
  {Metzner}},\ }\href {https://doi.org/10.1103/PhysRevB.43.8549} {\bibfield
  {journal} {\bibinfo  {journal} {Phys. Rev. B}\ }\textbf {\bibinfo {volume}
  {43}},\ \bibinfo {pages} {8549} (\bibinfo {year} {1991})}\BibitemShut
  {NoStop}%
\bibitem [{\citenamefont {Kutzelnigg}\ and\ \citenamefont
  {Mukherjee}(1997)}]{kutzelnigg1997normal}%
  \BibitemOpen
  \bibfield  {author} {\bibinfo {author} {\bibfnamefont {W.}~\bibnamefont
  {Kutzelnigg}}\ and\ \bibinfo {author} {\bibfnamefont {D.}~\bibnamefont
  {Mukherjee}},\ }\href@noop {} {\bibfield  {journal} {\bibinfo  {journal} {J.
  Chem. Phys.}\ }\textbf {\bibinfo {volume} {107}},\ \bibinfo {pages} {432}
  (\bibinfo {year} {1997})}\BibitemShut {NoStop}%
\bibitem [{\citenamefont {Hugenholtz}(1957)}]{hugenholtz_perturbation_1957}%
  \BibitemOpen
  \bibfield  {author} {\bibinfo {author} {\bibfnamefont {N.~M.}\ \bibnamefont
  {Hugenholtz}},\ }\href {https://doi.org/10.1016/S0031-8914(57)92950-6}
  {\bibfield  {journal} {\bibinfo  {journal} {Physica}\ }\textbf {\bibinfo
  {volume} {23}},\ \bibinfo {pages} {481} (\bibinfo {year} {1957})}\BibitemShut
  {NoStop}%
\bibitem [{\citenamefont {Rowe}(1968)}]{rowe_equations--motion_1968}%
  \BibitemOpen
  \bibfield  {author} {\bibinfo {author} {\bibfnamefont {D.~J.}\ \bibnamefont
  {Rowe}},\ }\href {https://doi.org/10.1103/RevModPhys.40.153} {\bibfield
  {journal} {\bibinfo  {journal} {Rev. Mod. Phys.}\ }\textbf {\bibinfo {volume}
  {40}},\ \bibinfo {pages} {153} (\bibinfo {year} {1968})}\BibitemShut
  {NoStop}%
\bibitem [{\citenamefont {Scuseria}\ \emph {et~al.}(2008)\citenamefont
  {Scuseria}, \citenamefont {Henderson},\ and\ \citenamefont
  {Sorensen}}]{scuseria_ground_2008}%
  \BibitemOpen
  \bibfield  {author} {\bibinfo {author} {\bibfnamefont {G.~E.}\ \bibnamefont
  {Scuseria}}, \bibinfo {author} {\bibfnamefont {T.~M.}\ \bibnamefont
  {Henderson}},\ and\ \bibinfo {author} {\bibfnamefont {D.~C.}\ \bibnamefont
  {Sorensen}},\ }\href {https://doi.org/10.1063/1.3043729} {\bibfield
  {journal} {\bibinfo  {journal} {J. Chem. Phys.}\ }\textbf {\bibinfo {volume}
  {129}},\ \bibinfo {pages} {231101} (\bibinfo {year} {2008})}\BibitemShut
  {NoStop}%
\bibitem [{\citenamefont {Fukuda}\ \emph {et~al.}(1964)\citenamefont {Fukuda},
  \citenamefont {Iwamoto},\ and\ \citenamefont
  {Sawada}}]{fukuda_linearized_1964}%
  \BibitemOpen
  \bibfield  {author} {\bibinfo {author} {\bibfnamefont {N.}~\bibnamefont
  {Fukuda}}, \bibinfo {author} {\bibfnamefont {F.}~\bibnamefont {Iwamoto}},\
  and\ \bibinfo {author} {\bibfnamefont {K.}~\bibnamefont {Sawada}},\ }\href
  {https://doi.org/10.1103/PhysRev.135.A932} {\bibfield  {journal} {\bibinfo
  {journal} {Phys. Rev.}\ }\textbf {\bibinfo {volume} {135}},\ \bibinfo {pages}
  {A932} (\bibinfo {year} {1964})}\BibitemShut {NoStop}%
\bibitem [{\citenamefont {Roos}\ \emph {et~al.}(1980)\citenamefont {Roos},
  \citenamefont {Taylor},\ and\ \citenamefont {Sigbahn}}]{roos_complete_1980}%
  \BibitemOpen
  \bibfield  {author} {\bibinfo {author} {\bibfnamefont {B.~O.}\ \bibnamefont
  {Roos}}, \bibinfo {author} {\bibfnamefont {P.~R.}\ \bibnamefont {Taylor}},\
  and\ \bibinfo {author} {\bibfnamefont {P.~E.}\ \bibnamefont {Sigbahn}},\
  }\href {https://doi.org/10.1016/0301-0104(80)80045-0} {\bibfield  {journal}
  {\bibinfo  {journal} {Chem. Phys.}\ }\textbf {\bibinfo {volume} {48}},\
  \bibinfo {pages} {157} (\bibinfo {year} {1980})}\BibitemShut {NoStop}%
\bibitem [{\citenamefont {Dyall}(1995)}]{dyall_choice_1995}%
  \BibitemOpen
  \bibfield  {author} {\bibinfo {author} {\bibfnamefont {K.~G.}\ \bibnamefont
  {Dyall}},\ }\href {https://doi.org/10.1063/1.469539} {\bibfield  {journal}
  {\bibinfo  {journal} {J. Chem. Phys.}\ }\textbf {\bibinfo {volume} {102}},\
  \bibinfo {pages} {4909} (\bibinfo {year} {1995})}\BibitemShut {NoStop}%
\bibitem [{\citenamefont {Peng}\ \emph {et~al.}(2013)\citenamefont {Peng},
  \citenamefont {Steinmann}, \citenamefont {van Aggelen},\ and\ \citenamefont
  {Yang}}]{peng_equivalence_2013}%
  \BibitemOpen
  \bibfield  {author} {\bibinfo {author} {\bibfnamefont {D.}~\bibnamefont
  {Peng}}, \bibinfo {author} {\bibfnamefont {S.~N.}\ \bibnamefont {Steinmann}},
  \bibinfo {author} {\bibfnamefont {H.}~\bibnamefont {van Aggelen}},\ and\
  \bibinfo {author} {\bibfnamefont {W.}~\bibnamefont {Yang}},\ }\href
  {https://doi.org/10.1063/1.4820556} {\bibfield  {journal} {\bibinfo
  {journal} {J. Chem. Phys.}\ }\textbf {\bibinfo {volume} {139}},\ \bibinfo
  {pages} {104112} (\bibinfo {year} {2013})}\BibitemShut {NoStop}%
\bibitem [{\citenamefont {Sun}\ \emph {et~al.}(2020)\citenamefont {Sun},
  \citenamefont {Zhang}, \citenamefont {Banerjee}, \citenamefont {Bao},
  \citenamefont {Barbry}, \citenamefont {Blunt}, \citenamefont {Bogdanov},
  \citenamefont {Booth}, \citenamefont {Chen}, \citenamefont {Cui},
  \citenamefont {Eriksen}, \citenamefont {Gao}, \citenamefont {Guo},
  \citenamefont {Hermann}, \citenamefont {Hermes}, \citenamefont {Koh},
  \citenamefont {Koval}, \citenamefont {Lehtola}, \citenamefont {Li},
  \citenamefont {Liu}, \citenamefont {Mardirossian}, \citenamefont {McClain},
  \citenamefont {Motta}, \citenamefont {Mussard}, \citenamefont {Pham},
  \citenamefont {Pulkin}, \citenamefont {Purwanto}, \citenamefont {Robinson},
  \citenamefont {Ronca}, \citenamefont {Sayfutyarova}, \citenamefont
  {Scheurer}, \citenamefont {Schurkus}, \citenamefont {Smith}, \citenamefont
  {Sun}, \citenamefont {Sun}, \citenamefont {Upadhyay}, \citenamefont {Wagner},
  \citenamefont {Wang}, \citenamefont {White}, \citenamefont {Whitfield},
  \citenamefont {Williamson}, \citenamefont {Wouters}, \citenamefont {Yang},
  \citenamefont {Yu}, \citenamefont {Zhu}, \citenamefont {Berkelbach},
  \citenamefont {Sharma}, \citenamefont {Sokolov},\ and\ \citenamefont
  {Chan}}]{sun_recent_2020}%
  \BibitemOpen
  \bibfield  {author} {\bibinfo {author} {\bibfnamefont {Q.}~\bibnamefont
  {Sun}}, \bibinfo {author} {\bibfnamefont {X.}~\bibnamefont {Zhang}}, \bibinfo
  {author} {\bibfnamefont {S.}~\bibnamefont {Banerjee}}, \bibinfo {author}
  {\bibfnamefont {P.}~\bibnamefont {Bao}}, \bibinfo {author} {\bibfnamefont
  {M.}~\bibnamefont {Barbry}}, \bibinfo {author} {\bibfnamefont {N.~S.}\
  \bibnamefont {Blunt}}, \bibinfo {author} {\bibfnamefont {N.~A.}\ \bibnamefont
  {Bogdanov}}, \bibinfo {author} {\bibfnamefont {G.~H.}\ \bibnamefont {Booth}},
  \bibinfo {author} {\bibfnamefont {J.}~\bibnamefont {Chen}}, \bibinfo {author}
  {\bibfnamefont {Z.-H.}\ \bibnamefont {Cui}}, \bibinfo {author} {\bibfnamefont
  {J.~J.}\ \bibnamefont {Eriksen}}, \bibinfo {author} {\bibfnamefont
  {Y.}~\bibnamefont {Gao}}, \bibinfo {author} {\bibfnamefont {S.}~\bibnamefont
  {Guo}}, \bibinfo {author} {\bibfnamefont {J.}~\bibnamefont {Hermann}},
  \bibinfo {author} {\bibfnamefont {M.~R.}\ \bibnamefont {Hermes}}, \bibinfo
  {author} {\bibfnamefont {K.}~\bibnamefont {Koh}}, \bibinfo {author}
  {\bibfnamefont {P.}~\bibnamefont {Koval}}, \bibinfo {author} {\bibfnamefont
  {S.}~\bibnamefont {Lehtola}}, \bibinfo {author} {\bibfnamefont
  {Z.}~\bibnamefont {Li}}, \bibinfo {author} {\bibfnamefont {J.}~\bibnamefont
  {Liu}}, \bibinfo {author} {\bibfnamefont {N.}~\bibnamefont {Mardirossian}},
  \bibinfo {author} {\bibfnamefont {J.~D.}\ \bibnamefont {McClain}}, \bibinfo
  {author} {\bibfnamefont {M.}~\bibnamefont {Motta}}, \bibinfo {author}
  {\bibfnamefont {B.}~\bibnamefont {Mussard}}, \bibinfo {author} {\bibfnamefont
  {H.~Q.}\ \bibnamefont {Pham}}, \bibinfo {author} {\bibfnamefont
  {A.}~\bibnamefont {Pulkin}}, \bibinfo {author} {\bibfnamefont
  {W.}~\bibnamefont {Purwanto}}, \bibinfo {author} {\bibfnamefont {P.~J.}\
  \bibnamefont {Robinson}}, \bibinfo {author} {\bibfnamefont {E.}~\bibnamefont
  {Ronca}}, \bibinfo {author} {\bibfnamefont {E.~R.}\ \bibnamefont
  {Sayfutyarova}}, \bibinfo {author} {\bibfnamefont {M.}~\bibnamefont
  {Scheurer}}, \bibinfo {author} {\bibfnamefont {H.~F.}\ \bibnamefont
  {Schurkus}}, \bibinfo {author} {\bibfnamefont {J.~E.~T.}\ \bibnamefont
  {Smith}}, \bibinfo {author} {\bibfnamefont {C.}~\bibnamefont {Sun}}, \bibinfo
  {author} {\bibfnamefont {S.-N.}\ \bibnamefont {Sun}}, \bibinfo {author}
  {\bibfnamefont {S.}~\bibnamefont {Upadhyay}}, \bibinfo {author}
  {\bibfnamefont {L.~K.}\ \bibnamefont {Wagner}}, \bibinfo {author}
  {\bibfnamefont {X.}~\bibnamefont {Wang}}, \bibinfo {author} {\bibfnamefont
  {A.}~\bibnamefont {White}}, \bibinfo {author} {\bibfnamefont {J.~D.}\
  \bibnamefont {Whitfield}}, \bibinfo {author} {\bibfnamefont {M.~J.}\
  \bibnamefont {Williamson}}, \bibinfo {author} {\bibfnamefont
  {S.}~\bibnamefont {Wouters}}, \bibinfo {author} {\bibfnamefont
  {J.}~\bibnamefont {Yang}}, \bibinfo {author} {\bibfnamefont {J.~M.}\
  \bibnamefont {Yu}}, \bibinfo {author} {\bibfnamefont {T.}~\bibnamefont
  {Zhu}}, \bibinfo {author} {\bibfnamefont {T.~C.}\ \bibnamefont {Berkelbach}},
  \bibinfo {author} {\bibfnamefont {S.}~\bibnamefont {Sharma}}, \bibinfo
  {author} {\bibfnamefont {A.~Y.}\ \bibnamefont {Sokolov}},\ and\ \bibinfo
  {author} {\bibfnamefont {G.~K.-L.}\ \bibnamefont {Chan}},\ }\href
  {https://doi.org/10.1063/5.0006074} {\bibfield  {journal} {\bibinfo
  {journal} {J. Chem. Phys.}\ }\textbf {\bibinfo {volume} {153}},\ \bibinfo
  {pages} {024109} (\bibinfo {year} {2020})}\BibitemShut {NoStop}%
\bibitem [{\citenamefont {White}(1992)}]{white_density_1992}%
  \BibitemOpen
  \bibfield  {author} {\bibinfo {author} {\bibfnamefont {S.~R.}\ \bibnamefont
  {White}},\ }\href {https://doi.org/10.1103/PhysRevLett.69.2863} {\bibfield
  {journal} {\bibinfo  {journal} {Phys. Rev. Lett.}\ }\textbf {\bibinfo
  {volume} {69}},\ \bibinfo {pages} {2863} (\bibinfo {year}
  {1992})}\BibitemShut {NoStop}%
\bibitem [{\citenamefont {Chan}\ and\ \citenamefont
  {Sharma}(2011)}]{chan_density_2011}%
  \BibitemOpen
  \bibfield  {author} {\bibinfo {author} {\bibfnamefont {G.~K.-L.}\
  \bibnamefont {Chan}}\ and\ \bibinfo {author} {\bibfnamefont {S.}~\bibnamefont
  {Sharma}},\ }\href {https://doi.org/10.1146/annurev-physchem-032210-103338}
  {\bibfield  {journal} {\bibinfo  {journal} {Annu. Rev. Phys. Chem.}\ }\textbf
  {\bibinfo {volume} {62}},\ \bibinfo {pages} {465} (\bibinfo {year}
  {2011})}\BibitemShut {NoStop}%
\bibitem [{\citenamefont {Xiang}\ \emph {et~al.}(2024)\citenamefont {Xiang},
  \citenamefont {Jia}, \citenamefont {Fang},\ and\ \citenamefont
  {Li}}]{xiang_distributed_2024}%
  \BibitemOpen
  \bibfield  {author} {\bibinfo {author} {\bibfnamefont {C.}~\bibnamefont
  {Xiang}}, \bibinfo {author} {\bibfnamefont {W.}~\bibnamefont {Jia}}, \bibinfo
  {author} {\bibfnamefont {W.-H.}\ \bibnamefont {Fang}},\ and\ \bibinfo
  {author} {\bibfnamefont {Z.}~\bibnamefont {Li}},\ }\href
  {https://doi.org/10.1021/acs.jctc.3c01228} {\bibfield  {journal} {\bibinfo
  {journal} {J. Chem. Theory Comput.}\ }\textbf {\bibinfo {volume} {20}},\
  \bibinfo {pages} {775} (\bibinfo {year} {2024})}\BibitemShut {NoStop}%
\bibitem [{\citenamefont {Zou}(2024)}]{zou_molecular_2024}%
  \BibitemOpen
  \bibfield  {author} {\bibinfo {author} {\bibfnamefont {J.}~\bibnamefont
  {Zou}},\ }\href {https://gitlab.com/jxzou/mokit} {\bibinfo {title} {Molecular
  {Orbital} {Kit} ({MOKIT})}} (\bibinfo {year} {2024})\BibitemShut {NoStop}%
\bibitem [{\citenamefont {Dunning}(1989)}]{dunning_gaussian_1989}%
  \BibitemOpen
  \bibfield  {author} {\bibinfo {author} {\bibfnamefont {T.~H.}\ \bibnamefont
  {Dunning}, \bibfnamefont {Jr.}},\ }\href {https://doi.org/10.1063/1.456153}
  {\bibfield  {journal} {\bibinfo  {journal} {J. Chem. Phys.}\ }\textbf
  {\bibinfo {volume} {90}},\ \bibinfo {pages} {1007} (\bibinfo {year}
  {1989})}\BibitemShut {NoStop}%
\bibitem [{\citenamefont {Nakatsuji}\ and\ \citenamefont
  {Yasuda}(1996)}]{nakatsuji_direct_1996}%
  \BibitemOpen
  \bibfield  {author} {\bibinfo {author} {\bibfnamefont {H.}~\bibnamefont
  {Nakatsuji}}\ and\ \bibinfo {author} {\bibfnamefont {K.}~\bibnamefont
  {Yasuda}},\ }\href {https://doi.org/10.1103/PhysRevLett.76.1039} {\bibfield
  {journal} {\bibinfo  {journal} {Phys. Rev. Lett.}\ }\textbf {\bibinfo
  {volume} {76}},\ \bibinfo {pages} {1039} (\bibinfo {year}
  {1996})}\BibitemShut {NoStop}%
\bibitem [{\citenamefont {Mazziotti}(1998)}]{mazziotti_approximate_1998}%
  \BibitemOpen
  \bibfield  {author} {\bibinfo {author} {\bibfnamefont {D.~A.}\ \bibnamefont
  {Mazziotti}},\ }\href {https://doi.org/10.1016/S0009-2614(98)00470-9}
  {\bibfield  {journal} {\bibinfo  {journal} {Chem. Phys. Lett.}\ }\textbf
  {\bibinfo {volume} {289}},\ \bibinfo {pages} {419} (\bibinfo {year}
  {1998})}\BibitemShut {NoStop}%
\bibitem [{\citenamefont {Kutzelnigg}\ and\ \citenamefont
  {Mukherjee}(1999)}]{kutzelnigg_cumulant_1999}%
  \BibitemOpen
  \bibfield  {author} {\bibinfo {author} {\bibfnamefont {W.}~\bibnamefont
  {Kutzelnigg}}\ and\ \bibinfo {author} {\bibfnamefont {D.}~\bibnamefont
  {Mukherjee}},\ }\href {https://doi.org/10.1063/1.478189} {\bibfield
  {journal} {\bibinfo  {journal} {J. Chem. Phys.}\ }\textbf {\bibinfo {volume}
  {110}},\ \bibinfo {pages} {2800} (\bibinfo {year} {1999})}\BibitemShut
  {NoStop}%
\bibitem [{\citenamefont {Hanauer}\ and\ \citenamefont
  {Köhn}(2012)}]{hanauer_meaning_2012}%
  \BibitemOpen
  \bibfield  {author} {\bibinfo {author} {\bibfnamefont {M.}~\bibnamefont
  {Hanauer}}\ and\ \bibinfo {author} {\bibfnamefont {A.}~\bibnamefont
  {Köhn}},\ }\href {https://doi.org/10.1016/j.chemphys.2011.09.024} {\bibfield
   {journal} {\bibinfo  {journal} {Chem. Phys.}\ }\textbf {\bibinfo {volume}
  {401}},\ \bibinfo {pages} {50} (\bibinfo {year} {2012})}\BibitemShut
  {NoStop}%
\bibitem [{\citenamefont {Misiewicz}\ \emph {et~al.}(2020)\citenamefont
  {Misiewicz}, \citenamefont {Turney},\ and\ \citenamefont
  {Schaefer}}]{misiewicz_reduced_2020}%
  \BibitemOpen
  \bibfield  {author} {\bibinfo {author} {\bibfnamefont {J.~P.}\ \bibnamefont
  {Misiewicz}}, \bibinfo {author} {\bibfnamefont {J.~M.}\ \bibnamefont
  {Turney}},\ and\ \bibinfo {author} {\bibfnamefont {H.~F.~I.}\ \bibnamefont
  {Schaefer}},\ }\href {https://doi.org/10.1021/acs.jctc.0c00422} {\bibfield
  {journal} {\bibinfo  {journal} {J. Chem. Theory Comput.}\ }\textbf {\bibinfo
  {volume} {16}},\ \bibinfo {pages} {6150} (\bibinfo {year}
  {2020})}\BibitemShut {NoStop}%
\end{thebibliography}%

\let\addcontentsline\oldaddcontentsline% Restore \addcontentsline

\pagebreak
\clearpage
\raggedbottom
\pagebreak
\widetext
\allowdisplaybreaks[4]
\begin{center}
\textbf{\large Supplemental material for \\
``A unified diagrammatic formulation of single-reference and multi-reference 
random phase approximations: the particle-hole and particle-particle channels"}\\
\vspace{2ex}
Yuqi Wang, Wei-Hai Fang, and Zhendong Li*

\vspace{2ex}
Key Laboratory of Theoretical and Computational Photochemistry, Ministry of Education, College of Chemistry, Beijing Normal University, Beijing 100875, China

% \vspace{2ex}
% zhendongli@bnu.edu.cn
\end{center}

\setcounter{secnumdepth}{3}
\setcounter{section}{0}
\setcounter{equation}{0}
\setcounter{figure}{0}
\setcounter{table}{0}
\setcounter{page}{1}
\makeatletter
\renewcommand{\theequation}{S\arabic{equation}}
\renewcommand{\thesection}{S\arabic{section}}
\renewcommand{\thefigure}{S\arabic{figure}}
\renewcommand{\thetable}{S\arabic{table}}

%\tableofcontents
\section{Details of the theoretical developments}
\subsection{Cumulant expansion of time-ordered Green's functions}
In probability theory and statistics, joint moments and cumulants of a multivariate distribution can be defined by generating functions. Let $\{X_i\}$ be a set of random variables, the
moment generating function for joint moments is defined as,
\begin{align}
    M(\{J_i\}) = \langle e^{\sum_i J_i X_i}\rangle = 1 + \sum_i J_i \langle X_i\rangle  + \frac{1}{2!}\sum_{ij}J_iJ_j\langle X_iX_j\rangle +\cdots, \label{eq:mmt}
\end{align}
where $\langle X_i\rangle$ and $\langle X_iX_j\rangle$ are moments of the distribution. 
Here, we use the notation $\langle X_i\rangle$ for the expection value of $X_i$, 
since we will generalize the definition of cumulants for expectation values of operators later. Joint cumulants are defined via the cumulant generating function
\begin{align}
    K(\{J_i\}) = \ln M(\{J_i\}) = \sum_i J_i \langle X_i \rangle_c + \frac{1}{2!}\sum_{ij}J_iJ_j\langle X_iX_j \rangle_c+\cdots.
\end{align}
The general relations between moments and cumulants are
\begin{align}
    \langle X_1\cdots X_n \rangle = \sum_{\pi}\prod_{I_k\in\pi}\langle X_i:i\in I_k\rangle_c,\label{eq:mmt-cmlt-var}
\end{align}
where $\pi$ represents a set partition of the set $\mathcal{I} = \{1,2,\cdots,n\}$ and $I_k$ represents the blocks of the partition
\begin{align}
    \pi = \{I_1,I_2,\cdots,I_{|\pi|}\},\quad 
    I_k = \{i^k_1,i^k_2,\cdots,i^k_{|I_k|}\}, 
    \quad i_j^k \in \mathcal{I},
\end{align}
where $|\pi|$ represents the length of the set partition. Explicit expressions of Eq. \eqref{eq:mmt-cmlt-var} for $n$ equal to 2 and 3 read
\begin{align}
    \langle X_1X_2\rangle &= \langle X_1X_2\rangle_c+\langle X_1\rangle_c\langle X_2\rangle_c,\\
    \langle X_1X_2X_3 \rangle &= \langle X_1X_2X_3\rangle_c\nonumber\\
                     &\quad + \langle X_1X_2\rangle_c\langle X_3\rangle_c + \langle X_1X_3\rangle_c\langle X_2\rangle_c+ \langle X_2X_3\rangle_c\langle X_1 \rangle_c\nonumber\\
                     &\quad + \langle X_1 \rangle_c\langle X_2 \rangle_c\langle X_3 \rangle_c.
\end{align}
These expressions can be inverted to express cumulants in terms of moments recursively. An important property of joint cumulants is that cumulants involving two or more
statistically independent random variables are zero.

We now generalize the definition of cumulants to expectation values of time-dependent second-quantized operators, viz., Green's functions. Now $\hat{X}_i$ is either a creation or annihilation operator, and $\langle\hat{O}\rangle$ is an expectation value over a given state with a fixed
particle number. $\hat{O}$ must contain an equal number of creation
and annihilation operators, otherwise, $\langle\hat{O}\rangle$ vanishes. We can generalize the definition of cumulants in Eq. \eqref{eq:mmt-cmlt-var} as
\begin{align}
    \langle \hat{X}_1\cdots\hat{X}_n\rangle  = \sum_{\pi_e}\epsilon(\pi_e)\prod_{I^e_k\in\pi_e}\langle\hat{X}_i:i\in I^e_k\rangle_c,\label{eq:mmt-cmlt-op}
\end{align}
where the set partition $\pi_e$ is defined as $\pi_e = \{I_1^e,I_2^e,\cdots,I_{|\pi_e|}^e\}$ with $I^e_k = \{i^k_1,i^k_2,\cdots,i^k_{|I^e_k|}\}$ ($i^k_1<i^k_2<\cdots<i^k_{|I^e_k|}$). 
The subscript/superscript `e’ indicates that the number of elements in $I^e_k$ is even. The ordering $i^k_1<i^k_2<\cdots<i^k_{|I^e_k|}$ reflects the operator nature of $\hat{X}_i$, and it allows
to uniquely determines the prefactor $\epsilon(\pi_e)$, which is given by the signature of the permutation obtained by flatten $\pi_e$. Since the numbers of elements in $I_e^k$ are all even, the ordering of $I_e$ in $\pi_e$ does not affect the value of $\epsilon(\pi_e)$. The generalization Eq. \eqref{eq:mmt-cmlt-op} is consistent with other ways of defining cumulants in the context of Green’s functions and reduced density matrix (RDM) theories\cite{nakatsuji_direct_1996,mazziotti_approximate_1998,kutzelnigg_cumulant_1999,hanauer_meaning_2012,misiewicz_reduced_2020}, e.g., obtained by generating functions with Grassmann variables ${J_i}$ in Eq. \eqref{eq:mmt}. Explicit expressions of Eq. \eqref{eq:mmt-cmlt-op} for $n$ equal to 2 and 4 are
\begin{align}
    \langle \hat{X}_1\hat{X}_2\rangle &= \langle \hat{X}_1\hat{X}_2\rangle_c,\\
    \langle \hat{X}_1\hat{X}_2\hat{X}_3\hat{X}_4\rangle &= \langle \hat{X}_1\hat{X}_2\hat{X}_3\hat{X}_4\rangle_c + \langle \hat{X}_1\hat{X}_2\rangle\langle\hat{X}_3\hat{X}_4\rangle - \langle \hat{X}_1\hat{X}_3\rangle\langle\hat{X}_2\hat{X}_4\rangle + \langle \hat{X}_1\hat{X}_4\rangle\langle\hat{X}_2\hat{X}_3\rangle.\label{eq:cmlt-op-4}
\end{align}
Some of the terms in Eq. \eqref{eq:cmlt-op-4} can be zero depending on the nature of $\hat{X}_i$, e.g.,
\begin{align}
    \langle \hp^\dagger(t_1)\hq^\dagger(t_2)\hr(t_3)\hs(t_4)\rangle &= \langle \hp^\dagger(t_1)\hq^\dagger(t_2)\hr(t_3)\hs(t_4)\rangle_c \nonumber\\
    &\quad + \langle \hp^\dagger(t_1)\hs(t_4)\rangle \langle\hq^\dagger(t_2)\hr(t_3)\rangle - \langle \hp^\dagger(t_1)\hr(t_3)\rangle \langle\hq^\dagger(t_2)\hs(t_4)\rangle,
\end{align}
and
\begin{align}
    \langle \hp^\dagger(t_1)\hr(t_2)\hq^\dagger(t_3)\hs(t_4)\rangle &= \langle \hp^\dagger(t_1)\hr(t_2)\hq^\dagger(t_3)\hs(t_4)\rangle_c\nonumber\\
    &\quad + \langle \hp^\dagger(t_1)\hr(t_2)\rangle\langle\hq^\dagger(t_3)\hs(t_4)\rangle + \langle \hp^\dagger(t_1)\hs(t_4)\rangle\langle \hr(t_2)\hq^\dagger(t_3)\rangle.
\end{align}
The restriction of even partition also enables a generalization to the time-ordered form, as each permutation introduces a factor of $\pm 1$ to all the terms simultaneously. The time-ordered form of the first examples reads
\begin{align}
    \langle \Tau{\hp^\dagger(t_1)\hq^\dagger(t_2)\hr(t_3)\hs(t_4)}\rangle &= \langle \Tau{\hp^\dagger(t_1)\hq^\dagger(t_2)\hr(t_3)\hs(t_4)}\rangle_c \nonumber\\
    &\quad + \langle \Tau{\hp^\dagger(t_1)\hs(t_4)}\rangle \langle\Tau{\hq^\dagger(t_2)\hr(t_3)}\rangle - \langle \Tau{\hp^\dagger(t_1)\hr(t_3)}\rangle \langle\Tau{\hq^\dagger(t_2)\hs(t_4)}\rangle,
\end{align}
which is equivalent to Eq. \eqref{eq:g2decomp} in the main text.
One important point is that the property of joint cumulants also holds for the generalization Eq. \eqref{eq:mmt-cmlt-op} in the sense that the generalized cumulant involving two or more operators corresponding to different noninteracting \emph{unentangled} subsystems vanishes.

\subsection{Detailed derivations for Eqs. \eqref{eq:pp-diag} - \eqref{eq:ecor-plas-pprpa}}
We start from Eq. \eqref{eq:ecor-pprpa-sum},
\begin{align}
    \Delta E^{\text{ppRPA}} = \sum_{n \ge 2}-\frac{1}{2\pi}\frac{1}{n}\int_{-\infty}^{\infty}d\omega\, \mathrm{tr}\left(\left[\frac{1}{4}\mathbf{\bar{g}}\mathbf{K}^0(\ii\omega)\right]^n\right).\nonumber
\end{align}
Using the identity $\frac{1}{n}=\int_0^1 \alpha^{n-1}d\alpha$, $\Delta E^{\text{ppRPA}}$ is rewritten as
\begin{align}
\Delta E^{\text{ppRPA}}=&-\sum_{n\ge 2}\int_0^1 d\alpha\,\int_{-\infty}^{\infty}\frac{d\omega}{2\pi}\,\alpha^{n-1}\mathrm{tr}\left(\left[\frac{1}{4}\mathbf{\bar{g}}\mathbf{K}^0(\ii\omega)\right]^n\right)\nonumber\\
=&-\int_0^1 d\alpha\int_{-\infty}^{\infty}\frac{d\omega}{2\pi}\,\mathrm{tr}\left(\frac{1}{4}\mathbf{\bar{g}}\mathbf{K}^0(\ii\omega)\sum_{n\ge 2}\left[\frac{\alpha}{4}\mathbf{\bar{g}}\mathbf{K}^0(\ii\omega)\right]^{n-1}\right)\nonumber\\
=&-\frac{1}{4}\int_0^1 d\alpha \int_{-\infty}^{\infty}\frac{d\omega}{2\pi}\,\mathrm{tr}\left(\mathbf{\bar{g}}\mathbf{K}^0(\ii\omega)\left[(\mathbf{I}-\frac{\alpha}{4}\mathbf{\bar{g}}\mathbf{K}^0(\ii\omega))^{-1}-\mathbf{I}\right]\right).\label{eq:ecor-pprpa-2}
\end{align}
By introducing an auxiliary variable $\mathbf{K}^\alpha$, which obeys a Dyson-like equation 
\begin{align}
\mathbf{K}^{\alpha}(z) \equiv \mathbf{K}^{0}(z)\left(\mathbf{I}-\frac{\alpha}{4}\mathbf{\bar{g}}\mathbf{K}^0(\ii\omega)\right)^{-1}=\mathbf{K}^{0}(z)+\frac{\alpha}{4}\mathbf{K}^{0}(z)\mathbf{\bar{g}}\mathbf{K}^{\alpha}(z),
\end{align}
we can convert Eq. \eqref{eq:ecor-pprpa-2} into a more compact form,
\begin{align}
\Delta E^{\text{ppRPA}}=&-\frac{1}{4}\int_0^1 d\alpha\int_{-\infty}^{\infty}\frac{d\omega}{2\pi}\,\mathrm{tr}\left(\mathbf{\bar{g}[K}^{\alpha}(\ii\omega)-\mathbf{K}^0(\ii\omega)]\right).\label{eq:ecorr-pprpa-lambda}
\end{align}
Further simplification can be made by defining two auxiliary matrices $\mathbf{V}$ and $\mathbf{D}$, analogous to those in phRPA\cite{wang_generalized_2025},
\begin{align}
   V^{11}_{PQ} &= \frac{1}{4}\langle \Phi^{N+2}_P| \hp^\dagger \hq^\dagger |\Phi_0\rangle \bar{g}_{pq,rs}\langle \Phi_0|\hs\hr|\Phi^{N+2}_Q\rangle,\\
   V^{12}_{PH} &= \frac{1}{4}\langle \Phi^{N+2}_P| \hp^\dagger \hq^\dagger |\Phi_0\rangle \bar{g}_{pq,rs}\langle \Phi^{N-2}_H|\hs\hr|\Phi_0\rangle,\\
   V^{21}_{HP} &= \frac{1}{4}\langle \Phi_0|\hp^\dagger \hq^\dagger |\Phi_H^{N-2}\rangle \bar{g}_{pq,rs}\langle \Phi_0|\hs\hr|\Phi_P^{N+2}\rangle,\\ 
   V^{22}_{HI} &= \frac{1}{4}\langle \Phi_0|\hp^\dagger \hq^\dagger |\Phi_H^{N-2}\rangle \bar{g}_{pq,rs} \langle \Phi_I^{N-2}|\hs\hr|\Phi_0\rangle,\\
\mathbf{V}&\equiv\begin{bmatrix}
  \mathbf{V}^{11} & \mathbf{V}^{12}\\
  \mathbf{V}^{21} & \mathbf{V}^{22}
\end{bmatrix},\\
  \mathbf{D}^0(z)&\equiv\left(z\begin{bmatrix}
  \mathbf{I}^+ & \mathbf{0} \\
  \mathbf{0} & -\mathbf{I}^-
\end{bmatrix}-
\begin{bmatrix}
  \boldsymbol{\omega}^+ & \mathbf{0}\\
  \mathbf{0} & \boldsymbol{\omega}^-
\end{bmatrix}\right)^{-1}, \label{eq:D0-def}
\end{align}
where $\boldsymbol{\omega}^{+}$ and $\boldsymbol{\omega}^{-}$ represent a collection of $\omega^{N+2}_{P}$ and $\omega^{N-2}_{H}$, respectively. With $N_{pp}$ and $N_{hh}$ denoting the numbers of $(N+2)$- and $(N-2)$-electron states, the sizes of $\mathbf{V}^{11}, \mathbf{V}^{12}, \mathbf{V}^{21}$ and $\mathbf{V}^{22}$ are $N_{pp}\times N_{pp}, N_{pp}\times N_{hh}, N_{hh}\times N_{pp}$ and $N_{hh}\times N_{hh}$, respectively. By definition, it is easy to see that $\mathbf{V}^{11}$ and $\mathbf{V}^{22}$ are Hermitian. A useful relation is obtained using the cyclic property of trace,
\begin{align}
   \mathrm{tr}\left(\left[\frac{1}{4}\mathbf{\bar{g}K}^0(z)\right]^n\right) = \mathrm{tr}\left(\left[\mathbf{VD}^0(z)\right]^n\right).
\end{align}
 Now we rewrite Eq. \eqref{eq:ecor-pprpa-sum} with the aid of $\mathbf{V}$ and $\mathbf{D}^0$,
\begin{align}
  \Delta E^{\text{ppRPA}} &=-\sum_{n\ge 2}\int_0^1 d\alpha\int_{-\infty}^{\infty}\frac{d\omega}{2\pi}\,\alpha^{n-1}\mathrm{tr}\left(\left[\frac{1}{4}\mathbf{\bar{g}}\mathbf{K}^0(\ii\omega)\right]^n\right)\nonumber\nonumber\\
  &=-\sum_{n\ge 2}\int_0^1 d\alpha\int_{-\infty}^{\infty}\frac{d\omega}{2\pi}\,\alpha^{n-1}\mathrm{tr}\left(\left[\mathbf{V}\mathbf{D}^0(\ii\omega)\right]^n\right)\nonumber\\
  &=-\int_0^1 d\alpha\int_{-\infty}^{\infty}\frac{d\omega}{2\pi}\,\mathrm{tr}\left(\mathbf{V}[\mathbf{D}^{\alpha}(\ii\omega)-\mathbf{D}^0(\ii\omega)]\right),\label{eq:ec-pp-intdw}
\end{align}
in which $\mathbf{D}^{\alpha}$ is defined as
\begin{align}
  \mathbf{D}^{\alpha}(z) \equiv \mathbf{D}^0(z) +\alpha \mathbf{D}^{0}(z)\mathbf{V}\mathbf{D}^{\alpha}(z).\label{eq:dyson-D}
\end{align}
Now we evaluate $\mathbf{D}^{\alpha}$ more explicitly in order to evaluate the double integral in Eq. \eqref{eq:ec-pp-intdw}. 
To this end, we substitute Eq. \eqref{eq:D0-def} into Eq. \eqref{eq:dyson-D},
\begin{align}
\left[\mathbf{D}^{\alpha}(z)\right]^{-1}+\alpha \mathbf{V} &=\left[\mathbf{D}^0(z)\right]^{-1} \nonumber\\
&=z\begin{bmatrix}
  \mathbf{I}^+ & \mathbf{0} \\
  \mathbf{0} & -\mathbf{I}^-
\end{bmatrix}-
\begin{bmatrix}
  \boldsymbol{\omega}^+ & \mathbf{0}\\
  \mathbf{0} & \boldsymbol{\omega}^-
\end{bmatrix}\nonumber\\
&\equiv z\mathbf{S}-\mathbf{\Delta},
\end{align}
such that
\begin{align}
  \mathbf{D^\alpha}(z) &=(z\mathbf{S}-\mathbf{\Delta}-\alpha\mathbf{V})^{-1}.
\end{align}
To evaluate the inverse, we solve an auxiliary eigenvalue problem
\begin{align}
    \mathbf{E}^\alpha \equiv \mathbf{\Delta}&+\alpha\mathbf{V} = \begin{bmatrix}
        \mathbf{A}^{+,\alpha} & \mathbf{C}^\alpha \\ \mathbf{C}^{\alpha,\dagger} & \mathbf{A}^{-,\alpha}
    \end{bmatrix},\label{eq:mat-ABC-lambda}\\
    \mathbf{E}^\alpha \begin{bmatrix}
        \mathbf{U}^{+,\alpha}  & \mathbf{U}^{-,\alpha}
    \end{bmatrix}&= \mathbf{S}\begin{bmatrix}
        \mathbf{U}^{+,\alpha} & \mathbf{U}^{-,\alpha}
    \end{bmatrix}\begin{bmatrix}
        \boldsymbol{\Omega}^{+,\alpha} & \mathbf{0} \\ \mathbf{0} & \boldsymbol{\Omega}^{-,\alpha}
    \end{bmatrix}.\label{eq:pp-diag-lambda}
\end{align}
where $\boldsymbol{\Omega}^{+(-),\alpha}$ contains all positive (negative) eigenvalues, 
which will be denoted as $\Omega_{P(H)}$, with corresponding eigenvectors as $\mathbf{u}_{P(H)}$, 
collected in $\mathbf{U}^{+(-),\alpha}$, viz., 
$\mathbf{U}^{+,\alpha}= [\cdots\mathbf{u}_P^\alpha\cdots], \mathbf{U}^{-,\alpha}= [\cdots\mathbf{u}_H^\alpha\cdots]$. 
$\mathbf{A}^{\pm}, \mathbf{C}$ and $\boldsymbol{\Omega}^{\pm}$ defined in Eqs. \eqref{eq:pp-diag}-\eqref{eq:mat-pp-2} 
correspond to the case with $\alpha=1$ in Eqs. \eqref{eq:mat-ABC-lambda} and \eqref{eq:pp-diag-lambda}. 
With $\mathbf{U}^\alpha \equiv [\mathbf{U}^{+,\alpha}\quad \mathbf{U}^{-,\alpha}]$, we have
\begin{align}
\mathbf{U}^{\alpha,\dagger}\mathbf{E}^\alpha \mathbf{U}^\alpha
= 
\begin{bmatrix}
\boldsymbol{\Omega}^{+,\alpha} & \mathbf{0} \\ 
\mathbf{0} & -\boldsymbol{\Omega}^{-,\alpha} \\
\end{bmatrix}.\label{eq:pp-diag-E}
\end{align}
The chemical potential $\mu$ is chosen to make Eq. \eqref{eq:mat-ABC-lambda} positive definite, 
so that we have the normalization condition
\begin{align}
\mathbf{U}^{\alpha,\dagger}\mathbf{S} \mathbf{U}^\alpha
=\begin{bmatrix}
\mathbf{I}^+ & \mathbf{0} \\ 
\mathbf{0} & -\mathbf{I}^- \\
\end{bmatrix}
=\mathbf{S},\label{eq:pp-diag-norm}
\end{align}
as has been used in the single reference theory\cite{peng_equivalence_2013}. With the positive definiteness of Eq. \eqref{eq:mat-ABC-lambda}, the numbers of its positive and negative eigenvalues (viz., $\Omega_P$ and $\Omega_H$) can be proven to be $N_{pp}$ and $N_{hh}$, respectively, following Ref. \cite{peng_equivalence_2013}.

\begin{figure}[hbt]
    \centering
    \includegraphics[width=0.4\textwidth]{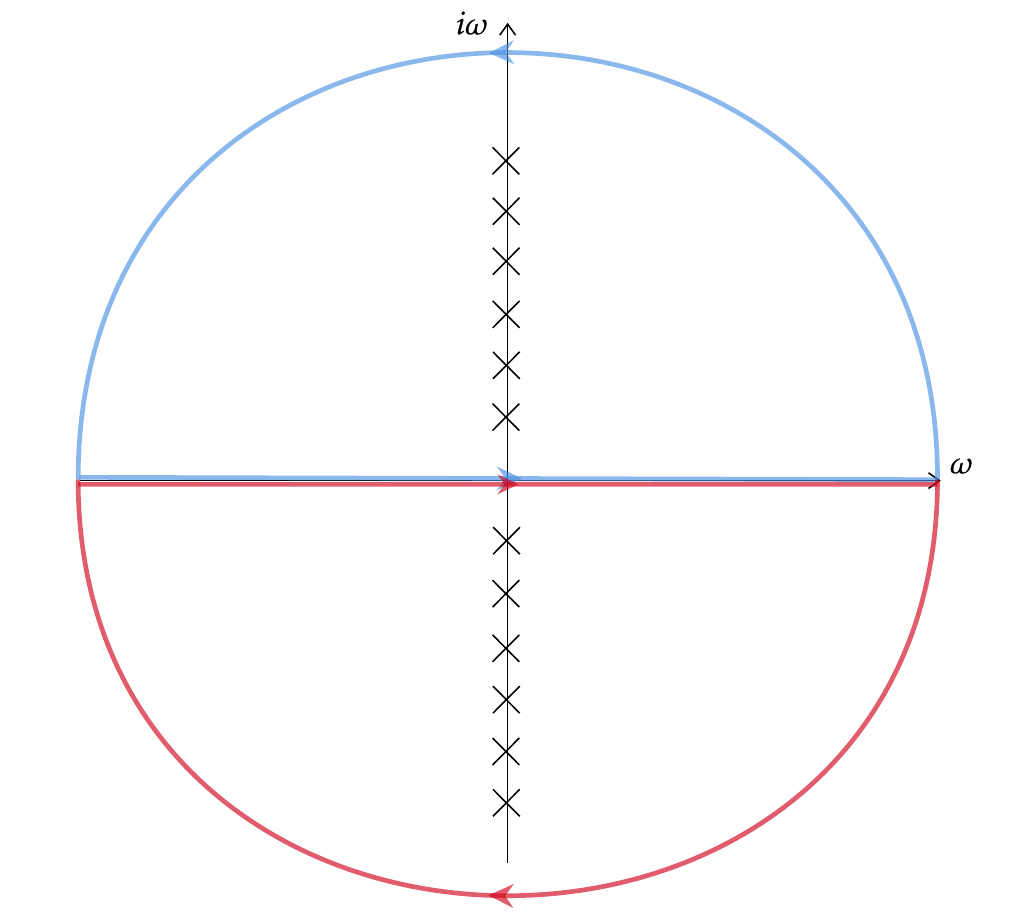}
    \caption{Contours for evaluating the MR-ppRPA correlation energy. Integrations along the red and blue contours result in Eqs. \eqref{eq:Epp-A} and \eqref{eq:Epp-I}, respectively.} 
    \label{fig:contour-pprpa}
\end{figure}

Using Eqs. \eqref{eq:pp-diag-E} and \eqref{eq:pp-diag-norm}, the spectral representation of $\mathbf{D}^\alpha(z)$ can now be expressed as
\begin{align}
\mathbf{D}^\alpha(z)=&-(\mathbf{E}^\alpha - z\mathbf{S})^{-1}=-\mathbf{U}^\alpha
\begin{bmatrix}
\boldsymbol{\Omega}^{+,\alpha}-z\mathbf{I}^+ & \mathbf{0} \\ 
\mathbf{0} & -\boldsymbol{\Omega}^{-,\alpha}+z\mathbf{I}^- \\
\end{bmatrix}^{-1}
\mathbf{U}^{\alpha,\dagger}\nonumber\\
=&-\sum_{P} \frac{\mathbf{u}_{P}^\alpha\mathbf{u}_{P}^{\alpha,\dagger}}{\Omega_{P}^\alpha - z}+
\sum_{H} \frac{\mathbf{u}_{H}^\alpha\mathbf{u}_{H}^{\alpha,\dagger}}{-\Omega_{H}^\alpha + z}.
\label{eq:DlambdaSpec}
\end{align}
We can use this expression to integrate out both $\alpha$ and $\omega$ in Eq. \eqref{eq:ec-pp-intdw} 
analytically. As Eq. \eqref{eq:ecor-pprpa-sum} indicates, $\Delta E^{\text{ppRPA}}$ starts at the second order, 
and thus first-order poles doesn't exist, enabling contour integrations. 
Substitute Eqs. \eqref{eq:D0-def} and \eqref{eq:DlambdaSpec} into Eq. \eqref{eq:ec-pp-intdw}, and we get
\begin{align}
    \Delta E^{\text{ppRPA}} =&\int_0^1 d\alpha\int_{-\infty}^{\infty}\frac{d\omega}{2\pi}\sum_{P}\left( \frac{\mathbf{u}_{P}^{\alpha,\dagger}\mathbf{V}\mathbf{u}_{P}^\alpha}{\Omega_{P}^\alpha - \ii \omega}+\frac{V^{11}_{PP}}{\ii\omega - (\omega^{N+2}_{P}-2\mu)}\right)+
\sum_{H}\left(\frac{\mathbf{u}_{H}^{\alpha,\dagger}\mathbf{V}\mathbf{u}_{H}^\alpha}{-\Omega_{H}^\alpha + \ii \omega}  -\frac{V^{22}_{HH}}{\ii\omega + (\omega^{N-2}_{H}+2\mu)}\right).
\end{align}
The first term has poles on the negative imaginary axis, 
while the second term has poles on the positive imaginary axis. 
Integrating along a contour enclosing the lower half-plane (shown in Fig. \ref{fig:contour-pprpa}), 
we get the final expression for $\Delta E^{\text{ppRPA}}$,
\begin{align}
    \Delta E^{\text{ppRPA}} &= \int_0^1 d\alpha\oint^{-}\frac{d\omega}{2\pi}\sum_{P}\left( \frac{\mathbf{u}_{P}^{\alpha,\dagger}\mathbf{V}\mathbf{u}_{P}^\alpha}{\Omega_{P}^\alpha - \ii \omega}+\frac{V^{11}_{PP}}{\ii\omega - (\omega^{N+2}_{P}-2\mu)}\right)\nonumber\\
    &=\int_0^1 d\alpha\sum_{P}\mathbf{u}_{P}^{\alpha,\dagger}\mathbf{V}\mathbf{u}_{P}^\alpha-\tr( \mathbf{V}^{11})\nonumber\\
    &=\int_0^1 d\alpha\sum_{P}\frac{d \Omega_{P}^\alpha}{d \alpha}-\tr( \mathbf{V}^{11})\nonumber\\
    &=\sum_P\left(\Omega_{P}^{\alpha=1} - \Omega_{P}^{\alpha=0}\right)-\tr( \mathbf{V}^{11})\nonumber\\
    &=\left(\sum_P \Omega_{P}\right) - \tr(\mathbf{A}^+),\label{eq:Epp-A}
\end{align}
where the Feynman-Hellman's theorem is applied in the third line. Another equivalent expression can be derived by choosing a contour enclosing the upper half-plane (shown in Fig. \ref{fig:contour-pprpa}), 
\begin{align}
\Delta E^{\text{ppRPA}} &= \int_0^1 d\alpha\oint^+\frac{d\omega}{2\pi}
\sum_{H}\left(\frac{\mathbf{u}_{H}^{\alpha,\dagger}\mathbf{V}\mathbf{u}_{H}^\alpha}{-\Omega_{H}^\alpha + \ii \omega}  -\frac{V^{22}_{HH}}{\ii\omega + (\omega^{N-2}_{H}+2\mu)}\right)\nonumber\\
&=\left(-\sum_H\Omega_{H}\right) - \tr(\mathbf{A}^-).\label{eq:Epp-I}
\end{align}
This completes the proof of Eq. \eqref{eq:ecor-plas-pprpa}.

\subsection{Matrix elements for MR-RPAx with a CASSCF reference}
Given a CAS($N_{\text{act}},M_{\text{act}}$) active space, $\hH_{\text{act}}$ is exactly diagonalized in the $(N_{\text{act}}+d)$-electron ($d \in \{0, \pm 1, \pm 2\}$) subspaces,
\begin{align}
    \hH_{\text{act}}|\Xi_\lambda^{N_{\text{act}}+d}\rangle = \mathcal{E}_\lambda^{N_{\text{act}}+d} |\Xi_\lambda^{N_{\text{act}}+d}\rangle. \label{eq:HA-eigen}
\end{align}

Introducing 3 transition density matrices for wavefunctions in the active space, 
\begin{align}
    \gamma^{[+1]}_{\lambda x} &= \langle \Xi_\lambda^{N_{\text{act}}+1} | \hx^\dagger|\Xi_0^{N_{\text{act}}}\rangle,\quad \gamma^{[-1]}_{\lambda x} = \langle \Xi_\lambda^{N_{\text{act}}-1} | \hx|\Xi_0^{N_{\text{act}}}\rangle,\quad \gamma^{[0]}_{\lambda xy} = \langle\Xi_\lambda^{N_{\text{act}}} |\hx^\dagger \hy|\Xi_0^{N_{\text{act}}}\rangle,
    \label{eq:tdm-1}
\end{align}
elements of $\mathbf{\bar{A}}$ and $\mathbf{\bar{B}}$ for the CASSCF wavefunction can be expressed more explicitly as
\begin{align}
\mathbf{\bar{A}} =
\begin{pmatrix}
[\bar{A}_{ai,bj}] & [\bar{A}_{ai,\sigma j}] & [\bar{A}_{ai,b\sigma}] & [\bar{A}_{ai,\sigma}] \\
[\bar{A}_{\lambda i,bj}] &[\bar{A}_{\lambda i,\sigma j}] & [\bar{A}_{\lambda i,b\sigma}] & [\bar{A}_{\lambda i,\sigma}] \\
[\bar{A}_{a\lambda,bj}] & [\bar{A}_{a\lambda,\sigma j}] & [\bar{A}_{a\lambda,b\sigma}] & [\bar{A}_{a\lambda,\sigma}] \\
[\bar{A}_{\lambda,bj}] &  [\bar{A}_{\lambda,\sigma j}] & [\bar{A}_{\lambda,b\sigma}] & [\bar{A}_{\lambda,\sigma}]\\
\end{pmatrix},\quad
% \end{eqnarray}
% \begin{eqnarray}
\mathbf{\bar{B}} =
\begin{pmatrix}
[\bar{B}_{ai,bj}] & [\bar{B}_{ai,\sigma j}] & [\bar{B}_{ai,b\sigma}] & [\bar{B}_{ai,\sigma}] \\
[\bar{B}_{\lambda i,bj}] &[\bar{B}_{\lambda i,\sigma j}] & [\bar{B}_{\lambda i,b\sigma}] & [\bar{B}_{\lambda i,\sigma}] \\
[\bar{B}_{a\lambda,bj}] & [\bar{B}_{a\lambda,\sigma j}] & [\bar{B}_{a\lambda,b\sigma}] & [\bar{B}_{a\lambda,\sigma}] \\
[\bar{B}_{\lambda,bj}] &  [\bar{B}_{\lambda,\sigma j}] & [\bar{B}_{\lambda,b\sigma}] & [\bar{B}_{\lambda,\sigma}]\\
\end{pmatrix}.\label{eq:AandB4by4}
\end{align}
By definition (see Eqs. \eqref{eq:mat-ph-A} and \eqref{eq:mat-ph-B}), $\mathbf{\bar{A}}$ is Hermitian while $\mathbf{\bar{B}}$ is symmetric. Thus, only 10 of 16 blocks (viz., upper/lower triangle blocks) are independent. Expressions for the lower triangle blocks of $\mathbf{\bar{A}}$ and $\mathbf{\bar{B}}$ are shown in Tab. \ref{tab:MR-RPAx}.

\begin{table}[htb]
    \centering
    \renewcommand{\arraystretch}{1.5}
    \setlength{\tabcolsep}{3.5mm}
    \caption{Matrix elements of $\mathbf{\bar{A}}$ and $\mathbf{\bar{B}}$ for MR-RPAx, where $\epsilon_\lambda^{[d]} = \mathcal{E}_{\lambda}^{N_{\text{act}}+d} - \mathcal{E}_0^{N_{\text{act}}}$ with $d \in \{+1,-1,0\}$.
    }
    \begin{tabular}{cccc}
    \hline\hline
         $|\Phi_L\rangle$ & $|\Phi_R\rangle$ & $\bar{A}_{LR}$ & $\bar{B}_{LR}$\\
    \hline
         $|\Theta_i^a\rangle|\Xi_{0}^{N_{\text{act}}}\rangle$ & $ |\Theta_j^b\rangle|\Xi_{0}^{N_{\text{act}}}\rangle$ & $\bar{A}_{ai,bj} = \erias{aj}{ib} + (\epsilon_a - \epsilon_i)\delta_{ij}\delta_{ab}$ & $\bar{B}_{ai,bj} = \erias{ab}{ij}$\\
    \hline
         \multirow{2}{*}{$|\Theta_i\rangle|\Xi_{\lambda}^{N_{\text{act}}+1}\rangle$} & 
         $|\Theta_j^b\rangle|\Xi_{0}^{N_{\text{act}}}\rangle$  & $\bar{A}_{\lambda i,bj}=(-1)\gamma^{[+1]}_{\lambda x}\erias{xj}{ib}$ & $\bar{B}_{\lambda i,bj}=(-1)\gamma^{[+1]}_{\lambda x}\erias{xb}{ij}$\\
         & $|\Theta_j\rangle|\Xi_{\sigma}^{N_{\text{act}}+1}\rangle$ & $\bar{A}_{\lambda i,\sigma j}=\gamma^{[+1]}_{\lambda x} \erias{xj}{iy} \gamma^{[+1]*}_{\sigma y}+(\epsilon_{\lambda}^{[+1]}-\epsilon_i)\delta_{ij}\delta_{\lambda\sigma}$ & $\bar{B}_{\lambda i,\sigma j} = \gamma^{[+1]}_{\lambda x}\erias{xy}{ij}\gamma^{[+1]}_{\sigma y}$\\
    \hline
        \multirow{3}{*}{$|\Theta^a\rangle|\Xi_{\lambda}^{N_{\text{act}}-1}\rangle$} & $|\Theta_j^b\rangle|\Xi_{0}^{N_{\text{act}}}\rangle$ & $\bar{A}_{a\lambda,bj}=\erias{aj}{xb}\gamma^{[-1]}_{\lambda x}$ & $\bar{B}_{a\lambda,bj}=\erias{ab}{xj}\gamma^{[-1]}_{\lambda x}$\\
        & $|\Theta_j\rangle|\Xi_{\sigma}^{N_{\text{act}}+1}\rangle$ & $\bar{A}_{a\lambda,\sigma j}=(-1)\gamma^{[-1]}_{\lambda x}\erias{aj}{xy}\gamma^{[+1]*}_{\sigma y}$  & $\bar{B}_{a\lambda,\sigma j}=(-1)\gamma^{[-1]}_{\lambda x}\erias{ay}{xj}\gamma^{[+1]}_{\sigma y}$\\
        & $|\Theta^b\rangle|\Xi_{\sigma}^{N_{\text{act}}-1}\rangle$ & $\bar{A}_{a\lambda,b\sigma}=\gamma^{[-1]}_{\lambda x}
        \erias{ay}{xb}\gamma^{[-1]*}_{\sigma y}+ (\epsilon_\lambda^{[-1]}+\epsilon_a)\delta_{ab}\delta_{\lambda\sigma}$ & $\bar{B}_{a\lambda,b\sigma}=\gamma^{[-1]}_{\lambda x}
        \erias{ab}{xy}\gamma^{[-1]}_{\sigma y}$\\
    \hline
        \multirow{4}{*}{$|\Theta_0\rangle|\Xi_{\lambda>0}^{N_{\text{act}}}\rangle$} & $|\Theta_j^b\rangle|\Xi_{0}^{N_{\text{act}}}\rangle$ & $\bar{A}_{\lambda,bj}=\gamma^{[0]}_{\lambda>0, xy}\erias{xj}{yb}$ & $\bar{B}_{\lambda,bj}=\gamma^{[0]}_{\lambda>0, xy}\erias{xb}{yj}$\\
        & $|\Theta_j\rangle|\Xi_{\sigma}^{N_{\text{act}}+1}\rangle$ & $\bar{A}_{\lambda,\sigma j}=\gamma^{[0]}_{\lambda>0, xy} \erias{xj}{yz}\gamma^{[+1]*}_{\sigma z}$ & $\bar{B}_{\lambda,\sigma j}=\gamma^{[0]}_{\lambda>0, xy} \erias{xz}{yj}\gamma^{[+1]}_{\sigma z}$ \\
        &$|\Theta^b\rangle|\Xi_{\sigma}^{N_{\text{act}}-1}\rangle$ & $\bar{A}_{\lambda,b\sigma}=\gamma^{[0]}_{\lambda>0,xy}\erias{xz}{yb}\gamma^{[-1]*}_{\sigma z}$ & $\bar{B}_{\lambda,b\sigma}=\gamma^{[0]}_{\lambda>0,xy}\erias{xb}{yz}\gamma^{[-1]}_{\sigma z}$\\
        &$|\Theta_0\rangle|\Xi_{\sigma>0}^{N_{\text{act}}}\rangle$ & $\bar{A}_{\lambda,\sigma}=\epsilon^{[0]}_{\lambda>0}\delta_{\lambda\sigma}$ & $\bar{B}_{\lambda,b\sigma}=0$\\
    \hline\hline
    \end{tabular} 
    \label{tab:MR-RPAx}
\end{table}

\subsection{Matrix elements for MR-ppRPA with a CASSCF reference}
MR-ppRPA matrix elements require two additional transition density matrices
\begin{align}
    \gamma^{[+2]}_{\lambda xy} &= \langle \Xi_\lambda^{N_{\text{act}}+2} | \hx^\dagger\hy^\dagger|\Xi_0^{N_{\text{act}}}\rangle,\quad \gamma^{[-2]}_{\lambda xy} = \langle \Xi_\lambda^{N_{\text{act}}-2} | \hx\hy|\Xi_0^{N_{\text{act}}}\rangle.
\end{align}
By definition (see Eqs. \eqref{eq:mat-pp-1} and \eqref{eq:mat-pp-3}), $\mathbf{A}^+$ and $\mathbf{A}^-$ are Hermitian. 
$\mathbf{A}^+$ has $3\times3$ blocks, but with both rows and columns corresponding to $(N+2)$-electron states, 
only 6 in 9 blocks are independent,
\begin{align}
    \mathbf{A}^+ = &\begin{bmatrix}
        [A^+_{ab,cd}] & [A^+_{ab,c\sigma}] & [A^+_{ab,\sigma}]\\
        [A^+_{a\lambda,cd}] & [A^+_{a\mu,c\sigma}] & [A^+_{a\lambda,\sigma}]\\
        [A^+_{\lambda,cd}] & [A^+_{\lambda,c\sigma}] & [A^+_{\lambda,\sigma}]\\
    \end{bmatrix}.
\end{align}
Explicit expressions of its lower triangle blocks are shown in Table \ref{tab:A}.
\begin{table}[hbt]
   \centering
    \renewcommand{\arraystretch}{1.5}
    \setlength{\tabcolsep}{5mm}
    \caption{Matrix elements of $\mathbf{A}^+$ in Eq. \eqref{eq:pp-diag} for MR-ppRPA, where $\epsilon_\lambda^{[d]} = \mathcal{E}_{\lambda}^{N_{\text{act}}+d} - \mathcal{E}_0^{N_{\text{act}}}$ with $d \in \{+1, +2\}$}
    \begin{tabular}{ccc}
    \hline\hline
       $|\Phi_P^{+2}\rangle$  & $|\Phi_Q^{+2}\rangle$ & $A^+_{PQ}$ \\
       \hline
       $|\Theta^{ab}\rangle|\Xi_0^{N_{\text{act}}}\rangle$ 
       & $|\Theta^{cd}\rangle|\Xi_0^{N_{\text{act}}}\rangle$ & $A^+_{ab,cd} = \erias{ab}{cd} + (\epsilon_a+\epsilon_b-2\mu)\delta_{ac}\delta_{bd}$\\
       \hline
       \multirow{2}{*}{$|\Theta^{a}\rangle|\Xi_\lambda^{N_{\text{act}}+1}\rangle$}
       & $|\Theta^{cd}\rangle|\Xi_0^{N_{\text{act}}}\rangle$ & $A^+_{a\lambda, cd} = \gamma^{[+1]}_{\lambda x}\erias{ax}{cd}$\\
       & $|\Theta^{c}\rangle|\Xi_\sigma^{N_{\text{act}}+1}\rangle$ & $A^+_{a\lambda,c\sigma} = \gamma^{[+1]}_{\lambda x}\erias{ax}{cy}\gamma^{[+1]*}_{\sigma y} + (\epsilon_\lambda^{[+1]}+\epsilon_a-2\mu)\delta_{ac}\delta_{\lambda\sigma}$\\
        \hline
       \multirow{3}{*}{$|\Theta_0\rangle|\Xi_\lambda^{N_{\text{act}}+2}\rangle$} 
       & $|\Theta^{cd}\rangle|\Xi_0^{N_{\text{act}}}\rangle$ & $A^+_{\lambda,cd} = \frac{1}{2}\gamma^{[+2]}_{\lambda xy}\erias{xy}{cd}$\\       
       & $|\Theta^{c}\rangle|\Xi_\sigma^{N_{\text{act}}+1}\rangle$ & $A^+_{\lambda,c\sigma} = \frac{1}{2}\gamma^{[+2]}_{\lambda xy}\erias{xy}{cz}\gamma^{[+1]*}_{\sigma z}$\\
       & $|\Theta_0\rangle|\Xi_\sigma^{N_{\text{act}}+2}\rangle$ & $A^+_{\lambda,\sigma} = (\epsilon_{\lambda}^{[+2]}-2\mu)\delta_{\lambda\sigma}$\\
       \hline\hline
    \end{tabular}
    \label{tab:A}
\end{table}

Similarly, $\mathbf{A}^-$ also have $3\times3$ blocks, but with both rows and columns corresponding to $(N-2)$-electron states, 
only 6 in 9 blocks are independent,
\begin{align}
    \mathbf{A}^-=&\begin{bmatrix}
        [A^-_{ij,kl}] & [A^-_{ij,\sigma k}] & [A^-_{ij,\sigma}]\\
        [A^-_{\lambda i,kl}] & [A^-_{\lambda i,\sigma k}] & [A^-_{\lambda i,\sigma}]\\
        [A^-_{\lambda,kl}] & [A^-_{\lambda,\sigma k}] & [A^-_{\lambda,\sigma}]\\
    \end{bmatrix}.
\end{align}
Explicit expressions of its lower triangle blocks are shown in Table \ref{tab:C}.

\begin{table}[hbt]
   \centering
    \renewcommand{\arraystretch}{1.5}
    \setlength{\tabcolsep}{5mm}
    \caption{Matrix elements of $\mathbf{A}^-$ in Eq. \eqref{eq:pp-diag} for MR-ppRPA, where $\epsilon_\lambda^{[d]} = \mathcal{E}_{\lambda}^{N_{\text{act}}+d} - \mathcal{E}_0^{N_{\text{act}}}$ with $d \in \{-1, -2\}$.}
    \begin{tabular}{ccc}
    \hline\hline
       $|\Phi_H^{-2}\rangle$  & $|\Phi_I^{-2}\rangle$ & $A^-_{HI}$ \\
    \hline
       $|\Theta_{ij}\rangle|\Xi_0^{N_{\text{act}}}\rangle$
       & $|\Theta_{kl}\rangle|\Xi_0^{N_{\text{act}}}\rangle$ & $A^-_{ij,kl} = \erias{ij}{kl} + (-\epsilon_i-\epsilon_j+2\mu)\delta_{ik}\delta_{jl}$\\
       \hline
       \multirow{2}{*}{$|\Theta_{i}\rangle|\Xi_\lambda^{N_{\text{act}}-1}\rangle$}
       & $|\Theta_{kl}\rangle|\Xi_0^{N_{\text{act}}}\rangle$& $A^-_{\lambda i,kl}=\gamma^{[-1]*}_{\lambda x}\erias{xi}{kl}$\\
       & $|\Theta_{k}\rangle|\Xi_\sigma^{N_{\text{act}}-1}\rangle$ & $A^-_{\lambda i,\sigma k} = \gamma^{[-1]*}_{\lambda x}\erias{xi}{yk} \gamma^{[-1]}_{\sigma y}+(\epsilon_\lambda^{[-1]}-\epsilon_i+2\mu)\delta_{\lambda\sigma}\delta_{ik}$\\
       \hline
       \multirow{3}{*}{$|\Theta_0\rangle|\Xi_\lambda^{N_{\text{act}}-2}\rangle$}
       &$|\Theta_{kl}\rangle|\Xi_0^{N_{\text{act}}}\rangle$ & $A^-_{\lambda,kl} = \frac{1}{2}\gamma^{[-2]*}_{\lambda yx}\erias{xy}{kl}$\\
       &$|\Theta_{k}\rangle|\Xi_\nu^{N_{\text{act}}-1}\rangle$ & $A^-_{\lambda,\sigma k} = \frac{1}{2}\gamma^{[-2]*}_{\lambda yx}\erias{xy}{zk}\gamma^{[-1]}_{\sigma z}$\\
       &$|\Theta_0\rangle|\Xi_\sigma^{N_{\text{act}}-2}\rangle$ & $A^-_{\lambda,\sigma} = (\epsilon_{\lambda}^{[-2]}+2\mu)\delta_{\lambda\sigma}$\\
    \hline\hline
    \end{tabular}
    \label{tab:C}
\end{table}

The $\mathbf{C}$ matrix in Eq. \eqref{eq:pp-diag} has $3 \times 3$ blocks, where the rows correspond to $(N+2)$-electron states (see Eq. \eqref{eq:CAS-ex-pp}), while the columns correspond to $(N-2)$ electron states (see Eq. \eqref{eq:CAS-ex-hh}),
\begin{align}
    \mathbf{C}=&\begin{bmatrix}
        [C_{ab,ij}]   & [C_{ab,\sigma i}]   & [C_{ab,\sigma}]\\
        [C_{a\lambda,ij}] & [C_{a\lambda,\sigma i}] & [C_{a\lambda,\sigma}]\\
        [C_{\lambda,ij}]  & [C_{\lambda,\sigma i}]  & [C_{\lambda,\sigma}]\\
    \end{bmatrix}.
\end{align}
Elements of $\mathbf{C}$ are summarized in Table \ref{tab:B}.
\begin{table}[hbt]
    \centering
    \renewcommand{\arraystretch}{1.5}
    \setlength{\tabcolsep}{5mm}
    \caption{Matrix elements of $\mathbf{C}$ in Eq. \eqref{eq:pp-diag} for MR-ppRPA.}
    \begin{tabular}{ccc}
    \hline\hline
    $|\Phi_P^{+2}\rangle$ & $|\Phi_H^{-2}\rangle$ & $C_{PH}$\\
    \hline
    \multirow{3}{*}{$|\Theta^{ab}\rangle|\Xi_0\rangle$} 
    & $|\Theta_{ij}\rangle|\Xi_0\rangle$         & $C_{ab,ij} = \langle ab || ij\rangle$\\
    & $|\Theta_{i}\rangle|\Xi_\sigma^{N_{\text{act}}-1}\rangle$  & $C_{ab,\sigma i} = \langle ab||xi\rangle \gamma^{[-1]}_{\sigma x} $\\
    & $|\Theta_0\rangle|\Xi_\sigma^{N_{\text{act}}-2}\rangle$      & $C_{ab,\sigma} =\frac{1}{2}\langle ab||yx\rangle \gamma^{[-2]}_{\sigma xy}$\\
    \hline
    \multirow{3}{*}{$|\Theta^{a}\rangle|\Xi_\lambda^{N_{\text{act}}+1}\rangle$}
    & $|\Theta_{ij}\rangle|\Xi_0\rangle $        & $C_{a\lambda,ij} = \gamma^{[+1]}_{\lambda x}\langle ax||ij\rangle$\\
    & $|\Theta_{i}\rangle|\Xi_\sigma^{N_{\text{act}}-1}\rangle$  & $C_{a\lambda,\sigma i} = \gamma^{[+1]}_{\lambda x}\langle ax||yi\rangle \gamma^{[-1]}_{\sigma y}$\\
    & $|\Theta_0\rangle|\Xi_\sigma^{N_{\text{act}}-2}\rangle$      & $C_{a\lambda,\sigma} = \frac{1}{2}\gamma^{[+1]}_{\lambda x}\langle ax||yz\rangle \gamma^{[-2]}_{\sigma zy}$\\
    \hline
    \multirow{3}{*}{$|\Theta_0\rangle|\Xi_\lambda^{N_{\text{act}}+2}\rangle$}
    & $|\Theta_{ij}\rangle|\Xi_0\rangle$         & $C_{\lambda,ij} = \frac{1}{2}\gamma^{[+2]}_{\lambda xy}\langle xy||ij\rangle$\\
    & $|\Theta_{i}\rangle|\Xi_\sigma^{N_{\text{act}}-1}\rangle$  & $C_{\lambda,\sigma i} = \frac{1}{2}\gamma^{[+2]}_{\lambda xy}\langle xy||zi\rangle \gamma^{[-1]}_{\sigma z}$\\
    & $|\Theta_0\rangle|\Xi_\sigma^{N_{\text{act}}-2}\rangle$      & $C_{\lambda,\sigma}=0$\\
    \hline\hline
    \end{tabular}
    \label{tab:B}
\end{table}

\newpage
\section{Numerical results of ph- and ppRPA}
\begin{table}[!hbt]
    \renewcommand{\arraystretch}{1.5}
    \setlength{\tabcolsep}{3mm}
\centering
\caption{Energies (in Hartree) for \ce{HF} calculated by different methods using the cc-pVDZ basis set. A CAS(2,2) active space is employed, which contains the \ce{\sigma} bonding orbital and its corresponding anti-bonding orbital.
DMRG and dRPA results are taken from Ref. \cite{wang_generalized_2025} for comparison.
}
\begin{tabular}{c|cccccc}
\hline\hline
\makecell{$R/R_{0}$\\($R_{0}=0.92$\,\AA)}  &  RHF & SR-dRPA & SR-RPAx & SR-ppRPA & DMRG($D=3000$) & CASSCF \\ 
\hline
\num{ 0.5} & \num{-99.037350} & \num{-99.247510} & \num{-99.303019} & \num{-99.180062} & \num{-99.226582} & \num{-99.045218} \\
\num{ 0.7} & \num{-99.846510} & \num{-100.065123} & \num{-100.132831} & \num{-99.995459} & \num{-100.045505} & \num{-99.859759} \\
\num{ 0.8} & \num{-99.965524} & \num{-100.187395} & \num{-100.262725} & \num{-100.116858} & \num{-100.168712} & \num{-99.981771} \\
\num{ 0.9} & \num{-100.011410} & \num{-100.236300} & \num{-100.320793} & \num{-100.164922} & \num{-100.218668} & \num{-100.031102} \\
\num{ 1.0} & \num{-100.019289} & \num{-100.247052} & \num{-100.342963} & \num{-100.174827} & \num{-100.230595} & \num{-100.042969} \\
\num{ 1.1} & \num{-100.007652} & \num{-100.238157} & \num{-100.349097} & \num{-100.165061} & \num{-100.223039} & \num{-100.035934} \\
\num{ 1.3} & \num{-99.960690} & \num{-100.196387} & \num{-100.365610} & \num{-100.121440} & \num{-100.184769} & \num{-100.000381} \\
\num{ 1.5} & \num{-99.906552} & \num{-100.147393} & / & \num{-100.070389} & \num{-100.141062} & \num{-99.961200} \\
\num{ 2.0} & \num{-99.791264} & \num{-100.046706} & / & \num{-99.963596} & \num{-100.064761} & \num{-99.898113} \\
\num{ 2.5} & \num{-99.712320} & \num{-99.985346} & / & \num{-99.894959} & \num{-100.037268} & \num{-99.877407} \\
\num{ 3.0} & \num{-99.660363} & \num{-99.952043} & / & \num{-99.854054} & \num{-100.030649} & \num{-99.872507} \\
\num{ 4.0} & \num{-99.605617} & \num{-99.929457} & / & \num{-99.818985} & \num{-100.028894} & \num{-99.871164} \\
\num{ 5.0} & \num{-99.582258} & \num{-99.928427} & / & \num{-99.810520} & \num{-100.028801} & \num{-99.871142} \\
\hline
       $R/R_{0}$  & MR-dRPA & MR-dRPA-e & MR-RPAx & MR-RPAx-e &MR-ppRPA & SC-NEVPT2\\
\hline
\num{ 0.5} & \num{-99.247139} & \num{-99.220070} & \num{-99.294864} & \num{-99.264554} & \num{-99.181907} & \num{-99.217297} \\
\num{ 0.7} & \num{-100.065515} & \num{-100.037276} & \num{-100.123016} & \num{-100.087499} & \num{-100.000282} & \num{-100.037388} \\
\num{ 0.8} & \num{-100.188955} & \num{-100.159593} & \num{-100.252339} & \num{-100.213143} & \num{-100.123990} & \num{-100.161545} \\
\num{ 0.9} & \num{-100.239403} & \num{-100.209400} & \num{-100.309396} & \num{-100.266367} & \num{-100.174941} & \num{-100.212514} \\
\num{ 1.0} & \num{-100.251928} & \num{-100.221892} & \num{-100.329002} & \num{-100.282171} & \num{-100.188265} & \num{-100.225299} \\
\num{ 1.1} & \num{-100.244971} & \num{-100.215570} & \num{-100.329251} & \num{-100.278740} & \num{-100.182414} & \num{-100.218264} \\
\num{ 1.3} & \num{-100.207681} & \num{-100.181391} & \num{-100.305332} & \num{-100.247813} & \num{-100.148125} & \num{-100.179693} \\
\num{ 1.5} & \num{-100.164728} & \num{-100.143239} & \num{-100.272895} & \num{-100.208618} & \num{-100.108865} & \num{-100.134262} \\
\num{ 2.0} & \num{-100.091087} & \num{-100.082038} & / & \num{-100.134984} & \num{-100.043724} & \num{-100.054030} \\
\num{ 2.5} & \num{-100.065490} & \num{-100.063000} & / & \num{-100.109144} & \num{-100.021888} & \num{-100.025955} \\
\num{ 3.0} & \num{-100.059323} & \num{-100.058774} & / & \num{-100.103388} & \num{-100.016584} & \num{-100.019335} \\
\num{ 4.0} & \num{-100.057630} & \num{-100.057614} & / & \num{-100.101955} & \num{-100.014990} & \num{-100.017564} \\
\num{ 5.0} & \num{-100.057594} & \num{-100.057594} & / & \num{-100.101966} & \num{-100.014884} & \num{-100.017523} \\
\hline\hline
\end{tabular}
\label{tab:HF-data}
\end{table}

\begin{table}[!hbt]
    \renewcommand{\arraystretch}{1.5}
    \setlength{\tabcolsep}{3mm}
\centering
\caption{Energies (in Hartree) for \ce{ScH} calculated by different methods using the cc-pVDZ basis set. A CAS(4,4) active space is employed, which contains four $\sigma$ orbitals with $4s$(Sc), $3d_{z^2}$(Sc), $4p_z$(Sc), and $1s$(H) characters.
DMRG and dRPA results are taken from Ref. \cite{wang_generalized_2025} for comparison.
}
\begin{tabular}{c|cccccc}
\hline\hline
\makecell{$R/R_{0}$\\($R_{0}=1.7754$\,\AA)}  &  RHF & SR-dRPA & SR-RPAx & SR-ppRPA & DMRG($D=5000$) & CASSCF\\ 
\hline
\num{ 0.5} & \num{-759.814970} & \num{-760.075719} & / & \num{-759.979273} & \num{-760.072540} & \num{-759.839407} \\
\num{ 0.7} & \num{-760.176429} & \num{-760.443046} & / & \num{-760.343855} & \num{-760.441998} & \num{-760.205446} \\
\num{ 0.8} & \num{-760.238776} & \num{-760.504323} & / & \num{-760.404235} & \num{-760.504011} & \num{-760.269098} \\
\num{ 0.9} & \num{-760.265992} & \num{-760.530390} & / & \num{-760.429702} & \num{-760.531006} & \num{-760.298274} \\
\num{ 1.0} & \num{-760.272795} & \num{-760.536114} & / & \num{-760.435075} & \num{-760.537945} & \num{-760.307941} \\
\num{ 1.1} & \num{-760.267976} & \num{-760.530204} & / & \num{-760.428993} & \num{-760.533548} & \num{-760.306926} \\
\num{ 1.3} & \num{-760.242574} & \num{-760.502830} & / & \num{-760.401490} & \num{-760.510792} & \num{-760.291380} \\
\num{ 1.5} & \num{-760.211464} & \num{-760.470236} & / & \num{-760.368764} & \num{-760.485791} & \num{-760.272084} \\
\num{ 2.0} & \num{-760.145156} & \num{-760.402671} & / & \num{-760.300222} & \num{-760.455003} & \num{-760.247383} \\
\num{ 2.5} & \num{-760.100375} & \num{-760.362400} & / & \num{-760.256825} & \num{-760.452157} & \num{-760.247097} \\
\num{ 3.0} & \num{-760.069926} & \num{-760.342147} & / & \num{-760.231089} & \num{-760.452048} & \num{-760.247364} \\
\num{ 4.0} & \num{-760.035322} & \num{-760.336212} & / & \num{-760.211704} & \num{-760.451755} & \num{-760.247415} \\
\num{ 5.0} & \num{-760.021261} & \num{-760.342312} & / & \num{-760.209453} & \num{-760.451720} & \num{-760.247408} \\
\hline
       $R/R_{0}$ & MR-dRPA & MR-dRPA-e & MR-RPAx & MR-RPAx-e & MR-ppRPA & SC-NEVPT2\\
\hline
\num{ 0.5} & \num{-760.077704} & \num{-760.059072} & \num{-760.226618} & \num{-760.188651} & \num{-759.991535} & \num{-760.037822} \\
\num{ 0.7} & \num{-760.447312} & \num{-760.429119} & \num{-760.644181} & \num{-760.597076} & \num{-760.360183} & \num{-760.407394} \\
\num{ 0.8} & \num{-760.509164} & \num{-760.492750} & \num{-760.710207} & \num{-760.661035} & \num{-760.422655} & \num{-760.469129} \\
\num{ 0.9} & \num{-760.536231} & \num{-760.521588} & \num{-760.732929} & \num{-760.681135} & \num{-760.450469} & \num{-760.495850} \\
\num{ 1.0} & \num{-760.543399} & \num{-760.530226} & \num{-760.739702} & \num{-760.676811} & \num{-760.458716} & \num{-760.502526} \\
\num{ 1.1} & \num{-760.539359} & \num{-760.527303} & / & \num{-760.660315} & \num{-760.456136} & \num{-760.498037} \\
\num{ 1.3} & \num{-760.517241} & \num{-760.506661} & / & \num{-760.617690} & \num{-760.437546} & \num{-760.475540} \\
\num{ 1.5} & \num{-760.492899} & \num{-760.483186} & / & \num{-760.581737} & \num{-760.416310} & \num{-760.451464} \\
\num{ 2.0} & \num{-760.463417} & \num{-760.458691} & / & \num{-760.571647} & \num{-760.391690} & \num{-760.424320} \\
\num{ 2.5} & \num{-760.463243} & \num{-760.458578} & / & \num{-760.574314} & \num{-760.390719} & \num{-760.422611} \\
\num{ 3.0} & \num{-760.463403} & \num{-760.458832} & / & \num{-760.573918} & \num{-760.390777} & \num{-760.422679} \\
\num{ 4.0} & \num{-760.463080} & \num{-760.458538} & / & \num{-760.573540} & \num{-760.390423} & \num{-760.422412} \\
\num{ 5.0} & \num{-760.463027} & \num{-760.458487} & / & \num{-760.573528} & \num{-760.390342} & \num{-760.422368} \\
\hline\hline
\end{tabular}
\label{tab:ScH-data}
\end{table}

\begin{table}[!hbt]
    \renewcommand{\arraystretch}{1.5}
    \setlength{\tabcolsep}{3mm}
\centering
\caption{Energies (in Hartree) for the symmetric dissociation of \ce{H2O} calculated by different methods using the cc-pVDZ basis set. A CAS(4,4) active space is employed, which contains two \ce{\sigma} bonding orbitals and their corresponding anti-bonding orbitals. The \ce{H-O-H} angle is set as $104.5^\circ$.
DMRG and dRPA results are taken from Ref. \cite{wang_generalized_2025} for comparison.
}
\begin{tabular}{c|cccccc}
\hline\hline
\makecell{$R/R_{0}$\\($R_{0}=0.98$\,\AA)}  &  RHF & SR-dRPA & SR-RPAx & SR-ppRPA & DMRG($D=4000$) & CASSCF\\ 
\hline
\num{ 0.5} & \num{-74.337842} & \num{-74.539306} & \num{-74.592882} & \num{-74.470209} & \num{-74.518465} & \num{-74.365099} \\
\num{ 0.7} & \num{-75.735199} & \num{-75.952082} & \num{-76.025301} & \num{-75.877552} & \num{-75.932779} & \num{-75.767150} \\
\num{ 0.8} & \num{-75.940396} & \num{-76.162834} & \num{-76.249059} & \num{-76.086208} & \num{-76.144987} & \num{-75.978368} \\
\num{ 0.9} & \num{-76.016417} & \num{-76.243882} & \num{-76.346645} & \num{-76.165336} & \num{-76.227878} & \num{-76.061334} \\
\num{ 1.0} & \num{-76.024735} & \num{-76.257168} & \num{-76.382176} & \num{-76.176737} & \num{-76.243453} & \num{-76.077771} \\
\num{ 1.1} & \num{-75.998350} & \num{-76.235872} & \num{-76.392919} & \num{-76.153520} & \num{-76.225042} & \num{-76.060842} \\
\num{ 1.3} & \num{-75.905126} & \num{-76.153397} & \num{-76.467740} & \num{-76.066993} & \num{-76.151054} & \num{-75.991498} \\
\num{ 1.5} & \num{-75.801910} & \num{-76.061905} & / & \num{-75.971275} & \num{-76.073200} & \num{-75.920454} \\
\num{ 2.0} & \num{-75.586476} & \num{-75.880576} & / & \num{-75.780594} & \num{-75.951410} & \num{-75.816722} \\
\num{ 2.5} & \num{-75.473027} & \num{-75.741577} & / & \num{-75.621427} & \num{-75.916998} & \num{-75.791230} \\
\num{ 3.0} & \num{-75.438091} & \num{-75.720454} & / & \num{-75.589970} & \num{-75.911884} & \num{-75.787125} \\
\num{ 4.0} & \num{-75.415806} & \num{-75.722055} & / & \num{-75.584101} & \num{-75.910369} & \num{-75.786129} \\
\num{ 5.0} & \num{-75.406816} & \num{-75.727292} & / & \num{-75.587644} & \num{-75.910307} & \num{-75.786072} \\
\hline
       $R/R_{0}$ & MR-dRPA & MR-dRPA-e & MR-RPAx & MR-RPAx-e & MR-ppRPA & SC-NEVPT2\\
\hline
\num{ 0.5} & \num{-74.542543} & \num{-74.504061} & \num{-74.582794} & \num{-74.534452} & \num{-74.480897} & \num{-74.505417} \\
\num{ 0.7} & \num{-75.954161} & \num{-75.914938} & \num{-76.007148} & \num{-75.948161} & \num{-75.888958} & \num{-75.915535} \\
\num{ 0.8} & \num{-76.165876} & \num{-76.125512} & \num{-76.225832} & \num{-76.160503} & \num{-76.101022} & \num{-76.127744} \\
\num{ 0.9} & \num{-76.248831} & \num{-76.207775} & \num{-76.316706} & \num{-76.243885} & \num{-76.184625} & \num{-76.210959} \\
\num{ 1.0} & \num{-76.264974} & \num{-76.223826} & \num{-76.342100} & \num{-76.260537} & \num{-76.201687} & \num{-76.227072} \\
\num{ 1.1} & \num{-76.247560} & \num{-76.207068} & \num{-76.335559} & \num{-76.243857} & \num{-76.185446} & \num{-76.209332} \\
\num{ 1.3} & \num{-76.176226} & \num{-76.139610} & \num{-76.289501} & \num{-76.174664} & \num{-76.117291} & \num{-76.136527} \\
\num{ 1.5} & \num{-76.100826} & \num{-76.070603} & \num{-76.237409} & \num{-76.100544} & \num{-76.046183} & \num{-76.058740} \\
\num{ 2.0} & \num{-75.983211} & \num{-75.970434} & / & \num{-75.979442} & \num{-75.940776} & \num{-75.935722} \\
\num{ 2.5} & \num{-75.950743} & \num{-75.947790} & / & \num{-75.948546} & \num{-75.914436} & \num{-75.902153} \\
\num{ 3.0} & \num{-75.945102} & \num{-75.944513} & / & \num{-75.944520} & \num{-75.909546} & \num{-75.896442} \\
\num{ 4.0} & \num{-75.943705} & \num{-75.943696} & / & \num{-75.944005} & \num{-75.907899} & \num{-75.895061} \\
\num{ 5.0} & \num{-75.943625} & \num{-75.943625} & / & \num{-75.944030} & \num{-75.907626} & \num{-75.894978} \\
\hline\hline
\end{tabular}
\label{tab:H2O-data}
\end{table}

\begin{table}[!hbt]
    \renewcommand{\arraystretch}{1.5}
    \setlength{\tabcolsep}{3mm}
\centering
\caption{Energies (in Hartree) for \ce{N2} calculated by different methods using the cc-pVDZ basis set A CAS(6,6) active space is employed, which contains one \ce{\sigma} bonding orbital, two \ce{\pi} bonding orbitals and their corresponding anti-bonding orbitals.
DMRG and dRPA results are taken from Ref. \cite{wang_generalized_2025} for comparison.}
\begin{tabular}{c|cccccc}
\hline\hline
\makecell{$R/R_{0}$\\$(R_{0}=1.095$\,\AA)}  &  RHF & SR-dRPA & SR-RPAx & SR-ppRPA & DMRG($D=5000$) & CASSCF\\ 
\hline
\num{ 0.5} & \num{-104.709478} & \num{-104.944935} & \num{-105.017612} & \num{-104.857453} & \num{-104.921795} & \num{-104.755679} \\
\num{ 0.7} & \num{-108.201116} & \num{-108.467782} & \num{-108.596127} & \num{-108.374073} & \num{-108.452883} & \num{-108.274242} \\
\num{ 0.8} & \num{-108.718875} & \num{-109.003033} & \num{-109.179260} & \num{-108.905375} & \num{-108.993482} & \num{-108.809885} \\
\num{ 0.9} & \num{-108.918011} & \num{-109.219711} & \num{-109.472692} & \num{-109.118841} & \num{-109.217148} & \num{-109.029622} \\
\num{ 1.0} & \num{-108.954475} & \num{-109.274228} & \num{-109.675594} & \num{-109.171173} & \num{-109.280520} & \num{-109.089749} \\
\num{ 1.1} & \num{-108.911368} & \num{-109.250078} & / & \num{-109.146112} & \num{-109.267179} & \num{-109.073551} \\
\num{ 1.3} & \num{-108.741863} & \num{-109.121579} & / & \num{-109.020689} & \num{-109.166253} & \num{-108.967708} \\
\num{ 1.5} & \num{-108.591863} & \num{-108.917784} & / & \num{-108.802407} & \num{-109.067669} & \num{-108.866501} \\
\num{ 2.0} & \num{-108.426530} & \num{-108.732518} & / & \num{-108.600492} & \num{-108.971457} & \num{-108.780466} \\
\num{ 2.5} & \num{-108.340021} & \num{-108.664534} & / & \num{-108.522683} & \num{-108.962430} & \num{-108.777293} \\
\num{ 3.0} & \num{-108.283688} & \num{-108.631855} & / & \num{-108.482127} & \num{-108.960981} & \num{-108.777144} \\
\num{ 4.0} & \num{-108.228319} & \num{-108.614918} & / & \num{-108.456011} & \num{-108.960281} & \num{-108.776848} \\
\num{ 5.0} & \num{-108.210453} & \num{-108.618005} & / & \num{-108.456204} & \num{-108.960236} & \num{-108.776829} \\
\hline
       $R/R_{0}$  &  MR-dRPA & MR-dRPA-e & MR-RPAx & MR-RPAx-e & MR-ppRPA & SC-NEVPT2\\
\hline
\num{ 0.5} & \num{-104.940086} & \num{-104.909255} & \num{-104.974238} & \num{-104.901196} & \num{-104.869753} & \num{-104.894212} \\
\num{ 0.7} & \num{-108.466240} & \num{-108.436042} & \num{-108.519960} & \num{-108.440017} & \num{-108.396123} & \num{-108.418145} \\
\num{ 0.8} & \num{-109.007164} & \num{-108.975927} & \num{-109.071773} & \num{-108.981642} & \num{-108.935728} & \num{-108.958446} \\
\num{ 0.9} & \num{-109.231450} & \num{-109.199275} & \num{-109.307501} & \num{-109.204651} & \num{-109.159092} & \num{-109.182184} \\
\num{ 1.0} & \num{-109.295442} & \num{-109.262474} & \num{-109.383980} & \num{-109.265318} & \num{-109.222805} & \num{-109.245822} \\
\num{ 1.1} & \num{-109.282650} & \num{-109.249106} & \num{-109.385350} & \num{-109.247088} & \num{-109.210460} & \num{-109.232890} \\
\num{ 1.3} & \num{-109.182427} & \num{-109.148843} & \num{-109.322855} & \num{-109.129613} & \num{-109.113975} & \num{-109.132800} \\
\num{ 1.5} & \num{-109.084088} & \num{-109.053318} & / & \num{-109.006109} & \num{-109.024732} & \num{-109.034361} \\
\num{ 2.0} & \num{-108.987507} & \num{-108.978330} & / & \num{-108.882454} & \num{-108.956603} & \num{-108.937432} \\
\num{ 2.5} & \num{-108.979241} & \num{-108.977609} & / & \num{-108.880393} & \num{-108.953239} & \num{-108.929422} \\
\num{ 3.0} & \num{-108.978091} & \num{-108.977821} & / & \num{-108.881630} & \num{-108.951927} & \num{-108.928351} \\
\num{ 4.0} & \num{-108.977514} & \num{-108.977510} & / & \num{-108.882418} & \num{-108.950485} & \num{-108.927791} \\
\num{ 5.0} & \num{-108.977478} & \num{-108.977479} & / & \num{-108.882661} & \num{-108.950170} & \num{-108.927753} \\
\hline\hline
\end{tabular}
\label{tab:N2-data}
\end{table}

\end{document}